\documentclass[useAMS,usenatbib,twocolumn]{mn2e}
\pdfoutput=1
\usepackage{amssymb}
\usepackage{amsmath}
\usepackage{natbib}
\usepackage[pdftex]{graphicx}
\usepackage{subfigure}
\usepackage{booktabs}
\usepackage{tabularx}
\usepackage{float}
\usepackage{afterpage}
\begin{document}
\label{firstpage}
\title[Elliptical instability]
{Nonlinear tides in a homogeneous rotating planet or star: global modes and elliptical instability} 
        \author[A. J. Barker, H. J. Braviner \& G. I. Ogilvie]
        {Adrian J. Barker\thanks{Email address: ajb268@cam.ac.uk}, Harry J. Braviner \& Gordon I. Ogilvie \\
Department of Applied Mathematics and Theoretical Physics, University of Cambridge, Centre for Mathematical Sciences, \\ Wilberforce Road, Cambridge CB3 0WA, UK}
\date{}
\pagerange{\pageref{firstpage}--\pageref{lastpage}} \pubyear{2016}
\maketitle

\begin{abstract}
We revisit the global modes and instabilities of homogeneous rotating ellipsoidal fluid masses, which are the simplest global models of rotationally and tidally deformed gaseous planets or stars. The tidal flow in a short-period planet may be unstable to the elliptical instability, a hydrodynamic instability that can drive tidal evolution. We perform a global (and local WKB) analysis to study this instability using the elegant formalism of Lebovitz \& Lifschitz. We survey the parameter space of global instabilities with harmonic orders $\ell\leq 5$, for planets with spins that are purely aligned (prograde) or anti-aligned (retrograde) with their orbits. In general, the instability has a much larger growth rate if the planetary spin and orbit are anti-aligned rather than aligned. We have identified a violent instability for anti-aligned spins outside of the usual frequency range for the elliptical instability (when $\frac{n}{\Omega}\lesssim -1$, where $n$ and $\Omega$ are the orbital and spin angular frequencies, respectively) if the tidal amplitude is sufficiently large. We also explore the instability in a rigid ellipsoidal container, which is found to be quantitatively similar to that with a realistic free surface. Finally, we study the effect of rotation and tidal deformation on mode frequencies. We find that larger rotation rates and larger tidal deformations both decrease the frequencies of the prograde sectoral surface gravity modes. This increases the prospect of their tidal excitation, potentially enhancing the tidal response over expectations from linear theory. In a companion paper, we use our results to interpret global simulations of the elliptical instability.
\end{abstract}
	
\section{Introduction}

Gravitational tidal interactions are thought to play a crucial role in shaping the properties of short-period extrasolar planetary systems and close binary stars (e.g.~\citealt{Ogilvie2014}). Tidal dissipation inside hot Jupiters (with orbital periods shorter than 10 days) is thought to play an important role in their formation (e.g.~\citealt{WuLithwick2011,Naoz2011,Petrovich2015,Anderson2015}) as well as in explaining their preferentially circular orbits compared with planets that orbit more distantly, which have a wide range of eccentricities (e.g.~\citealt{Rasio1996,WF2015}). The mechanisms responsible for tidal dissipation are not fully understood, though much progress has been made over the past decade \citep{Gio2004,Wu2005b,IvanovPap2007,Gio2007,GoodmanLackner2009,Remus2012,FBBO2014,Ogilvie2014}.

Tidal deformations of the shortest-period extrasolar planets can have much larger amplitudes than those of the giant planets of our Solar system. For example, the dimensionless amplitude of the tide in WASP-19 b (\citealt{Hebb2010}; a roughly Jupiter-mass planet that orbits a Sun-like star in only 0.79 days) due to its host star ($0.05$, using Eq.~\ref{deftidalamp} below) is nearly 5 orders of magnitude larger than the tide in Jupiter due to Io ($2\times10^{-7}$). This can no longer be considered to be a small parameter, meaning that we must consider nonlinear tidal effects. This is unfortunate, because previous work has primarily focused on studying tides in the linear regime (e.g.~\citealt{Gio2004,Wu2005b,IvanovPap2007,Ogilvie2013}). While these calculations are a necessary first step, and are likely to provide important information to help us to understand the tidal response in real bodies, they do not probe any nonlinear fluid effects which could significantly modify the outcome of tidal forcing.
Nonlinear tides are particularly difficult to study because they tend to require numerical simulations to quantify, so that some progress towards understanding them has only been made very recently (e.g.~\citealt{GoodmanLackner2009,Barker2010,Weinberg2012,BL2013,BL2014,FBBO2014}).
 
Given the complexity of the tidal response in the fluid layers of rotating planets and stars, it is instructive to consider simplified models that capture the most important elements, before attempting to understand a realistic model. In this paper, and its companion \citep{Barker2015b}, we present a detailed study of the linear and nonlinear evolution of tidal flows in the simplest global model of a planet or star: a rotating homogeneous fluid body that is subjected to the tidal gravity of its companion. This model has the significant advantage that the basic non-wavelike tidal flow is a nonlinear solution that is valid for any amplitude (and which is steady in the case of a circular orbit with an aligned, or anti-aligned, spin). This allows us to study its stability for finite tidal deformations --  indeed, there already exists a significant body of work on the stability of ellipsoidal fluid bodes that we can utilise \citep{L1989a,L1989b,LL1996,LL1996a}. This simple model is also unique in that if we consider only the lowest order (quadrupolar i.e. $\ell=2$) tidal potential, there is no direct tidal forcing of inertial modes \citep{GoodmanLackner2009,Ogilvie2009,Ogilvie2013}, which makes it the cleanest way in which to study the elliptical instability in a global model.

The elliptical instability is a hydrodynamic instability of elliptical streamlines that excites inertial waves through parametric resonance \citep{Kerswell2002}. This is a nonlinear mechanism of tidal dissipation that requires a finite tidal deformation. The outcome of this instability has been studied in laboratory experiments \citep{Lacaze2004,LeBars2007,LeBars2010}, as well as idealised local \citep{BL2013,BL2014} and global numerical simulations \citep{Cebron2010,Cebron2013}. These works suggest that the elliptical instability could contribute to tidal dissipation at sufficiently short orbital periods. However, global aspects of the elliptical instability that are relevant for tidal dissipation in gaseous planets or stars have not been explored in detail, and the nonlinear outcome of the instability when global modes are excited is not yet clear. We study the properties of this instability in this paper and its nonlinear evolution in the companion paper \citep{Barker2015b}. Our aim is to quantify the astrophysical importance of this instability as a mechanism for tidal dissipation.

Previous work by \cite{Wu2005b} and \cite{Ogilvie2009,Ogilvie2013} has studied linear tides in homogeneous spherical fluid bodies (neglecting centrifugal effects) using linear theory, and more recently \cite{Braviner2014,BravinerOgilvie2015} studied linear tides in homogeneous spheroidal bodies with finite rotational deformations. Here we focus on (nonlinear) tides in ellipsoidal bodies with finite tidal and rotational deformations, which is motivated by the large-amplitude tides inside the shortest-period hot Jupiters. In this paper, we present the results of a global analysis of the modes and instabilities of our model planet (or star). In a companion paper \citep{Barker2015b}, we perform the first global hydrodynamical simulations to study tidal flows in an ellipsoidal fluid body with a realistic free surface. We first present our simplified model in \S~\ref{model}, before presenting the methods used for our global analysis in \S~\ref{LSA}. Our main results are presented in \S~\ref{elliptical} \& \ref{MS}, before we finish with a discussion and conclusion. A complementary local WKB analysis is presented in Appendix \ref{WKB}.

\section{Simplified model}
\label{model}

We study tides in a rotationally and tidally deformed homogeneous rotating planet (or star) consisting of incompressible fluid. We adopt Cartesian coordinates $(x,y,z)$ centred on the planet of mass $m_p$ and radius\footnote{Of an equivalent isolated non-rotating planet, or alternatively, the mean equatorial radius $R=\sqrt{\frac{a^2+b^2}{2}}$.} $R$ such that the $z$-axis is aligned with its spin axis, about which the angular velocity of the fluid is $\Omega$. The planet orbits the star in a circular orbit with orbital angular velocity $\boldsymbol{n}$. Here we consider the orbit and spin to be aligned or purely anti-aligned, so that $\boldsymbol{n}=n\boldsymbol{e}_z$, where $n$ can take either sign. This restriction is made so that there exists a frame in which the equilibrium shape of the ellipsoid is fixed (in the absence of instabilities), which is convenient for numerical simulations, as well as the analysis in this paper. (For completeness, we note that certain stationary configurations can also exist with a misaligned spin and orbit, as pointed out by \citealt{Aizenman1968}.) We work in the frame that rotates with the orbit at the rate $\boldsymbol{n}$, in which the tidal bulge is stationary. Our governing equations are (taking the density $\rho\equiv1$),
\begin{eqnarray}
\label{Eq1}
&&\left(\partial_{t} + \boldsymbol{u}\cdot \nabla\right)\boldsymbol{u} + 2\boldsymbol{n}\times \boldsymbol{u} = -\nabla \Pi, \\
&& \nabla \cdot \boldsymbol{u} = 0, \\
&& \Pi = p+\Phi-\frac{1}{2}|\boldsymbol{n}\times\boldsymbol{x}|^{2}+\Psi,
\label{Eq3}
\end{eqnarray}
where $p$ is a pressure, $\Phi$ is the gravitational potential of the planet and $\Psi$ is an imposed tidal potential due to the star.

In the absence of a companion and any internal motions, the planet is in equilibrium if it is spherical (with radius $R$) and is maintained by a balance between fluid pressure and a radial gravitational acceleration due to the potential
\begin{eqnarray}
\Phi(\boldsymbol{x})=\frac{1}{2}\omega_{d}^{2}r^2,
\end{eqnarray}
where $r^2=x^2+y^2+z^2$, and the dynamical frequency is $\omega_d= \sqrt{\frac{Gm_p}{R^3}}$. This is our adopted \textit{constant} spherically-symmetric gravitational potential. With rotation, the tidally unperturbed body is effectively a Maclaurin spheroid, but we neglect self-gravity for computational convenience. The omission of self-gravity makes the model algebraically simpler and should only lead to moderate quantitative (but not qualitative) differences, which we briefly discuss in Appendix \ref{SG}. We have adopted this model for simplicity, but many of its qualitative properties are expected to carry over to more realistic interior models.

The tidal potential (restricted to the lowest order terms, i.e., those arising from the the second order terms in Eq.2 in \citealt{Ogilvie2014}) is
\begin{eqnarray}
\Psi = \frac{A_{\psi}}{2} \left( r^2 - 3 (\hat{\boldsymbol{a}_\star}\cdot \boldsymbol{x})^2\right),
\end{eqnarray}
where $\hat{\boldsymbol{a}_\star}=(1,0,0)$ defines the direction to the star, which is stationary in the bulge frame because the orbit is circular. (Note that there is no linear term because we are working in the centre of mass frame of the planet.) Its amplitude is $A_\psi = \frac{Gm_{\star}}{a_\star^{3}}$, where $a_\star$ is the distance to the secondary. 
We define
\begin{eqnarray}
\label{deftidalamp}
A&=& \frac{m_{\star}}{m_{p}}\left(\frac{R}{a_\star}\right)^3=\frac{A_\psi}{\omega_d^2},
\end{eqnarray}
which is a measure of the dimensionless tidal amplitude. Throughout this paper we adopt units of length and time such that $R=1$ and $\omega_d=1$.

Our planet occupies a volume $V$, whose surface is deformed by centrifugal forces and tidal gravity into a triaxial ellipsoid described by
\begin{eqnarray}
\label{bdry}
\frac{x^2}{a^2}+\frac{y^2}{b^2}+\frac{z^2}{c^2}=1,
\end{eqnarray}
where $a,b$ and $c$ are the semi-axes, which are stationary in the bulge ($\boldsymbol{n}$) frame. The tidal flow in this frame is
\begin{eqnarray}
\label{basicflow}
\boldsymbol{U}_{0}(\boldsymbol{x})=\gamma\left(-\frac{a}{b}y,\frac{b}{a}x,0\right), 
\end{eqnarray}
where $\gamma=\Omega-n$, which is an exact inviscid solution that is steady (in the absence of instabilities) and results from the non-synchronous rotation of the fluid (whenever $\Omega\ne n$) in its ellipsoidal volume. This flow has elliptical streamlines, so an infinitesimal perturbation will drive the elliptical instability for certain choices of $A,\Omega$ and $n$. Studying this instability is the primary aim of this paper. 

For a given choice of the three input parameters ($A,\Omega,n$), the equilibrium shape of the body is determined by (see Appendix~\ref{ST})
\begin{eqnarray}
\label{shape1}
\epsilon &=& \frac{3A}{2(1-\gamma^2-n^2)-A}, \\
c^2 &=& \frac{2\left[ (2A+\gamma^2+n^2-1)(A-\gamma^2-n^2+1)+f\right]}{(A+1)(A+2(\gamma^2+n^2-1))},
\label{shape2}
\end{eqnarray}
with
\begin{eqnarray}
f=2\gamma n \sqrt{(1-2A-\gamma^2-n^2)(1+A-\gamma^2-n^2)},
\end{eqnarray}
and where $\epsilon$ is a measure of the tidal deformation defined by $a=\sqrt{1+\epsilon}$ and $b=\sqrt{1-\epsilon}$. Note that $\epsilon\approx \frac{3A}{2}$ for small $\gamma$, $n$ and $A$.
The tidally unperturbed body ($A=n=0$) is ``Maclaurin spheroid"-like with
\begin{eqnarray}
c^2=1-\Omega^2,
\end{eqnarray}
so that in these units $\Omega$ is equal to the eccentricity of the spheroid. When $A\ne 0$ (and $\gamma\ne 0$), our body is similar to a Roche-Riemann ellipsoid \citep{C1987}.

While this model is highly simplified and will not accurately represent the tidal response in a giant planet or star, it has the enormous benefit that Eq.~\ref{basicflow} is an exact nonlinear solution. This allows its stability to be studied for any tidal amplitude, which would not be feasible  for realistic models of stellar and planetary tidal flows. Studying this simplified model helps us to understand processes that are likely to play a role in real bodies. In addition, there is a significant body of work that has explored the stability of such flows which we now turn to exploit. The equilibrium and some aspects of the stability of a similar model have been analysed by \cite{Aizenman1968} and an elegant formalism that can be used to study its stability to global modes has been presented by \cite{L1989a,L1989b}. The corresponding global and local stability of a similar model (without a tidal deformation but with self-gravity) has been previously studied by \cite{LL1996,LL1996a}.

\section{Linear stability analysis}
\label{LSA}

If we introduce small perturbations to the tidal flow (Eq.~\ref{basicflow}), it will become unstable for many choices of the input parameters. In this section we revisit the linear stability of this flow. Our motivation is to provide a more detailed study of the instability than has been presented by \cite{LL1996}, focusing on those aspects that are relevant for astrophysical tides. We also aim to study the modes that we might obtain in our global simulations that we present in the companion paper \citep{Barker2015b}. We will map out the parameter space for the global instabilities with $\ell\leq 5$, since these modes might be thought to be the most important for controlling the tidal dissipation (by providing an ``outer scale" for the turbulence, as evidenced by the local numerical simulations of \citealt{BL2013}).

\subsection{Method}
We follow \cite{LL1996} in adopting a Lagrangian perturbation theory, rather than an Eulerian one, since this is ideally suited to capturing motions that perturb the free surface\footnote{A Lagrangian perturbation theory has the disadvantage that it contains physically insignificant particle-relabelling modes. However, these do not trouble our stability analysis because they have zero growth rates \citep{LL1996}.}. This also has the significant advantage that much hard work in devising the formalism has already been carried out. The Lagrangian theory is elucidated in \cite{L1989a,L1989b} and has been applied to the stability of Riemann S-type ellipsoids (with no tidal potential) by \cite{LL1996}. The equations describing linearised Lagrangian perturbations (working in the bulge frame) are
\begin{eqnarray}
\label{LPE}
&& \partial_t^2 \boldsymbol{\xi}+\mathrm{A}\, \partial_t \boldsymbol{\xi} + \mathrm{B}\, \boldsymbol{\xi} + \nabla \Delta p = 0, \\
&&\nabla \cdot \boldsymbol{\xi}=0,
\end{eqnarray} 
where $\boldsymbol{\xi}$ is the Lagrangian displacement, $\Delta p$ is the Lagrangian pressure perturbation (which vanishes on the boundary), and
\begin{eqnarray}
\mathrm{A}\,\partial_t \boldsymbol{\xi} &=& 2\left(\boldsymbol{U}_{0}\cdot \nabla\partial_t \boldsymbol{\xi}+\boldsymbol{n}\times \partial_t\boldsymbol{\xi}\right), \\
\nonumber
\mathrm{B}\,\boldsymbol{\xi} &=& \left(\boldsymbol{U}_{0}\cdot \nabla\right)^2  \boldsymbol{\xi} + 2\left(\boldsymbol{U}_{0}\cdot \nabla\right) \boldsymbol{n}\times \boldsymbol{\xi} - \nabla \left(\boldsymbol{\xi}\cdot \nabla p\right) \\ && -\nabla \delta \Phi - \boldsymbol{\xi}\cdot \nabla \left(\boldsymbol{U}_{0}\cdot\nabla\boldsymbol{U}_{0} + 2\boldsymbol{n}\times \boldsymbol{U}_{0} \right).
\end{eqnarray}
The Eulerian velocity perturbation $\boldsymbol{u}$ is related to the Lagrangian displacement $\boldsymbol{\xi}$ by
\begin{eqnarray}
\boldsymbol{u}=\partial_t \boldsymbol{\xi} + \boldsymbol{U}_0\cdot \nabla \boldsymbol{\xi}-\boldsymbol{\xi}\cdot \nabla \boldsymbol{U}_0.
\end{eqnarray}

We neglect the Eulerian gravitational potential perturbation by setting $\delta \Phi=0$. Since our primary aim is to study the elliptical instability, which excites inertial modes, this restriction is unlikely to significantly modify any of our results. This is because inertial modes only weakly perturb the gravitational potential (see Appendix~\ref{SG}). In addition, this restriction allows us to directly compare with the numerical simulations of \cite{Barker2015b}.

For $\boldsymbol{\xi}^{\prime},\boldsymbol{\xi}\in \mathbb{C}^3$, we define the inner product 
\begin{eqnarray}
\label{innerproduct}
\langle \boldsymbol{\xi}^{\prime},\boldsymbol{\xi}\rangle = \int_{V} (\boldsymbol{\xi}^{\prime})^{*}\cdot \boldsymbol{\xi}\;\mathrm{d}V,
\end{eqnarray}
under which A and B are respectively anti-Hermitian and Hermitian operators. 

\subsubsection{Basis functions}
We seek solutions of Eq.~\ref{LPE} in the form of solenoidal vector fields whose components are polynomials in the Cartesian coordinates up to a specified degree $\ell_\mathrm{max}-1$ ($\ell_\mathrm{max}$ can be viewed as the equivalent ``spherical harmonic degree" for the pressure perturbation). \cite{L1989a,L1989b} has shown that it is possible to construct a basis consisting of $N(\ell_\mathrm{max})=\ell_\mathrm{max} (\ell_\mathrm{max}+1)(2\ell_\mathrm{max}+7)/6$ such vectors that exactly represent all possible solutions with harmonic degrees $\ell\leq\ell_\mathrm{max}$. The operators A and B can be shown to act invariantly on such a basis\footnote{An alternative viewpoint is that the couplings are ``one-way" so that they only couple components of degree $\ell$ with components of degree $\ell-2$ (but not to components with degree $>\ell$). To describe a mode of degree $\ell_\mathrm{max}$, we therefore need only consider basis functions up to degree $\ell_\mathrm{max}$ to exactly represent such a mode.}. That is, for a given $\ell_\mathrm{max}$, we seek solutions of the form
\begin{eqnarray}
\boldsymbol{\xi}(\boldsymbol{x},t)= \sum_{i=1}^{N(\ell_\mathrm{max})}\alpha_i(t) \boldsymbol{\xi}_{i}(\boldsymbol{x}),
\end{eqnarray}
where $\alpha_i$ is a complex amplitude. The real vector fields $\left\{\boldsymbol{\xi}_1,\dots ,\boldsymbol{\xi}_{N(\ell_{\mathrm{max}})}\right\}$ are chosen so that they span the space of allowable perturbations that are solenoidal vector polynomials up to degree $\ell_{\mathrm{max}}$. For purely incompressible perturbations, \cite{L1989a,L1989b} has demonstrated that an appropriate choice of basis consists of the sum of the bases for each of two subspaces U$_\ell$ and V$_\ell$. U$_\ell$ is the subspace of irrotational motions that represent boundary perturbations up to degree $\ell$, and a convenient choice of basis for this subspace consist of the gradients of solid ellipsoidal harmonics -- this set is required to study surface gravity modes (hereafter SGMs). V$_\ell$ is the subspace of purely vortical perturbations up to degree $\ell$ that do not move the boundary, for which a convenient choice of basis can be constructed analytically \citep{GP1977,L1989a,GP1992} -- this set is required to study inertial modes (hereafter IMs). The resulting basis depends on the shape of the ellipsoid. More details regarding its construction is relegated to Appendix~\ref{basisfunctions}. (An example basis up to degree $\ell=2$ is also listed in Appendix D of \citealt{L1989a}.)

\subsubsection{Quadratic eigenvalue problem}

After we have constructed the basis, we project Eq.~\ref{LPE} onto this basis using the inner product defined by Eq.~\ref{innerproduct} to obtain\footnote{Given that the construction of the basis vectors and the calculation of the resulting integrals of Cartesian polynomials over the volume of the ellipsoid is rather involved, we automate all of the above processes using the symbolic algebra package in Matlab (Matlab is also used for the numerical computation of the eigenvalues), speeding up the calculation of integrals of polynomials over the ellipsoid by using the exact integral formula in \cite{L1989a}, Eq 50.}, for each basis function labelled by $i$,
\begin{eqnarray}
\label{quadeig}
\langle \boldsymbol{\xi}_i ,\boldsymbol{\xi}_j \rangle \ddot{\alpha}_j + \langle \boldsymbol{\xi}_i ,\mathrm{A}\,\boldsymbol{\xi}_j \rangle \dot{\alpha}_j +\langle \boldsymbol{\xi}_i ,\mathrm{B}\,\boldsymbol{\xi}_j \rangle \alpha_j=0,
\end{eqnarray}
where a sum over $j$ is implied, $\boldsymbol{\alpha}$ is a column vector of length $N(\ell_\mathrm{max})$ and the coefficients are the components of $N(\ell_\mathrm{max})\times N(\ell_\mathrm{max})$ matrices. Note that $\langle\boldsymbol{\xi},\nabla\Delta p\rangle=0$, so the last term in Eq.~\ref{LPE} does not contribute to Eq.~\ref{quadeig}.

We seek solutions with time dependence $\propto e^{-\mathrm{i}\hat{\omega} t}$, which converts the system to a quadratic eigenvalue problem for the complex frequency $\hat{\omega}$ and mode amplitude $\boldsymbol{\alpha}$, which can be solved using standard methods\footnote{We use Matlab's polyeig function.} (alternatively, the system can be solved after it has been converted into a linear eigenvalue problem). Note that this solution involves no truncation up to a given $\ell_\mathrm{max}$ and the solutions are therefore exact (to the numerical precision of the eigenvalue solution and the computation of the basis functions for the U$_\ell$ subspace). For future reference we define $\hat{\omega} = \omega + \mathrm{i} \sigma$, where $\omega$ is the (real) frequency of the mode and $\sigma$ is the (real) growth rate.

A further advantage of this formalism is that it is straightforward to determine the effect of changing the boundary to be rigid rather than free. This allows the modes and instabilities of the tidal flow in a triaxial ellipsoid with a rigid boundary to be determined, which is relevant for understanding the results of laboratory experiments, as well as to possible applications to the liquid cores of terrestrial planets \citep{Cebron2012,Vant2014}. To accomplish this, we simply omit from our basis the vectors in the U$_\ell$ subspace, and solely consider those of the V$_\ell$ subspace. This means that we can no longer capture surface gravity modes. Note that $\boldsymbol{\xi}_{i}\cdot \boldsymbol{n}_s=0$ on the boundary ($\boldsymbol{n}_s=(x/a^2,y/b^2,z/c^2)$ is a normal vector) for each element of the V$_\ell$ basis, which therefore satisfy the boundary conditions on a rigid boundary. In addition, $\langle \boldsymbol{\xi}_i,\nabla \Delta p\rangle=0$ for each element of such a basis, which allows Eq.~\ref{quadeig} to determine the stability, as in the case with a free surface.

We have thoroughly checked the resulting code in several ways. The matrix coefficients in Eq.~\ref{quadeig} have the same symmetry properties as the operators A and B (only spoiled by tiny numerical errors). The basis functions are solenoidal with the maximum value of $|\nabla \cdot \boldsymbol{\xi}|\lesssim 10^{-4}$, in all cases. We also compare the solutions produced by our code in detail against an independent analysis for a Maclaurin spheroid for all modes with $\ell\leq 4$ in \S~\ref{MS}, and we have checked our results for the $\ell=2$ inertial modes against the predictions of \cite{Kerswell1994} and \cite{Vant2014} for the case of a rigid boundary. Finally, we have compared our results in some cases with the numerical simulations that will be presented in \cite{Barker2015b}.

In the next two sections, we present the results of our global analysis. We begin by analysing the excitation of inertial modes (hereafter IMs) by the elliptical instability. We then move on to study the effect of rotation and tidal deformation on the mode frequencies, focusing on surface gravity modes (hereafter SGMs). In Appendix \ref{WKB} we describe a complementary local WKB analysis of the elliptical instability.

\section{Elliptical instability}
\label{elliptical}

In this section, we turn to the main purpose of this paper, which is to analyse the properties of the elliptical instability. We focus on instabilities involving global modes, which are probably the most important for tidal dissipation. 

\subsection{Basic mechanism}

The elliptical instability excites pairs of IMs through a parametric resonance \citep{Kerswell2002}, and is driven by the time-variation of fluid properties around an elliptical streamline. It draws upon the kinetic energy associated with the tidal flow whenever $\gamma\ne 0$ and $A\ne 0$. If we consider the tidal synchronisation problem, the pair of fastest growing modes typically have frequencies in the fluid ($\Omega$) frame approximately equal to half the frequency of tidal forcing, so that temporal resonance occurs if the frequencies in the fluid frame $\omega_{\Omega,1}\sim \omega_{\Omega,2}\sim \gamma$ (more generally we require $\omega_{\Omega,1}\pm\omega_{\Omega,2}=2\gamma$). Spatial resonance requires the modes to have azimuthal wavenumbers $m_{1}\pm m_{2}=2$ (since the tidal deformation has $m=2$), and also $\ell_1=\ell_2$ \citep{Kerswell1994,LL1996}. Modes with different $\ell$ do not couple, as has been proved by \cite{Kerswell1993} and \cite{LL1996}. In the limit of small $A$, the elliptical instability is only possible if $n/\Omega\in[-1,3]$, since IMs are restricted to $\omega_\Omega \in [-2\Omega,2\Omega]$, but as we will show below (and further demonstrate using a local WKB analysis in Appendix \ref{WKB}), this can be violated when $A$ is sufficiently large. 

An upper bound on the growth rate of the elliptical instability is equal to the maximum strain rate
\begin{eqnarray}
\label{maxgrowth}
\sigma_\mathrm{max}&=&\frac{|\gamma|}{2}\left(\frac{a}{b}-\frac{b}{a}\right) \\
&\approx& \epsilon |\gamma| \approx \frac{3}{2}A|\gamma|,
\label{smalleps}
\end{eqnarray}
where we have taken the limit of small tidal deformation and no background rotation ($n=0$) for the second line \citep{LL1996,LL1996a}. Eq.~\ref{maxgrowth} provides an upper bound on the growth rate of short-wavelength IMs\footnote{The canonical elliptical instability of an unbounded strained vortex, whose nonlinear evolution was studied by \cite{BL2013}, has a maximum growth rate of $\frac{9}{16}\epsilon |\gamma|$ when $n=0$ \citep{Waleffe1990}. This is smaller than Eq.~\ref{smalleps} by approximately a factor of 2 because the horizontal velocity components of the unstable IMs are not perfectly correlated in general (i.e.~$\langle u_x u_y\rangle \leq \langle |\boldsymbol{u}|^2\rangle$, where angled brackets denote temporal averages of the perturbations).}
\citep{LL1996a}, but also holds for the global modes considered here (as we will show in Fig.~\ref{7}).

\subsection{Parameter survey}

\begin{figure*}
  \begin{center}
      \subfigure[$\ell\leq2$]{\includegraphics[trim=3cm 0cm 6cm 0cm, clip=true,width=0.24\textwidth]{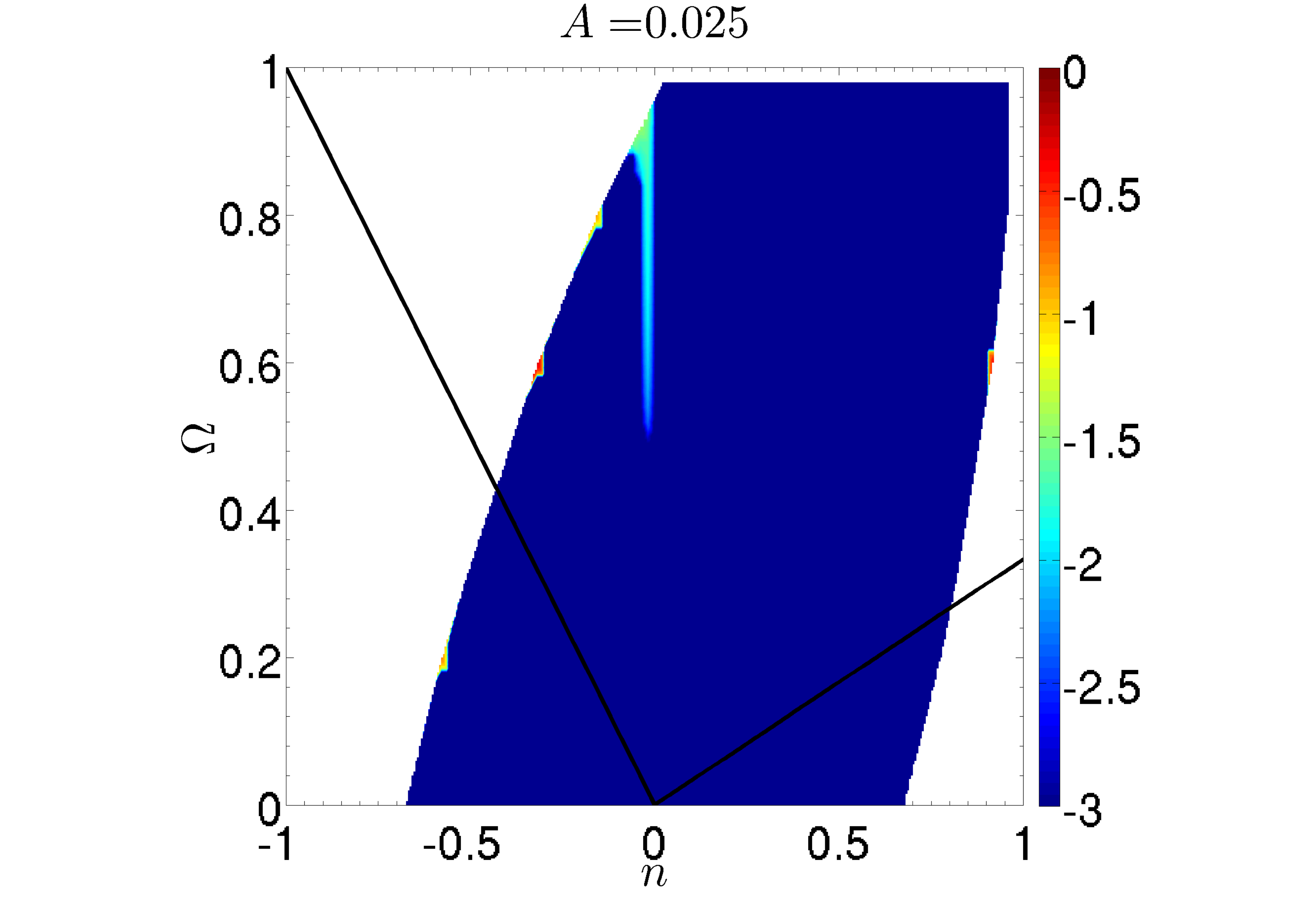} } 
      \subfigure[$\ell\leq2$]{\includegraphics[trim=3cm 0cm 6cm 0cm, clip=true,width=0.24\textwidth]{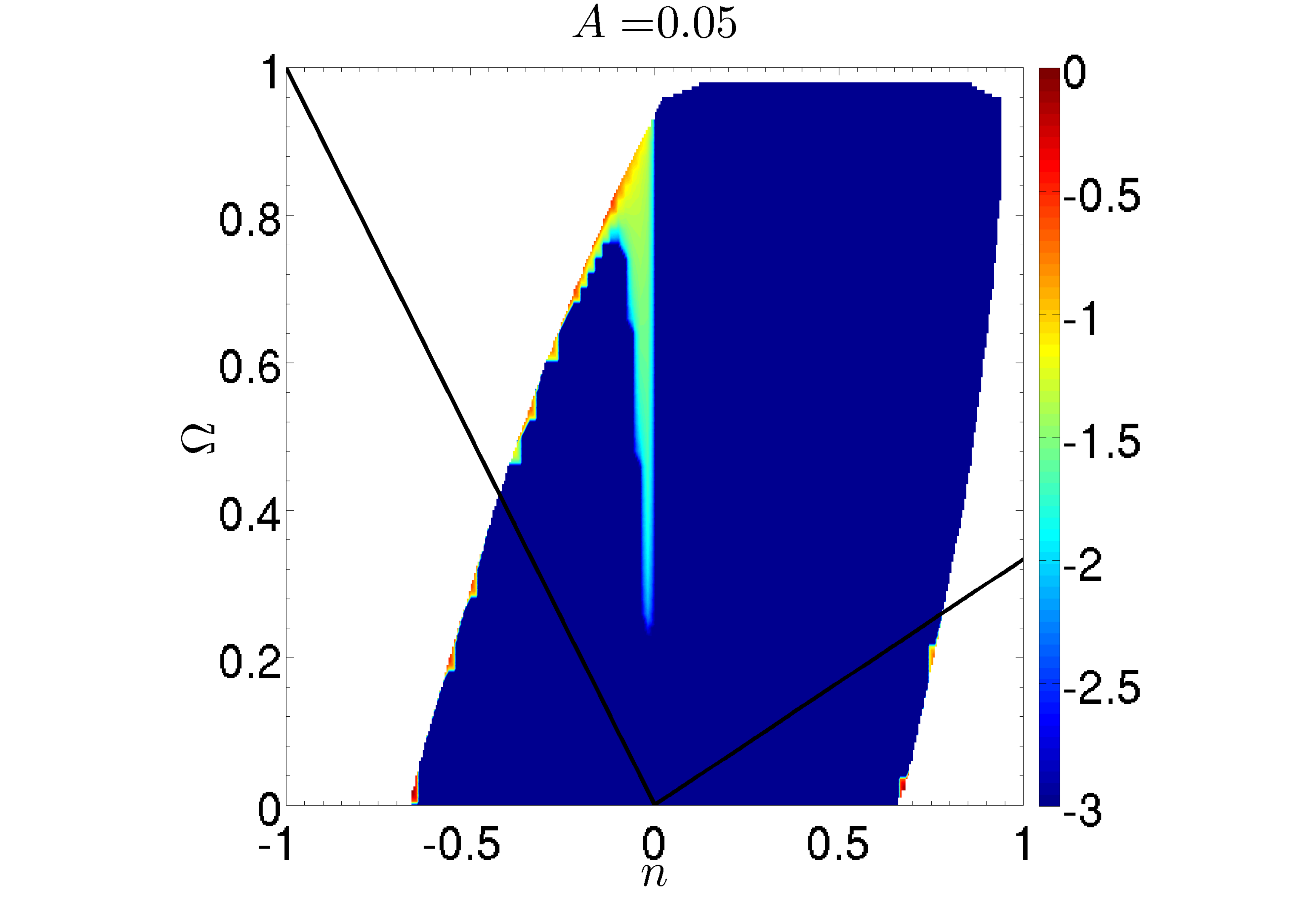} } 
      \subfigure[$\ell\leq2$]{\includegraphics[trim=3cm 0cm 6cm 0cm, clip=true,width=0.24\textwidth]{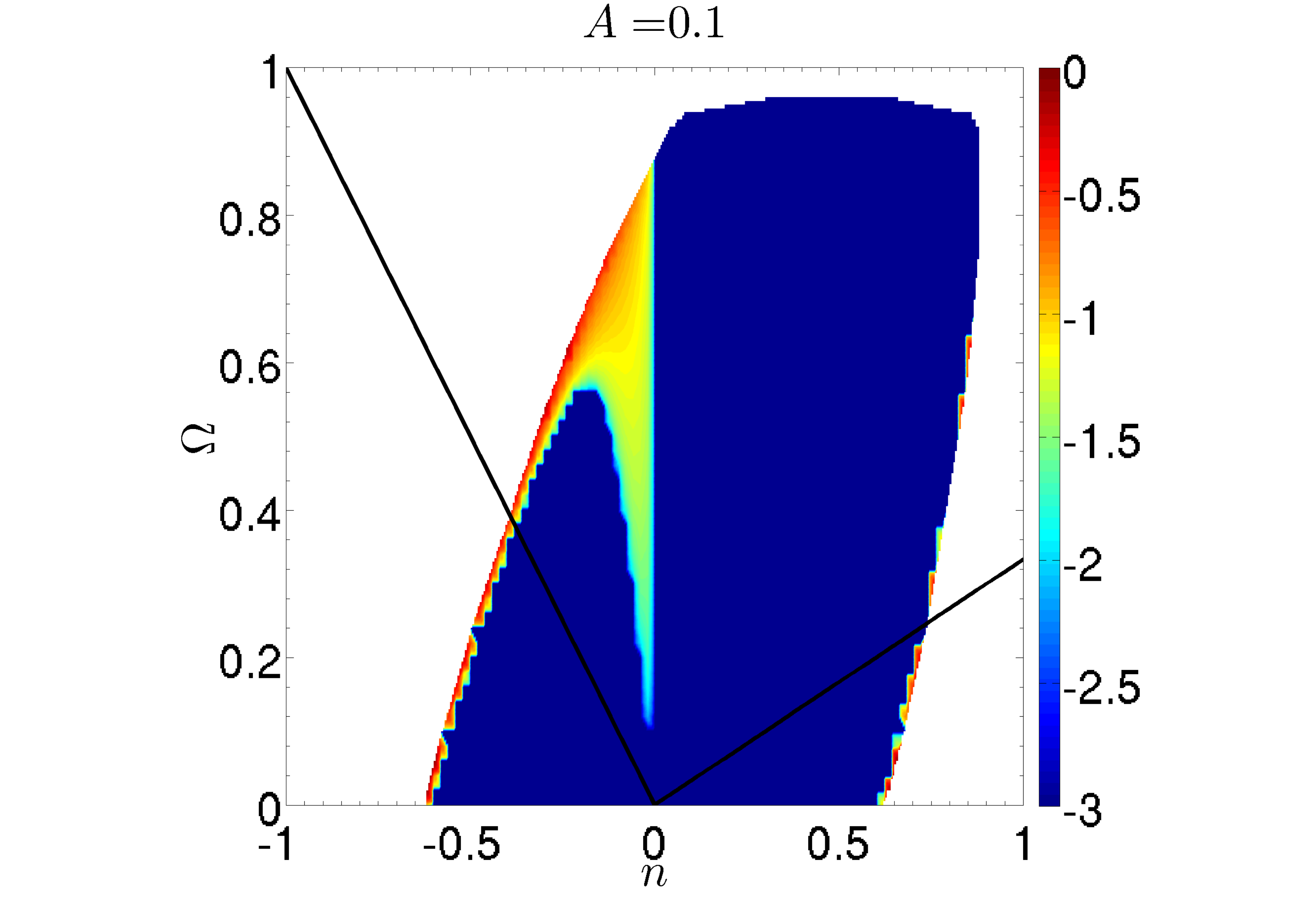} } 
      \subfigure[$\ell\leq2$]{\includegraphics[trim=3cm 0cm 6cm 0cm, clip=true,width=0.24\textwidth]{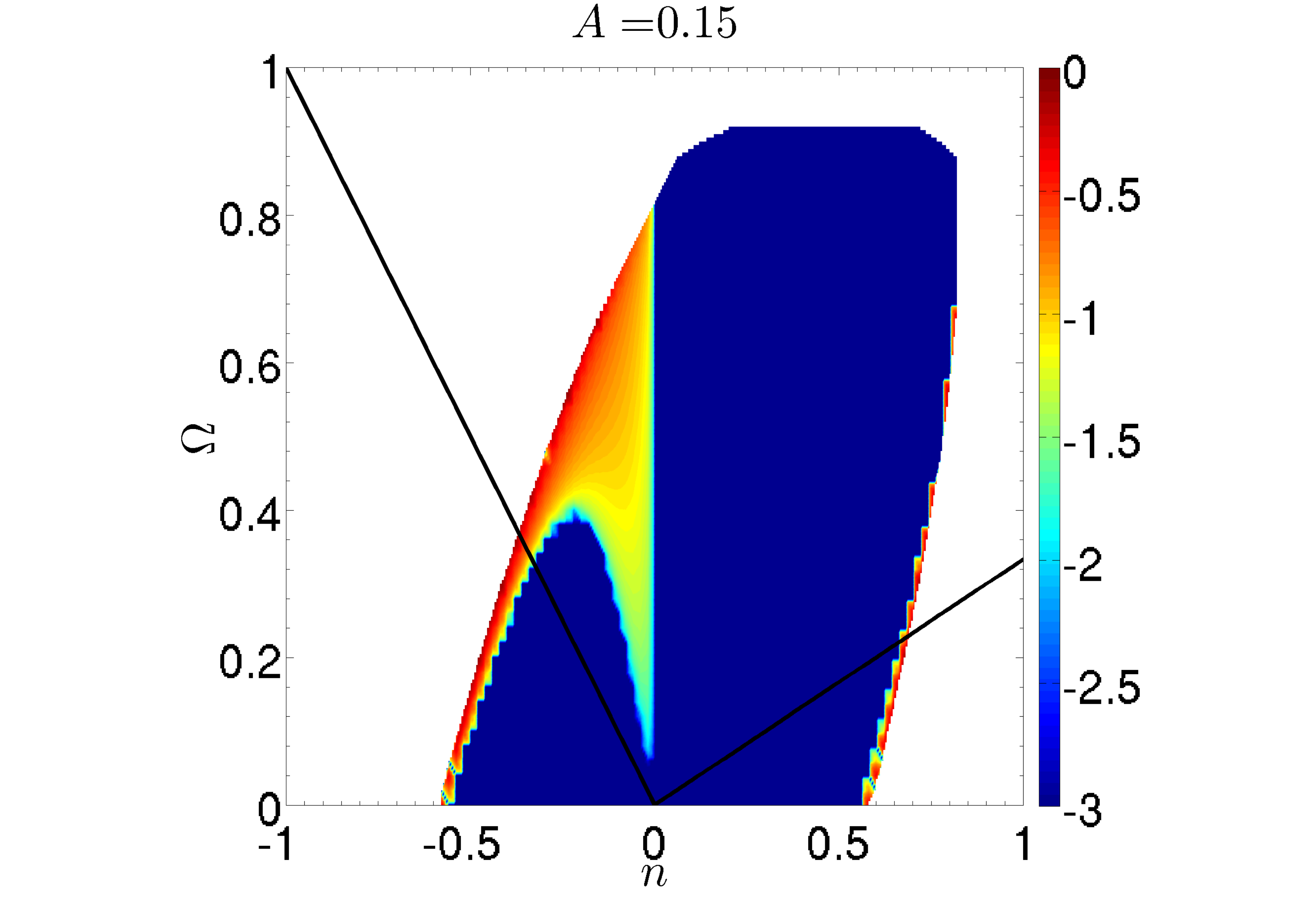} } 
      \subfigure[$\ell\leq3$]{\includegraphics[trim=3cm 0cm 6cm 0cm, clip=true,width=0.24\textwidth]{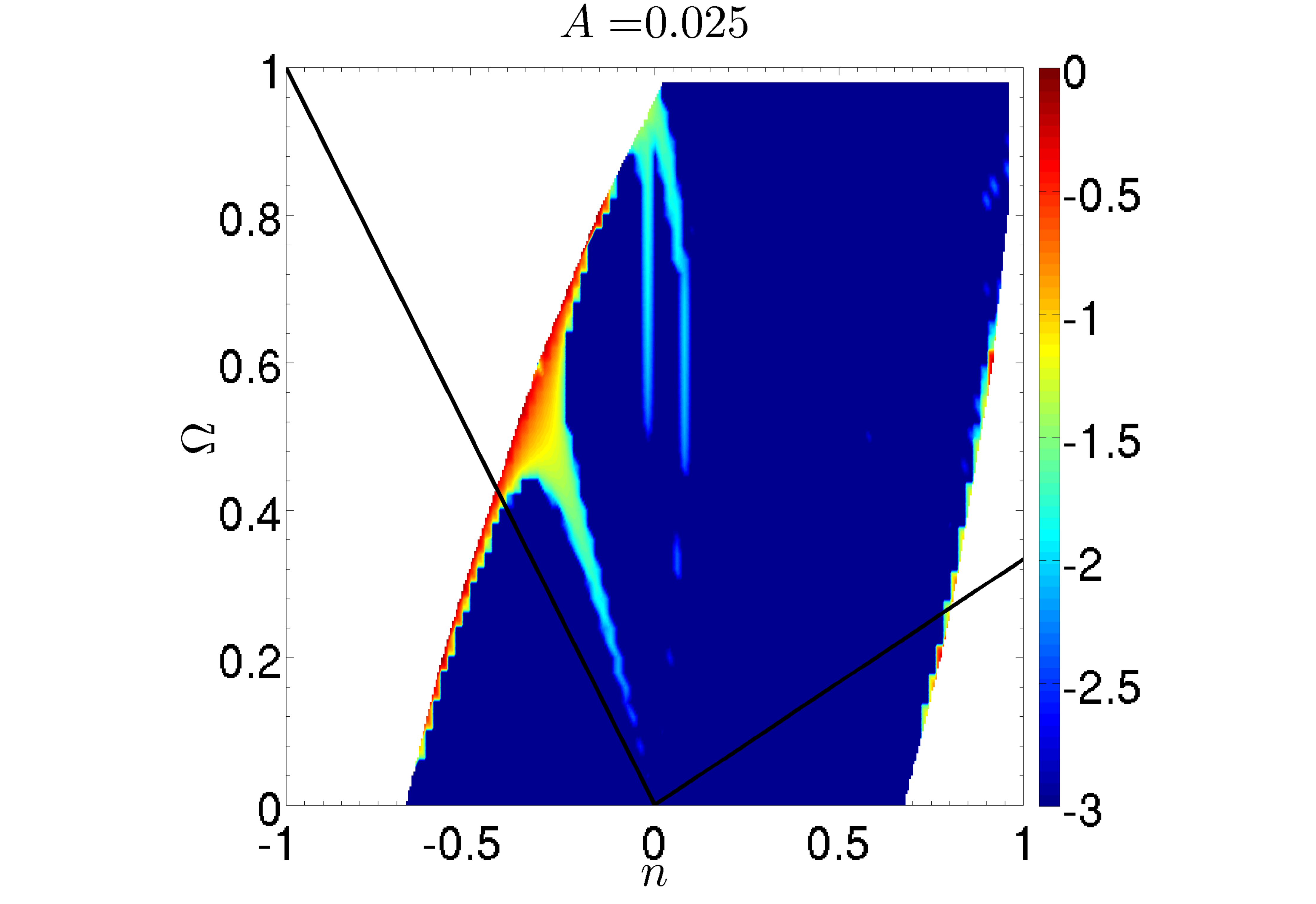} } 
      \subfigure[$\ell\leq3$]{\includegraphics[trim=3cm 0cm 6cm 0cm, clip=true,width=0.24\textwidth]{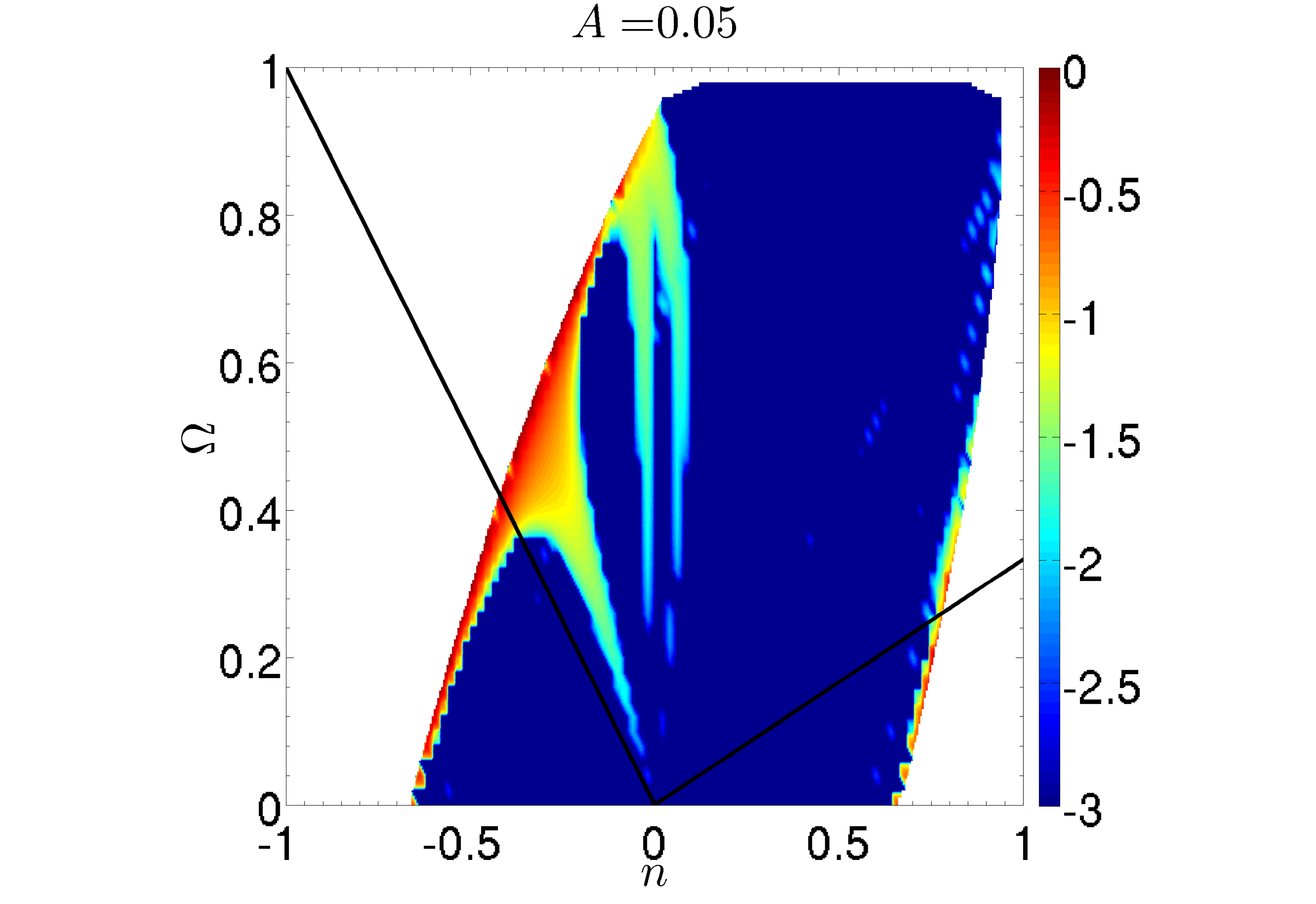} } 
      \subfigure[$\ell\leq3$]{\includegraphics[trim=3cm 0cm 6cm 0cm, clip=true,width=0.24\textwidth]{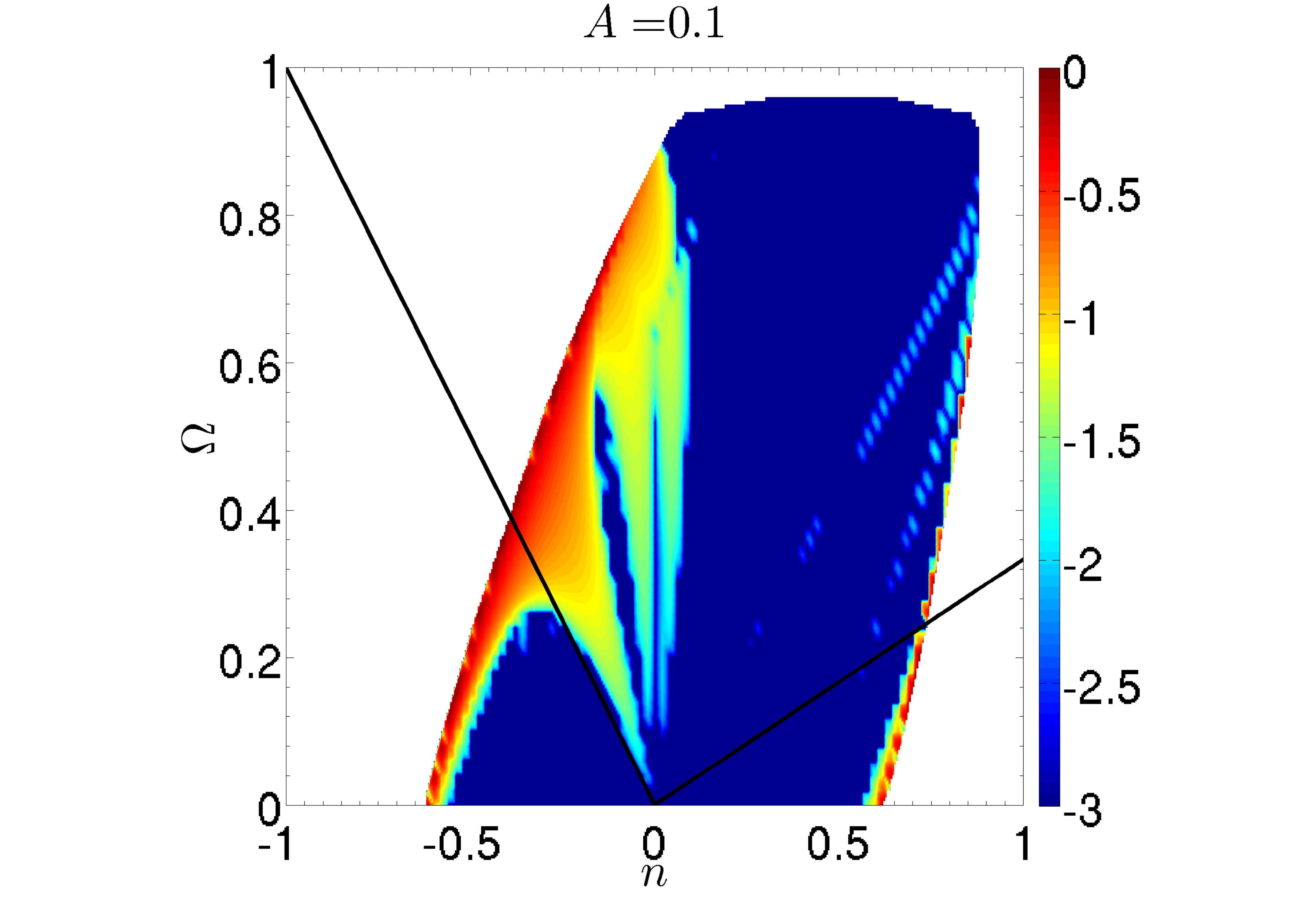} }
      \subfigure[$\ell\leq3$]{\includegraphics[trim=3cm 0cm 6cm 0cm, clip=true,width=0.24\textwidth]{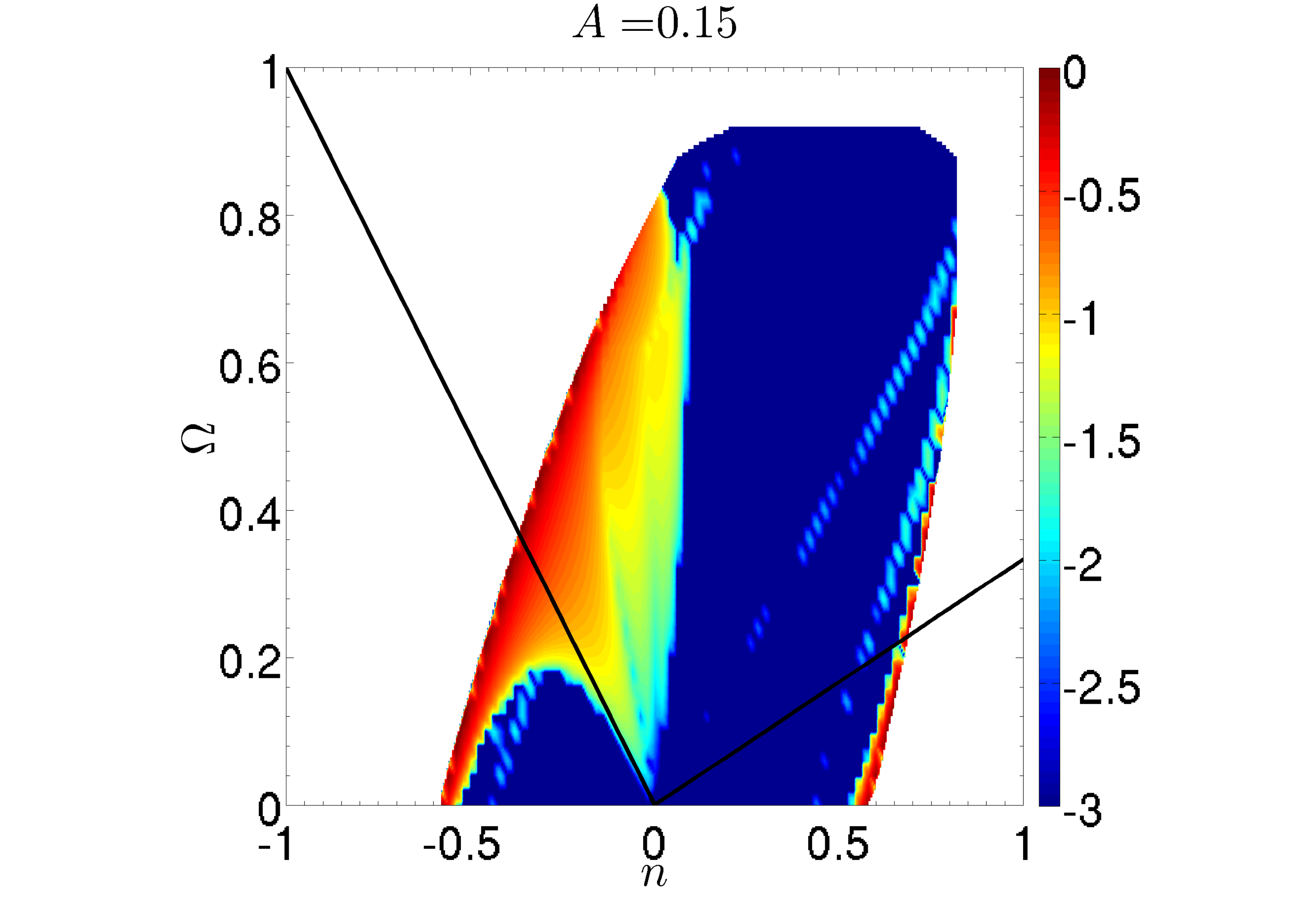} }
      \subfigure[$\ell\leq5$]{\includegraphics[trim=3cm 0cm 6cm 0cm, clip=true,width=0.24\textwidth]{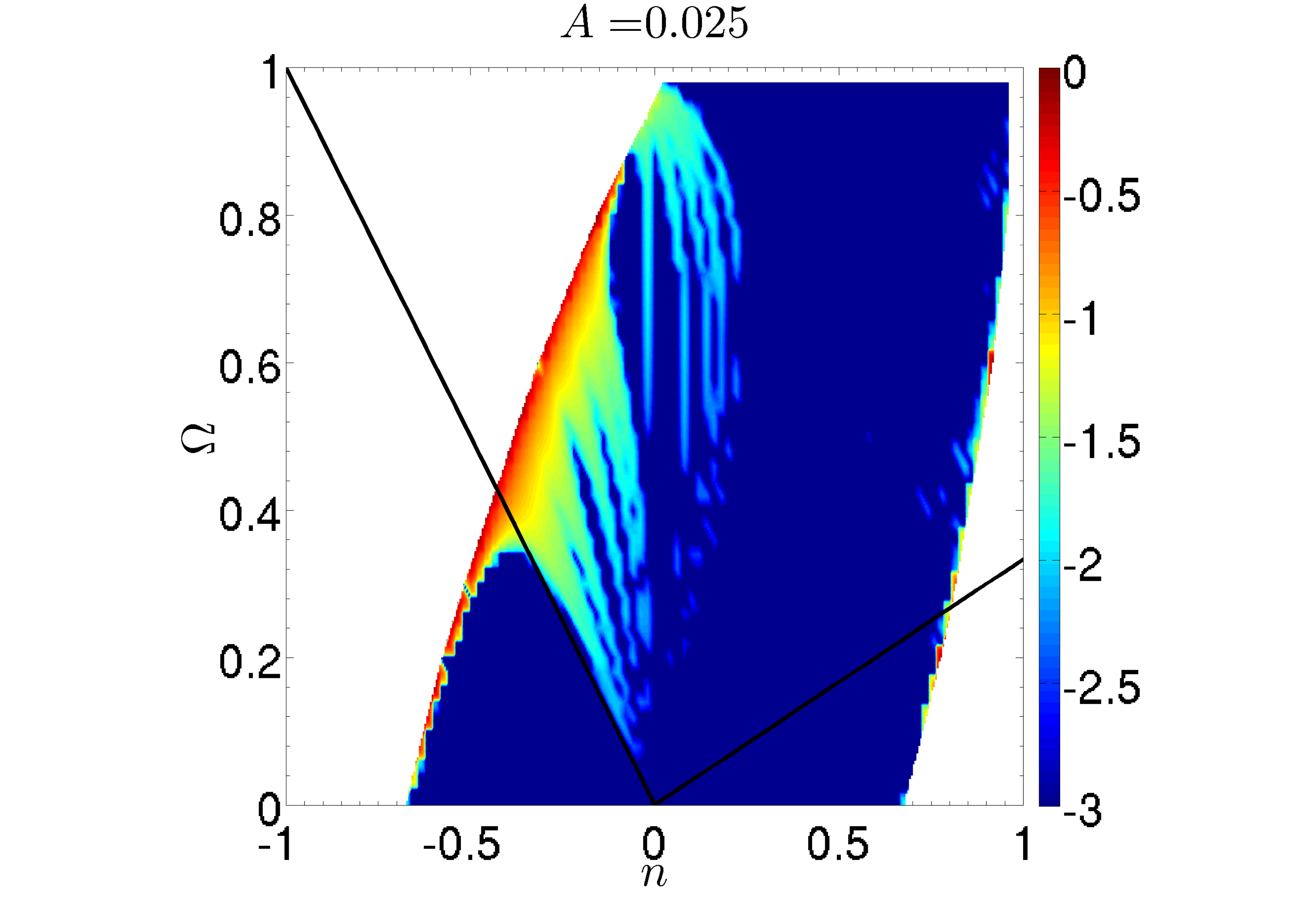} }
      \subfigure[$\ell\leq5$]{\includegraphics[trim=3cm 0cm 6cm 0cm, clip=true,width=0.24\textwidth]{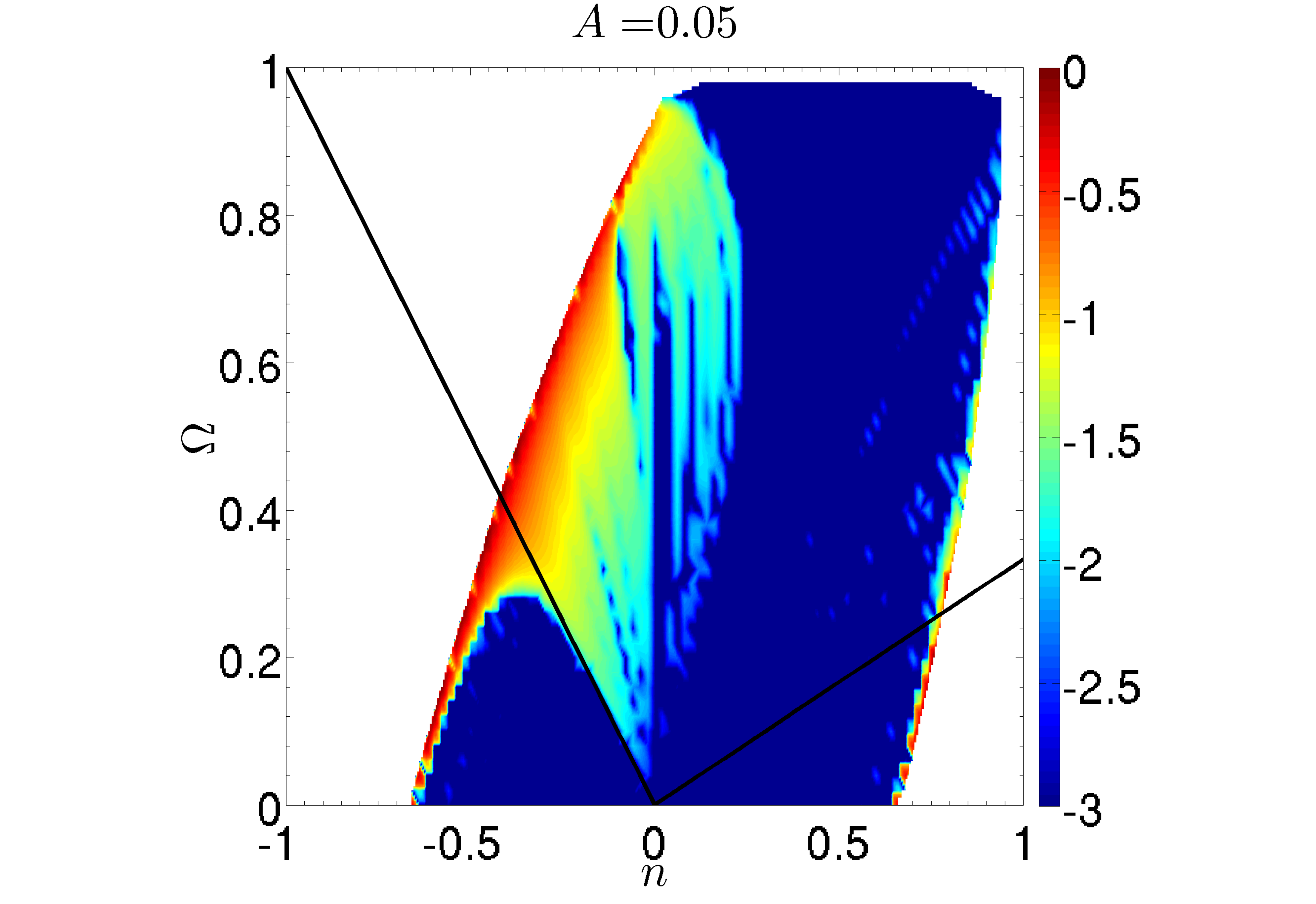} }
      \subfigure[$\ell\leq5$]{\includegraphics[trim=3cm 0cm 6cm 0cm, clip=true,width=0.24\textwidth]{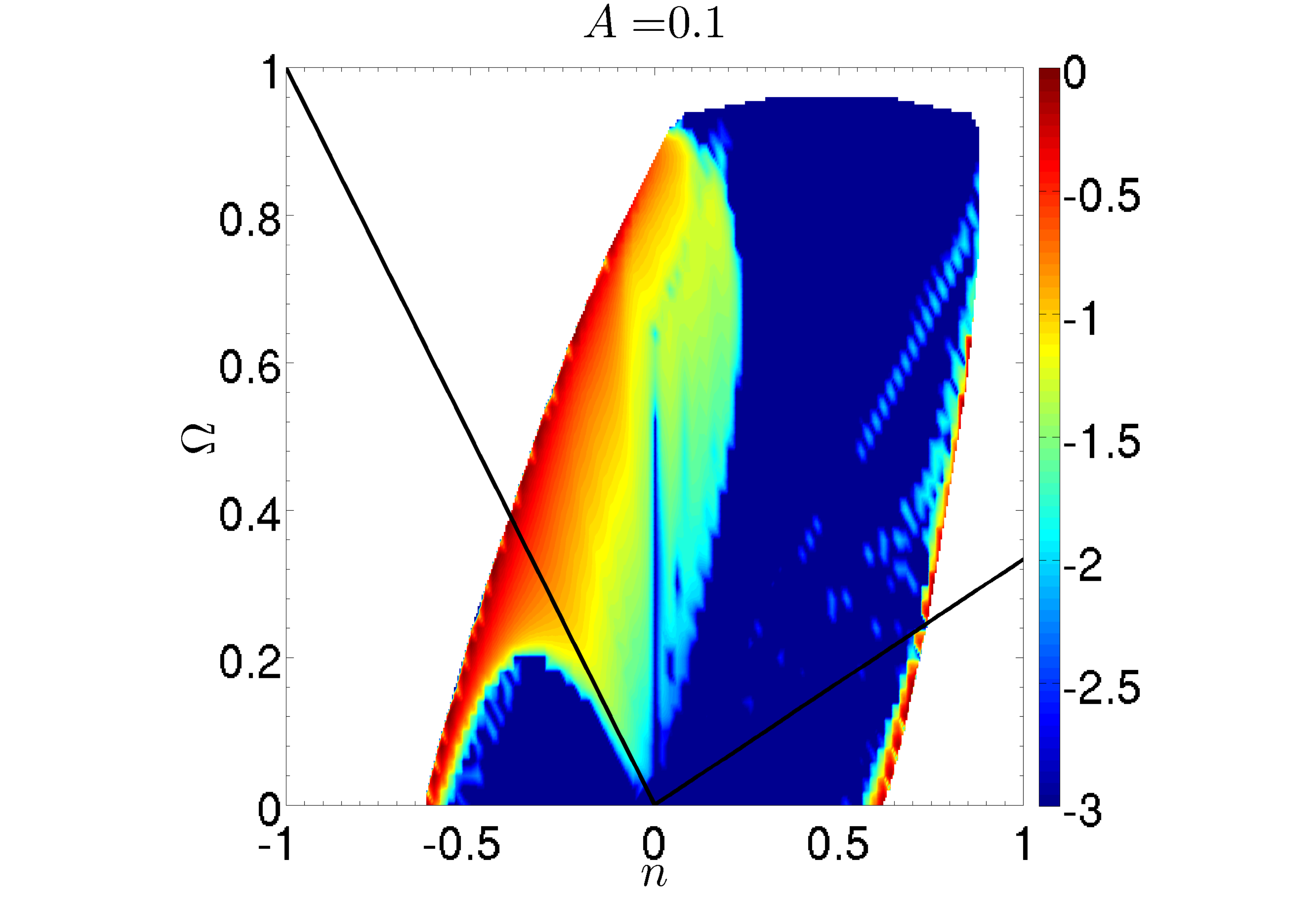} }
      \subfigure[$\ell\leq5$]{\includegraphics[trim=3cm 0cm 6cm 0cm, clip=true,width=0.24\textwidth]{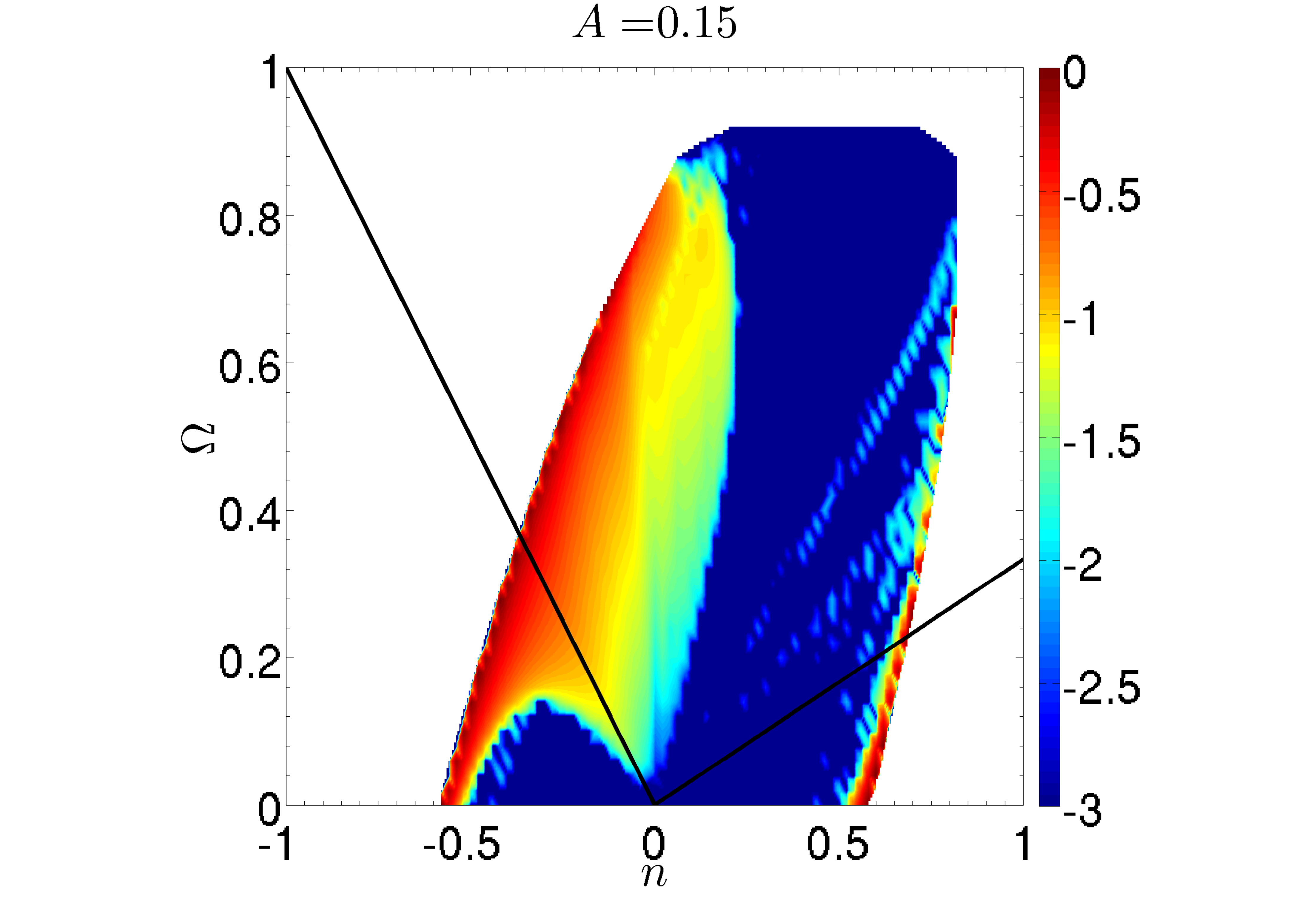} }
      \subfigure[$\ell\leq5$,RB]{\includegraphics[trim=3cm 0cm 6cm 0cm, clip=true,width=0.24\textwidth]{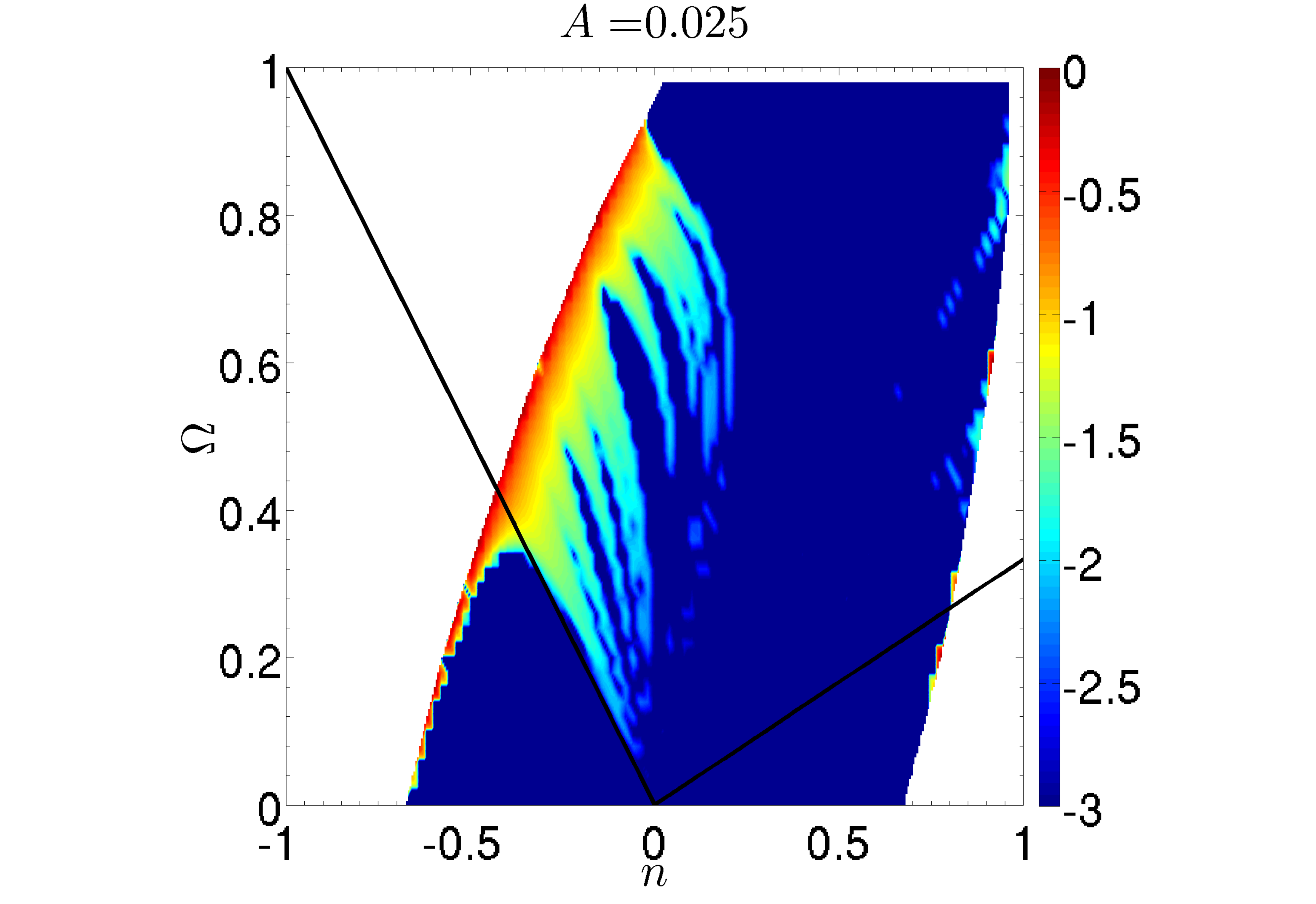} }
      \subfigure[$\ell\leq5$,RB]{\includegraphics[trim=3cm 0cm 6cm 0cm, clip=true,width=0.24\textwidth]{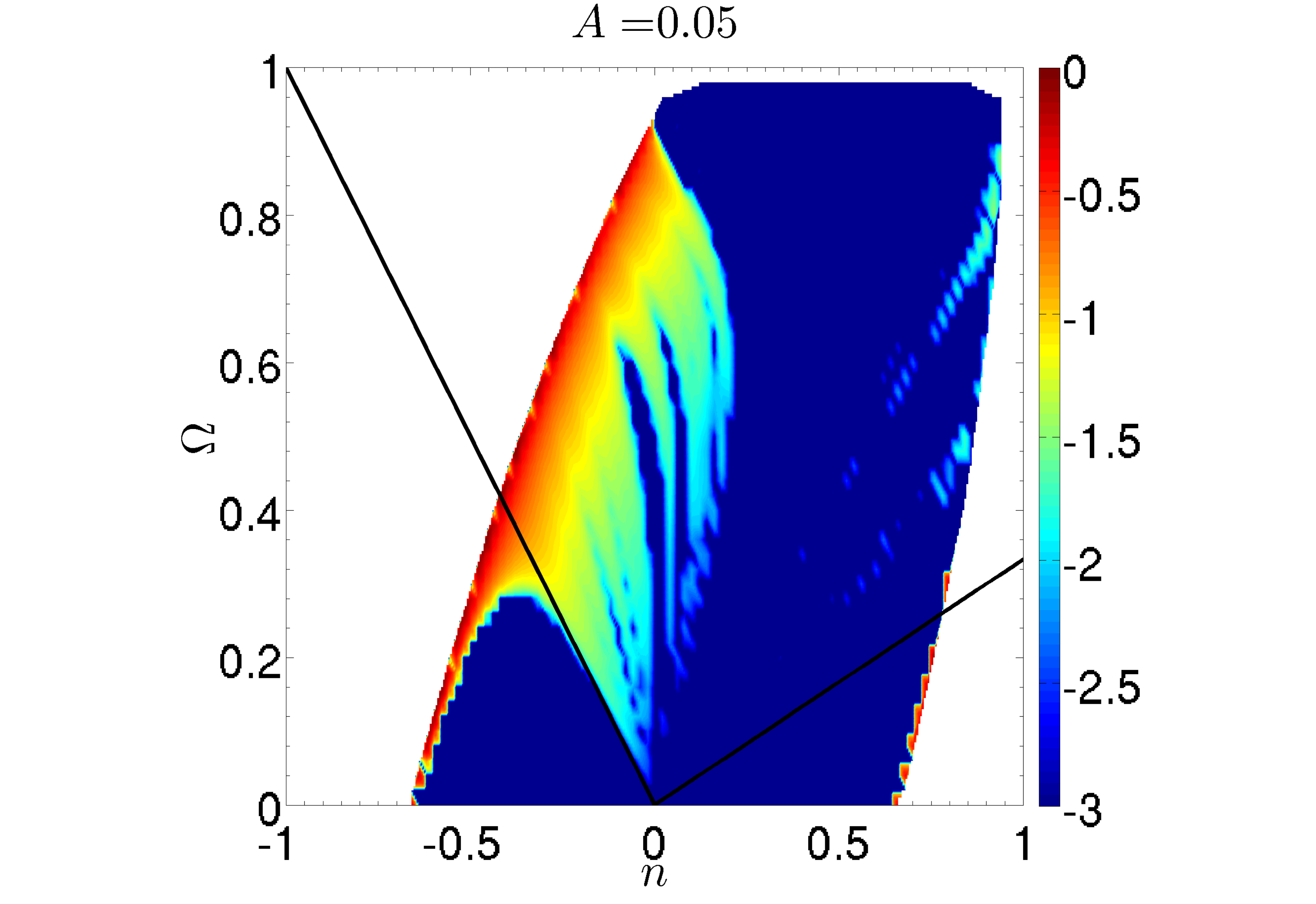} }
      \subfigure[$\ell\leq5$,RB]{\includegraphics[trim=3cm 0cm 6cm 0cm, clip=true,width=0.24\textwidth]{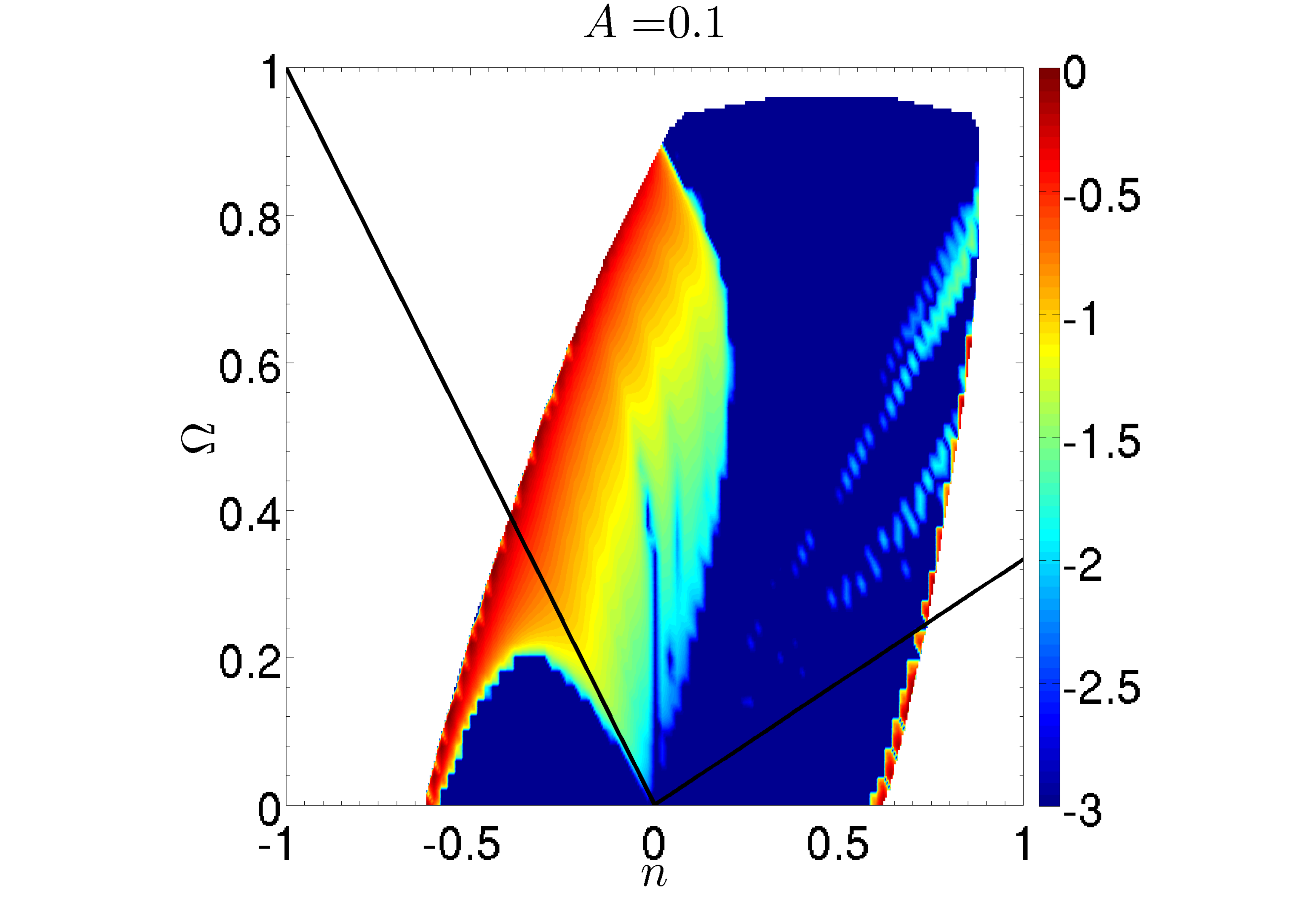} }
      \subfigure[$\ell\leq5$,RB]{\includegraphics[trim=3cm 0cm 6cm 0cm, clip=true,width=0.24\textwidth]{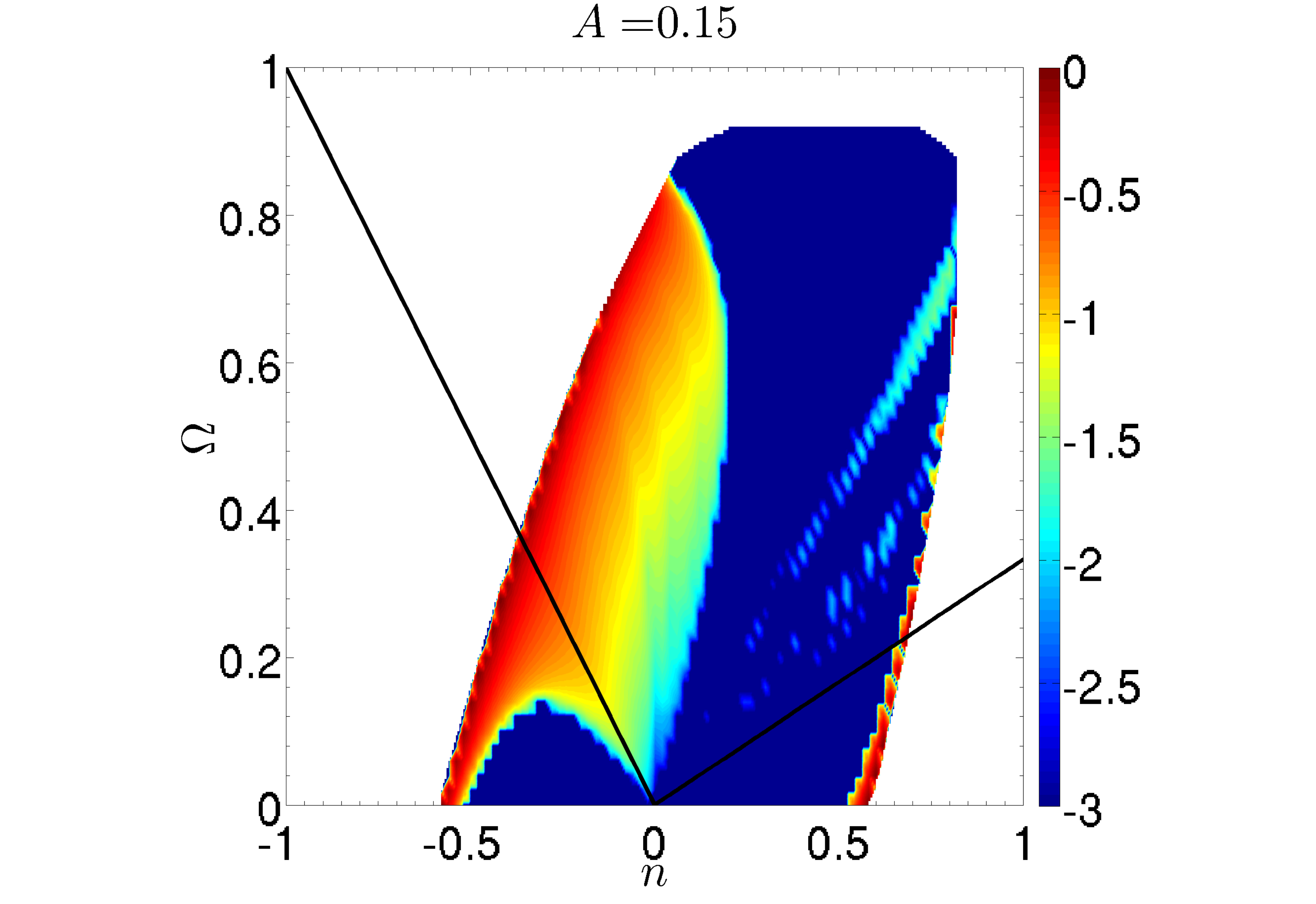} }
      \subfigure[Max]{\includegraphics[trim=3cm 0cm 6cm 0cm, clip=true,width=0.24\textwidth]{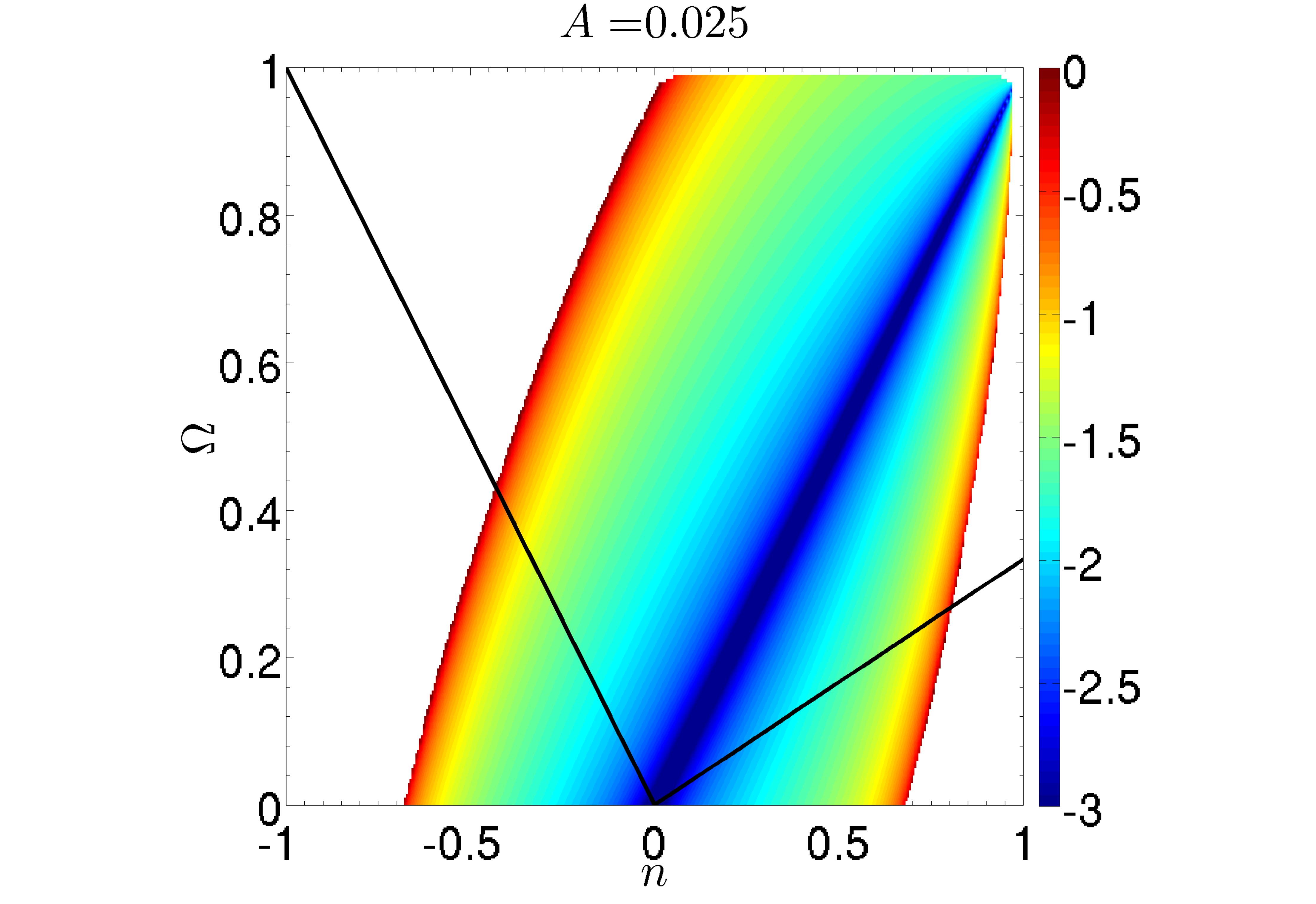} } 
      \subfigure[Max]{\includegraphics[trim=3cm 0cm 6cm 0cm, clip=true,width=0.24\textwidth]{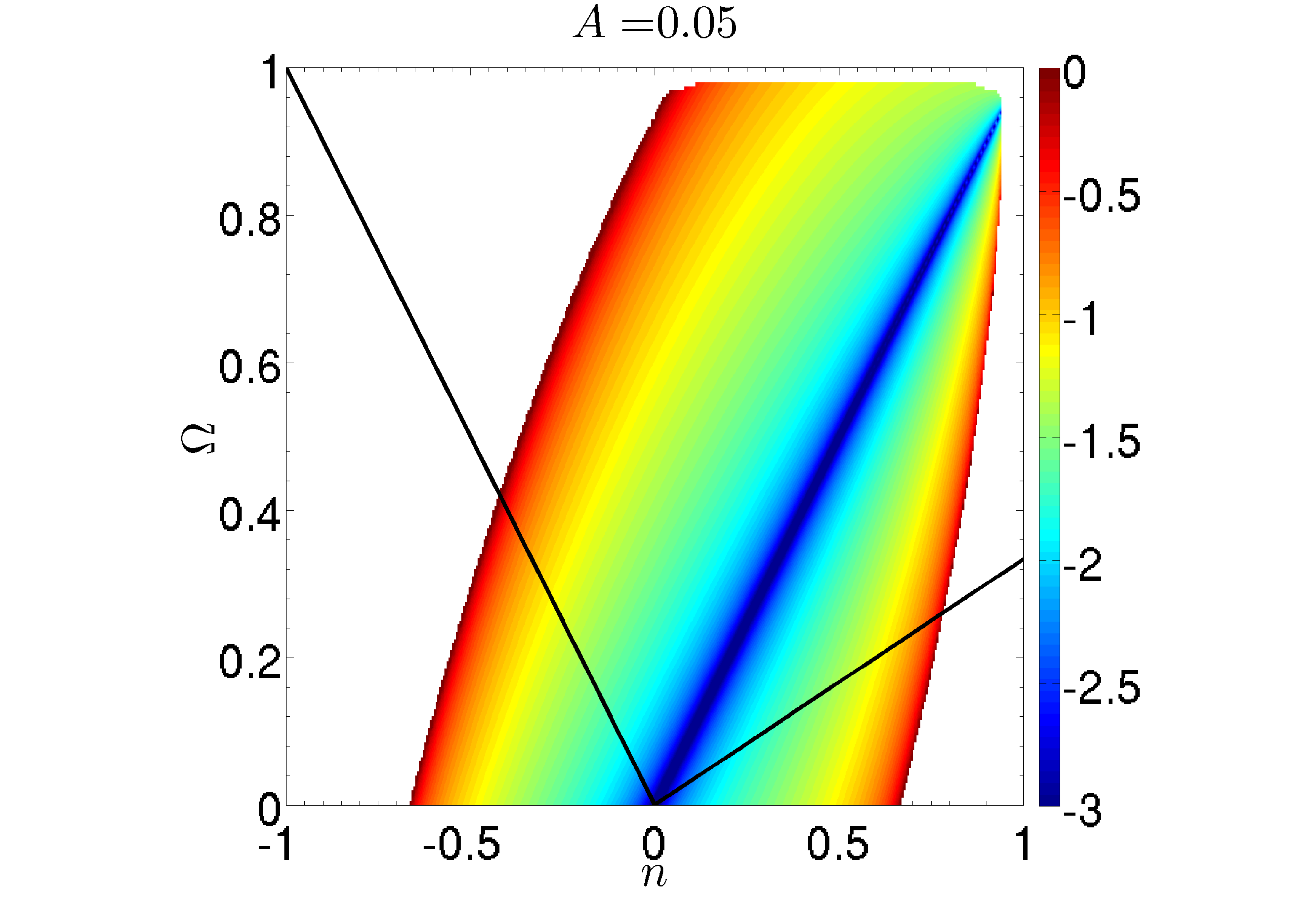} } 
     \subfigure[Max]{\includegraphics[trim=3cm 0cm 6cm 0cm, clip=true,width=0.24\textwidth]{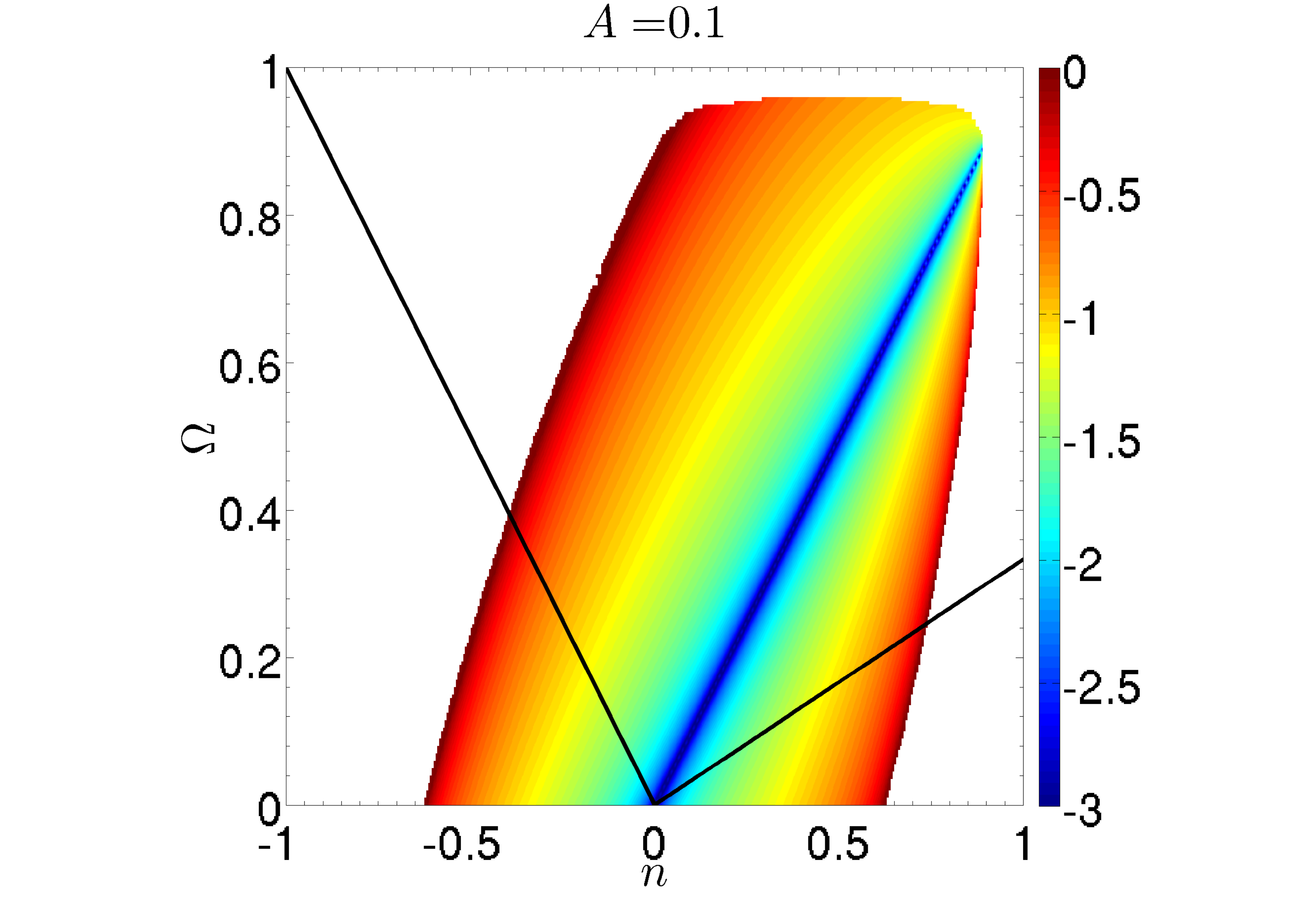} }
      \subfigure[Max]{\includegraphics[trim=3cm 0cm 6cm 0cm, clip=true,width=0.24\textwidth]{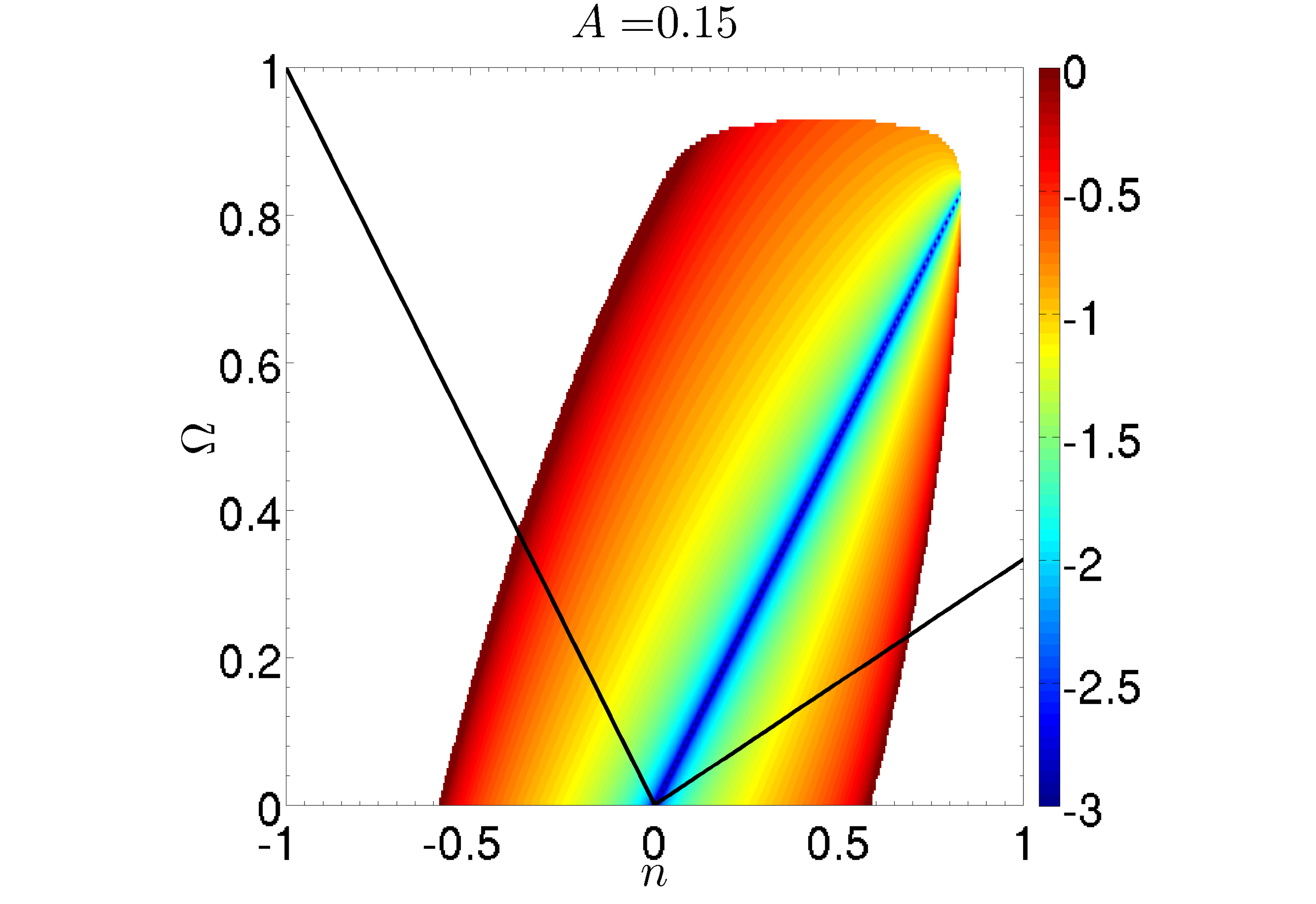} }
       \end{center}
  \caption{Surveying the parameter space for the elliptical instability.  The top three panels plot $\log\sigma$ for the maximum growth rate on the $(n,\Omega)$-plane for several $A$ from the solution of Eq.~\ref{quadeig} for global modes with $\ell_\mathrm{max}=2$ (first row), $\ell_\mathrm{max}=3$ (second row) and $\ell_\mathrm{max}=5$ (third row). The fourth row shows the same results with a rigid outer boundary for $\ell_\mathrm{max}=5$. The bottom row plots an upper bound on the growth rate given by Eq.~\ref{maxgrowth}. The usual elliptical instability of IMs is obtained above the solid black lines for $n\in[-\Omega,3\Omega]$, but instability is also observed for $n\lesssim -\Omega$ for sufficiently large $A$. White regions represent where the magnitude of the velocity divergence is greater than $10^{-4}$ (top four panels), or when the basic ellipsoid becomes undefined (bottom panel). The good correspondence between the third and fourth rows indicates that the elliptical instability is well captured by assuming a rigid outer boundary. For reference, the tide in WASP-19 b has $A\sim 0.05$ and $|n|\sim 0.2$. (Note that the ``striping" behaviour observed in the instability bands for $n>0$ for $A\geq 0.1$ are not physical but are due to the finite resolution adopted.)}
  \label{7}
\end{figure*}

Fig.~\ref{7} is the main result of this paper, and surveys the parameter space of the elliptical instability. This figure illustrates the variation of the maximum growth rate of the global instabilities as each parameter in the set $\left\{ n,\Omega,A,\ell_{\mathrm{max}}\right\}$ is varied. A complementary figure showing results from a local WKB analysis (effectively considering perturbations with $\ell_{\mathrm{max}}\rightarrow \infty$) is presented in Fig.~\ref{D1} in Appendix \ref{WKB}. Each point in each panel of Fig.~\ref{7} represents the maximum growth rate for a particular configuration, resulting from the solution of Eq.~\ref{quadeig}. We have computed results on a uniform grid containing 50 points for $\Omega\in[0,1]$ and 100 points for $n\in [-1,1]$ (these results have been interpolated and smoothed on a uniform grid with equal spacing 0.005 for the purposes of this figure). The top row shows $\log \sigma$ for the maximum growth rate of the unstable modes with $\ell\leq 2$ on the $(n,\Omega)$-plane for $A=0.025, 0.05, 0.1$ and $0.15$, respectively from left to right. The second row shows the same for $\ell \leq 3$, followed by $\ell \leq 5$ in the third row. The shortest wavelength of the unstable modes therefore decreases from the top row as we move downwards.

The fourth row of Fig.~\ref{7} shows the maximum growth rate for unstable modes with $\ell\leq 5$ for the elliptical instability in a rigid ellipsoidal container, which we will describe in more detail in \S~\ref{RB}. Finally, the bottom row of Fig.~\ref{7} shows Eq.~\ref{maxgrowth}, which provides an upper bound on the growth rate derived from energetic considerations \citep{LL1996a}. White regions in the bottom row represent points where there is no ellipsoidal equilibrium configuration (i.e.~with $a,b,c\in\mathbb{R}^+$), and in the top four rows represent a choice of parameters where the polynomial basis (for the U$_\ell$-subspace) is no longer accurately solenoidal and $\mathrm{max}[|\nabla \cdot \boldsymbol{\xi}|]\gtrsim 10^{-4}$, which in practice is more restrictive. In the limit of small $A$, we expect the elliptical instability to occur above the black slanted lines, defined as the region satisfying $n\in [-\Omega,3\Omega]$. (To see the results represented in a different way for the case without a tidal deformation and including self-gravity, see \citealt{LL1996}, Figs.~2 \& 3.)

Several aspects of Fig.~\ref{7} are worthy of comment. Firstly, the maximum growth rate in each panel tends to increase with increasing departure from synchronism, and vanishes along the line $\Omega=n$, as predicted by Eq.~\ref{maxgrowth}. The increase in the growth rate for faster anti-aligned (retrograde) rotations $n\leq 0$ (but above the black lines) can be partly understood by the unbounded plane wave analysis of \cite{Craik1989} as being due to rotation of the bulge shifting the frequencies of IMs in the fluid frame, which modifies the conditions for resonance (also given in Eq.~4 and plotted in Fig.~1 of \citealt{BL2013}). This explains why the growth rate is smaller for positive $n$ (aligned) compared with negative $n$ (anti-aligned) for the same departure from synchronism $\gamma$. We obtain growth rates that approach the upper bound given by Eq.~\ref{maxgrowth} when $n\approx -\Omega$ e.g.~when $A=0.025$, we obtain $\sigma/\sigma_{\mathrm{max}}\sim 0.91$ when $\Omega=0.42$ and $n=-0.36$, and when $A=0.15$, we obtain $\sigma/\sigma_{\mathrm{max}}\sim 0.92$ when $\Omega=0.26$ and $n=-0.26$ (this occurs if a free surface or rigid boundary is adopted)\footnote{This can also be understood from the local analysis of \cite{Craik1989} for the unbounded elliptical vortex, which predicts a maximum value of $\frac{\sigma}{\epsilon |\gamma |}=1$ when $\Omega=-n$ (see Eq.~4 in \citealt{BL2013}), where the correlation between $u_x$ and $u_y$ is maximal.}. We discuss these instabilities further when presenting our local WKB analysis in Appendix \ref{WKB}.

Fig.~\ref{7} shows that for a given $A$, an increasing fraction of the parameter space becomes unstable as we go to larger $\ell$. This is because the resonance criteria are easier to satisfy for shorter-wavelength waves (larger $\ell$), simply because there exist more waves with shorter wavelengths (and the IM spectrum is dense). On the other hand, global modes may be the most important for the energetics of the instability and for the resulting tidal dissipation (e.g.~\citealt{BL2013,BL2014}), which is why we have focussed on them here. 
The widths of the instability bands are $O(\epsilon\gamma)=O(A\gamma)$, meaning that it is possible to excite a pair of modes that are $O(\epsilon\gamma)$ from satisfying an exact resonance condition. Since a fortuitous exact frequency match is not required if we have a sufficiently large $A$, it is then more likely that longer-wavelength global modes could be excited in reality. This is shown clearly in Fig.~\ref{7} by the increasing region of each instability band as we move from left to right panels in each of the first four rows.

One consequence of the finite widths of the instability bands is that IMs can be excited outside of the usual frequency range ($n\in [-\Omega,3\Omega]$) if the tidal deformation is not small. This is shown most clearly as the violent instability band that exists for $n\lesssim -\Omega$,  below the black lines in the second through to fourth rows, in which $A=0.1$ or $A=0.15$. This instability is also present in the local WKB calculations that we present in Appendix \ref{WKB} (see Fig.~\ref{D1}). There, we explain analytically the occurrence of instability for $n\lesssim -\Omega$, and the precise region in the ($n,\Omega$)-plane in which it can operate, resulting from the finite width of the ``stack of pancakes"-type instabilities that are centred on $n=-\Omega$. In the local WKB limit, these instabilities represent modes in which the solution at each $z$ undergoes horizontal epicyclic oscillations independently of all other $z$ (though in the global model the modes are not as simple because of the boundaries, as we will show in \S~\ref{lgeq3}) and were analysed by \cite{Craik1989} and \cite{LL1996a}.

It might be thought that SGMs could interact with IMs to become unstable, since rotation and tidal deformation both reduce the frequencies of the prograde sectoral SGMs to have frequencies close to those of IMs (as we will discuss in \S~\ref{MS}). However, we have not observed such an instability in practice. Indeed, the strong agreement between the third and fourth rows, where results with a free surface and a rigid boundary (see \S~\ref{RB}), respectively, are compared, conclusively demonstrates that the instability is one of IMs. SGMs are only excited at the boundaries of the coloured regions, where the basic ellipsoid configuration has long been known to be unstable (e.g.~\citealt{C1987}).

For reference, note that the shortest-period observed hot Jupiters, such as WASP-19 b \citep{Hebb2010} or WASP-121 b \citep{WASP121}, have $A\sim 0.05$ and $|n|\sim 0.2$ (unfortunately, the sign of $n$ and the magnitude of $\Omega$ are not currently known). The ranges of $A$ and $n$ that we have considered contain the observationally relevant parameter regime, so our results may be important for the tidal synchronisation and spin-orbit alignment (and indirectly, the circularisation) of such planets.

In the next few subsections, we plot several illustrative examples of the instability, showing the spectrum of the ellipsoid and the spatial structure of the unstable modes. 

\subsection{Instability with $\ell=2$: spin-over mode}

\begin{figure}
  \begin{center}
      \subfigure{\includegraphics[trim=5cm 0cm 5cm 0cm, clip=true,width=0.23\textwidth]{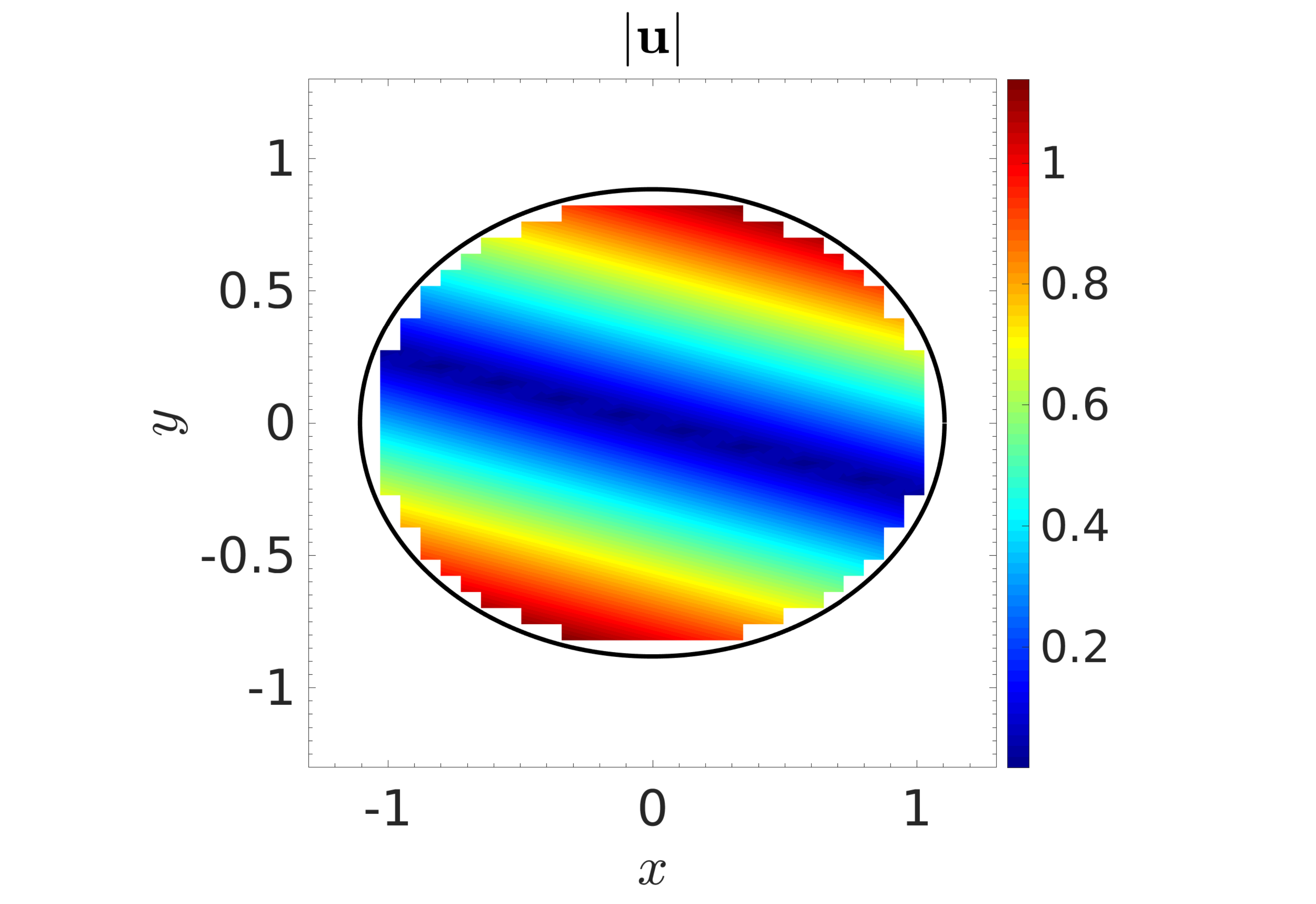} } 
      \subfigure{\includegraphics[trim=5cm 0cm 5cm 0cm, clip=true,width=0.23\textwidth]{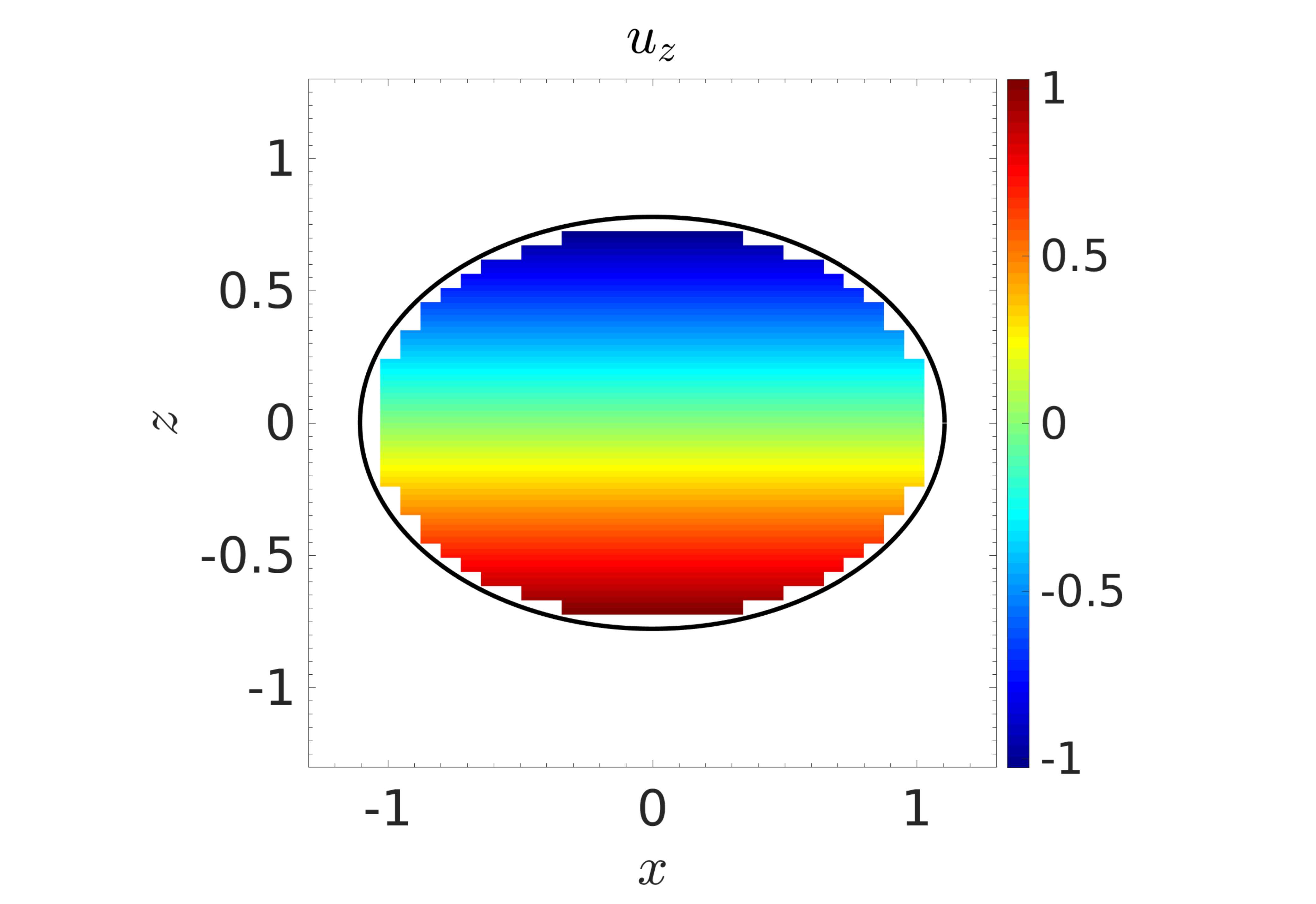} } 
      \end{center}
       \caption{Eigenfunction of the most unstable ``spin-over" mode with $\Omega=0.5,n=-0.02,A=0.1$ and $\ell\leq2$. This mode has $\omega=0$ (in the bulge frame) and $\sigma=0.0404$. This corresponds with the coupling of $\ell=2$, $m=\pm 1$ modes of the unperturbed sphere, which is effectively a rigid tilting of the spin axis of the fluid. Left: $|\boldsymbol{u}|$ on the $xy$-plane. Right: $u_z$ on the $xz$-plane. Only the relative mode amplitudes are meaningful because this is a linear calculation.}
  \label{4a}
\end{figure}

The only instability band (not lying along boundaries with white regions) in the top row of Fig.~\ref{7} represents the excitation of the ``spin-over" mode, which is related to the ``middle-moment-of-inertia" instability of rigid bodies \citep{Kerswell1994}. (A different instability also occurs in a narrow region adjacent to the white regions, which results from the excitation of $\ell=m=2$ SGM, near to where an equilibrium shape for the ellipsoid can no longer be found.) For $A=0.025$, this occupies a very narrow region of parameter space, which grows as $A$ is increased. This instability results from a subharmonic resonance involving a pair of (physically identical) modes with $\ell=2$ and azimuthal wavenumbers $m=\pm 1$ (one is the complex conjugate of the other). It occurs in a spheroid ($a=b$) when \citep{Kerswell1994}
\begin{eqnarray}
\gamma = \frac{2\Omega}{1+c^2} \Rightarrow \frac{n}{\Omega}=\frac{c^2-1}{c^2+1},
\end{eqnarray}
when the phase velocity of the mode matches the angular velocity of the orbiting companion, so that $\omega=0$ in the bulge frame. A finite tidal deformation weakly affects the frequencies of this mode \citep{Vant2014}. However, note that this instability is only excited by \textit{retrograde} companions ($n<0$), because $c^2 \leq 1$. This can be seen in the top panel of Fig.~\ref{7} where unstable modes are clearly absent if $n\geq 0$. We plot an example of the eigenfunction for such a mode in Fig.~\ref{4a}, for $\Omega=0.5, n=-0.02$ and $A=0.1$. 

The excitation of this mode inside exoplanet host stars by the elliptical instability has been suggested to provide an explanation for the observed spin-orbit misalignments of hot Jupiters \citep{Cebron2013}. However, this cannot be the case, since this mode is only excited when the spin and orbit are already misaligned ($n<0$). Note also that the finite width of the resonance does \textit{not} appear to allow the spin-over mode to be excited for prograde orbits $(n\geq 0)$, indicated by the sharp cut-off in the growth rate as $n$ passes through zero. However, the excitation of this mode, and its subsequent damping through secondary instabilities, may play an important role in tidal spin-orbit alignment (e.g.~\citealt{Lai2012}), and we return to this point in \cite{Barker2015b}.

\subsection{Further instabilities with $\ell\geq3$}
\label{lgeq3}

\begin{figure}
  \begin{center}
      \subfigure[]{\includegraphics[trim=5cm 0cm 6cm 0cm, clip=true,width=0.23\textwidth]{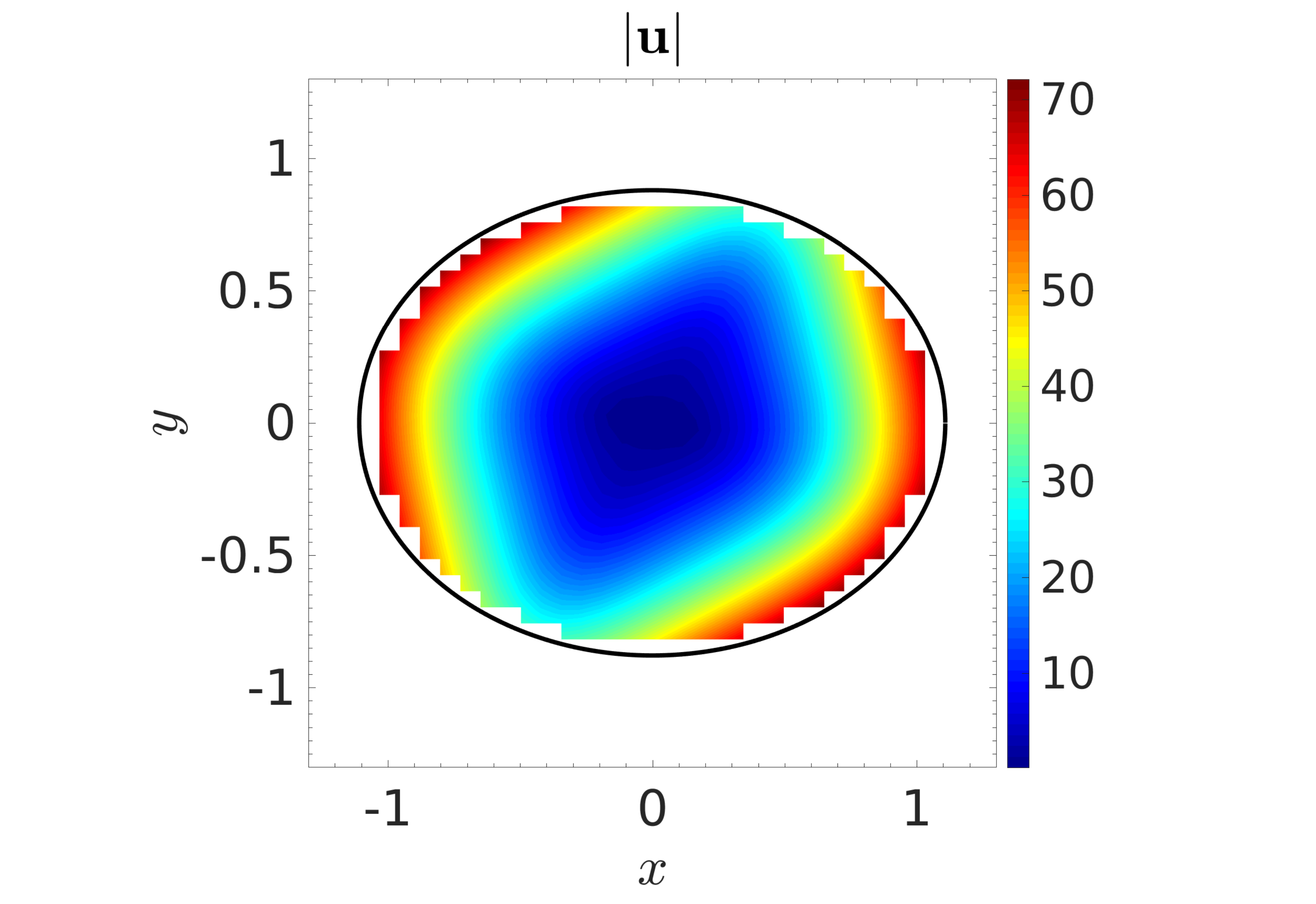} } 
      \subfigure[]{\includegraphics[trim=5cm 0cm 6cm 0cm, clip=true,width=0.23\textwidth]{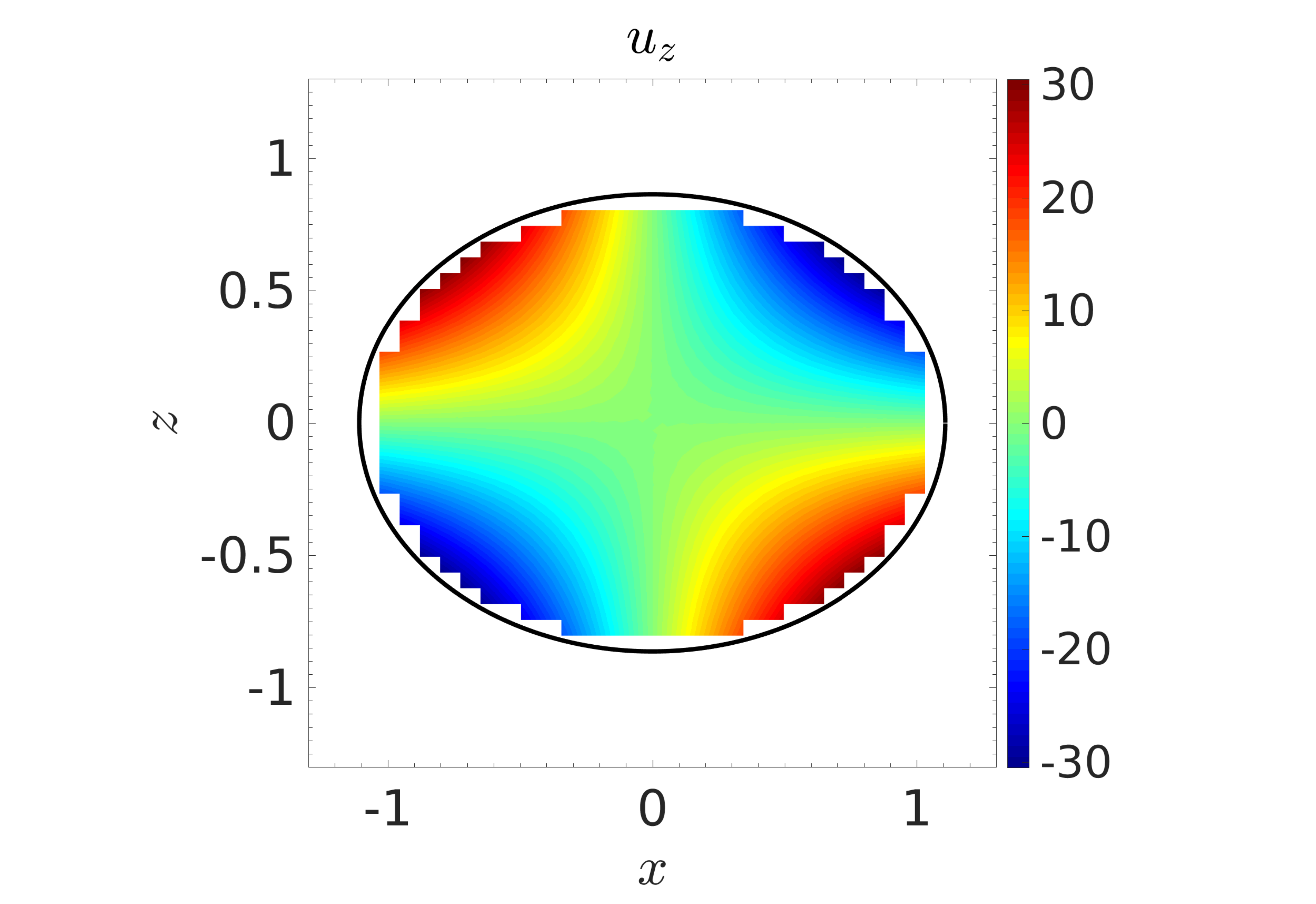} } 
      \end{center}
       \caption{Eigenfunction of the most unstable mode with $\ell\leq 3$, $\Omega=0.3,n=-0.2$ and $A=0.1$. This mode has has $\omega=0$ (in the bulge frame) and $\sigma=0.0886$. This is the finite-$A$ manifestation of the instability consisting of the interaction of the $\ell=3$, $m=2$ and $m=0$ mode (but because $A$ is finite, more components are involved). Left: $|\boldsymbol{u}|$ on the $xy$-plane. Right: $u_z$ on the $xz$-plane. Only the relative mode amplitudes are meaningful because this is a linear calculation.}
  \label{4b}
\end{figure}

The second row of Fig.~\ref{7} shows that additional instability bands (in addition to the spin-over mode) appear when $\ell=3$ modes are considered. The widest of these corresponds with the excitation of $\ell=3$, $m=0$ and $m=2$ IMs (e.g.~\citealt{Gledzer1992}). We plot an example of an unstable mode with $\Omega=0.3, n=-0.2$ and $A=0.1$ in Fig.~\ref{4b}. Once again, this instability band widens as $A$ is increased, and by $A\gtrsim0.1$, most of the parameter space for anti-aligned ($n<0$) spins is unstable. The strongest instability bands for aligned ($n>0$) spins are much narrower, occupying far less of the parameter space, illustrating that anti-aligned spins are much more strongly unstable to global modes than aligned spins, in general.

\begin{figure}
  \begin{center}
      \subfigure[]{\includegraphics[trim=5cm 0cm 6cm 0cm, clip=true,width=0.23\textwidth]{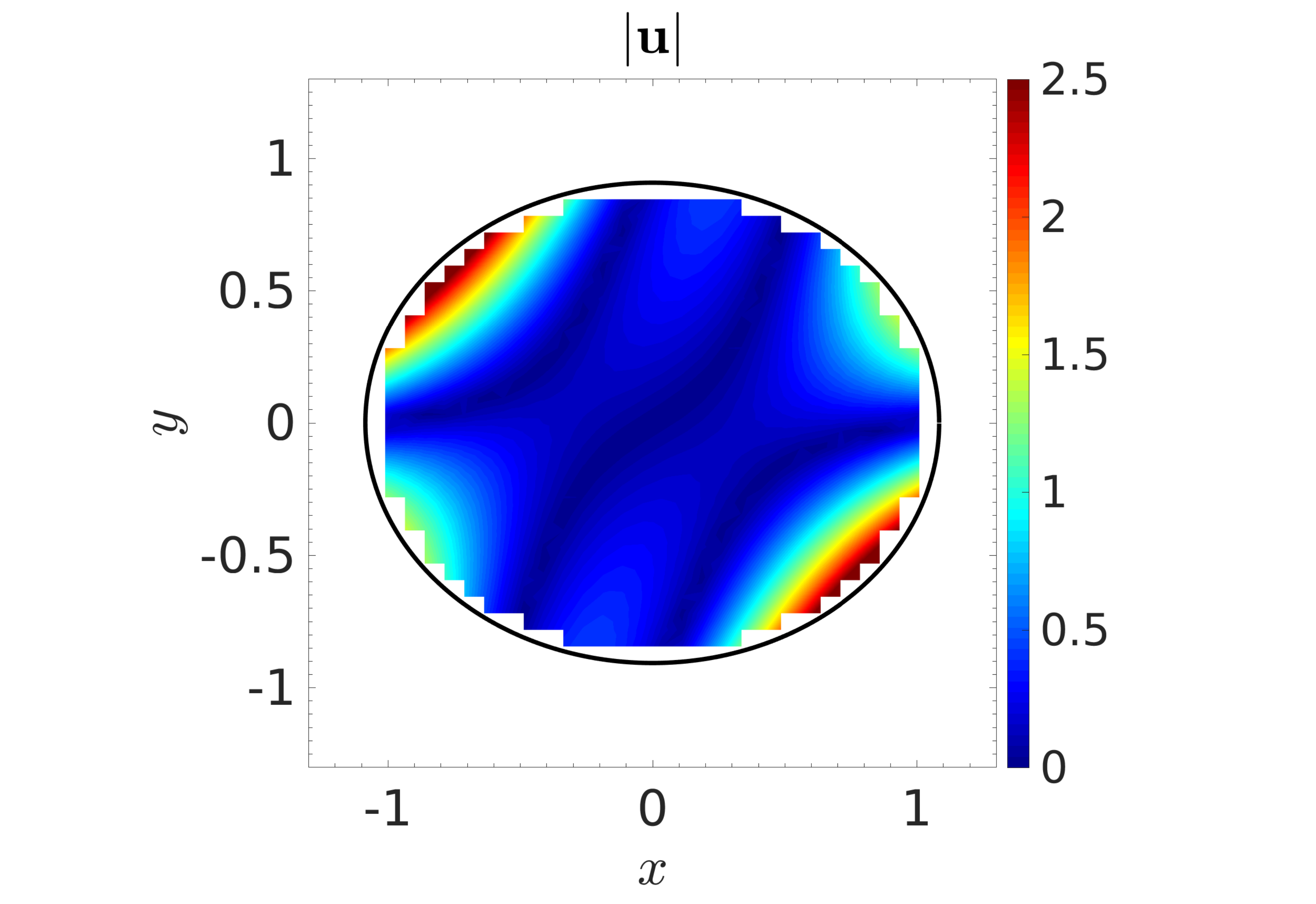} } 
      \subfigure[]{\includegraphics[trim=5cm 0cm 6cm 0cm, clip=true,width=0.23\textwidth]{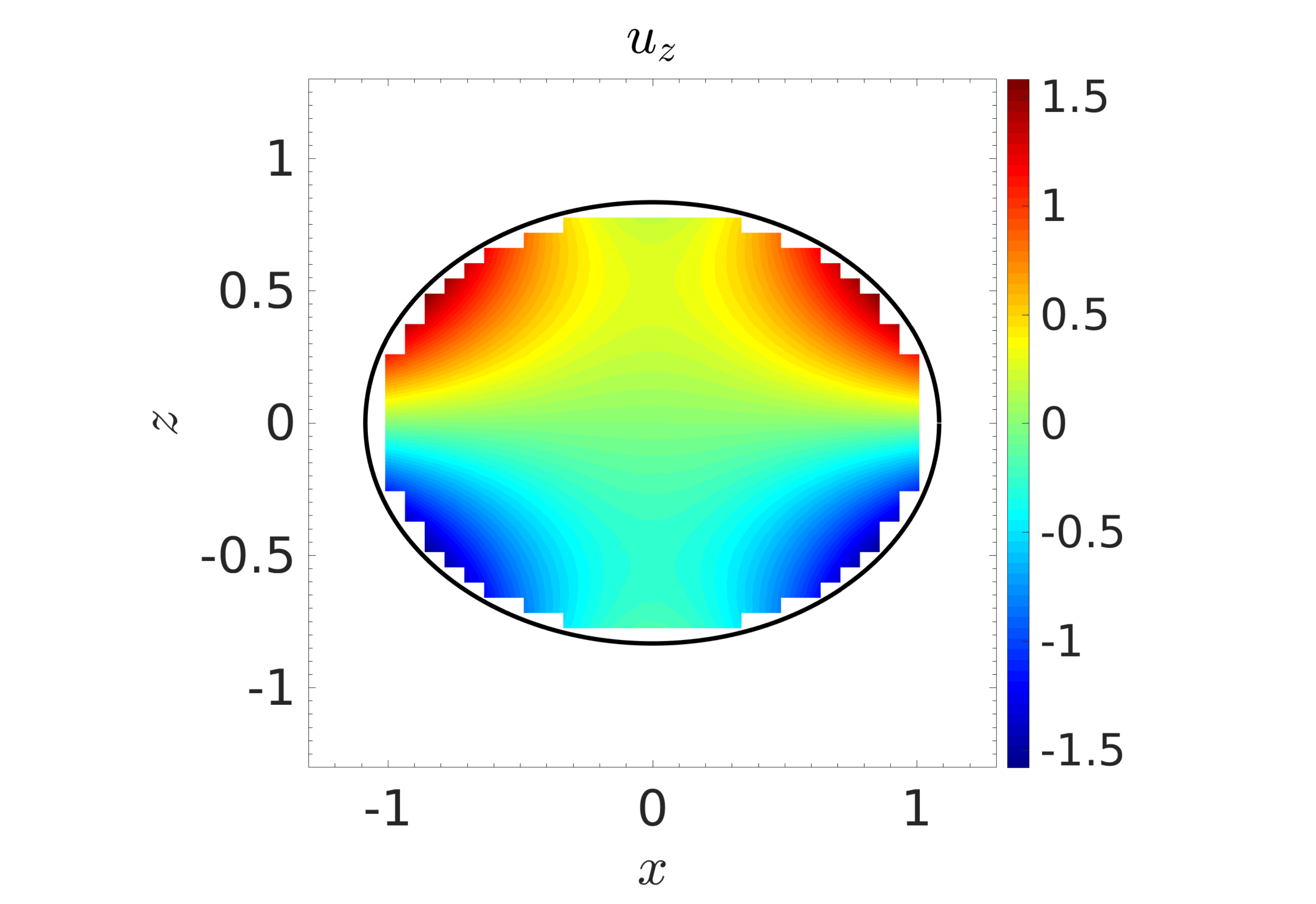} } 
      \end{center}
       \caption{Eigenfunction of the most unstable mode with $\ell \leq 4$, $\Omega=0.4,n=0.1$ and $A=0.1$. This mode has has $\omega=0.6566$ (in the bulge frame) and $\sigma=0.0195$. Left: $|\boldsymbol{u}|$ on the $xy$-plane. Right: $u_z$ on the $xz$-plane. Only the relative mode amplitudes are meaningful because this is a linear calculation.}
  \label{4c}
\end{figure}

\begin{figure}
  \begin{center}
      \subfigure[]{\includegraphics[trim=5cm 0cm 6cm 0cm, clip=true,width=0.23\textwidth]{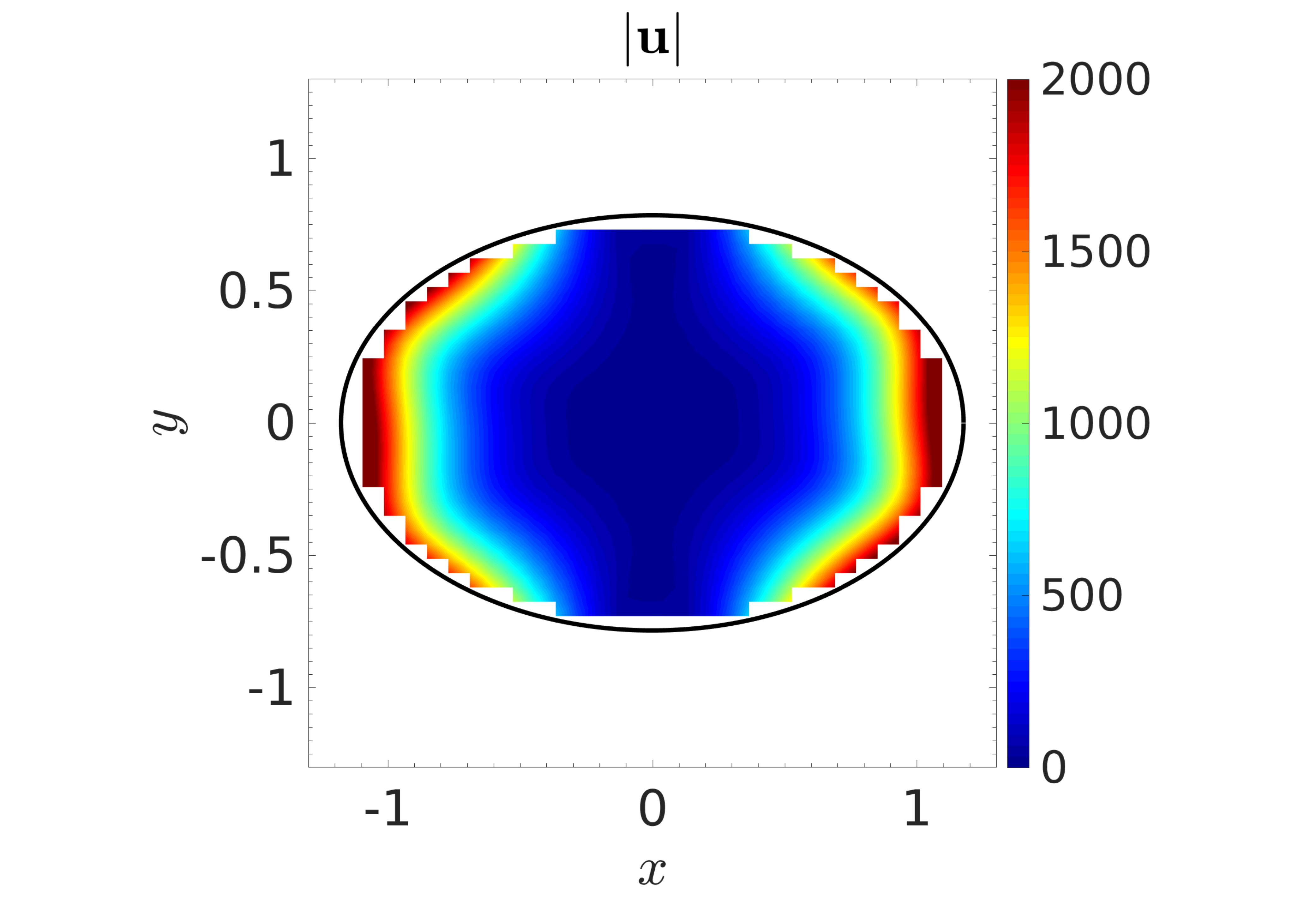} } 
      \subfigure[]{\includegraphics[trim=5cm 0cm 6cm 0cm, clip=true,width=0.23\textwidth]{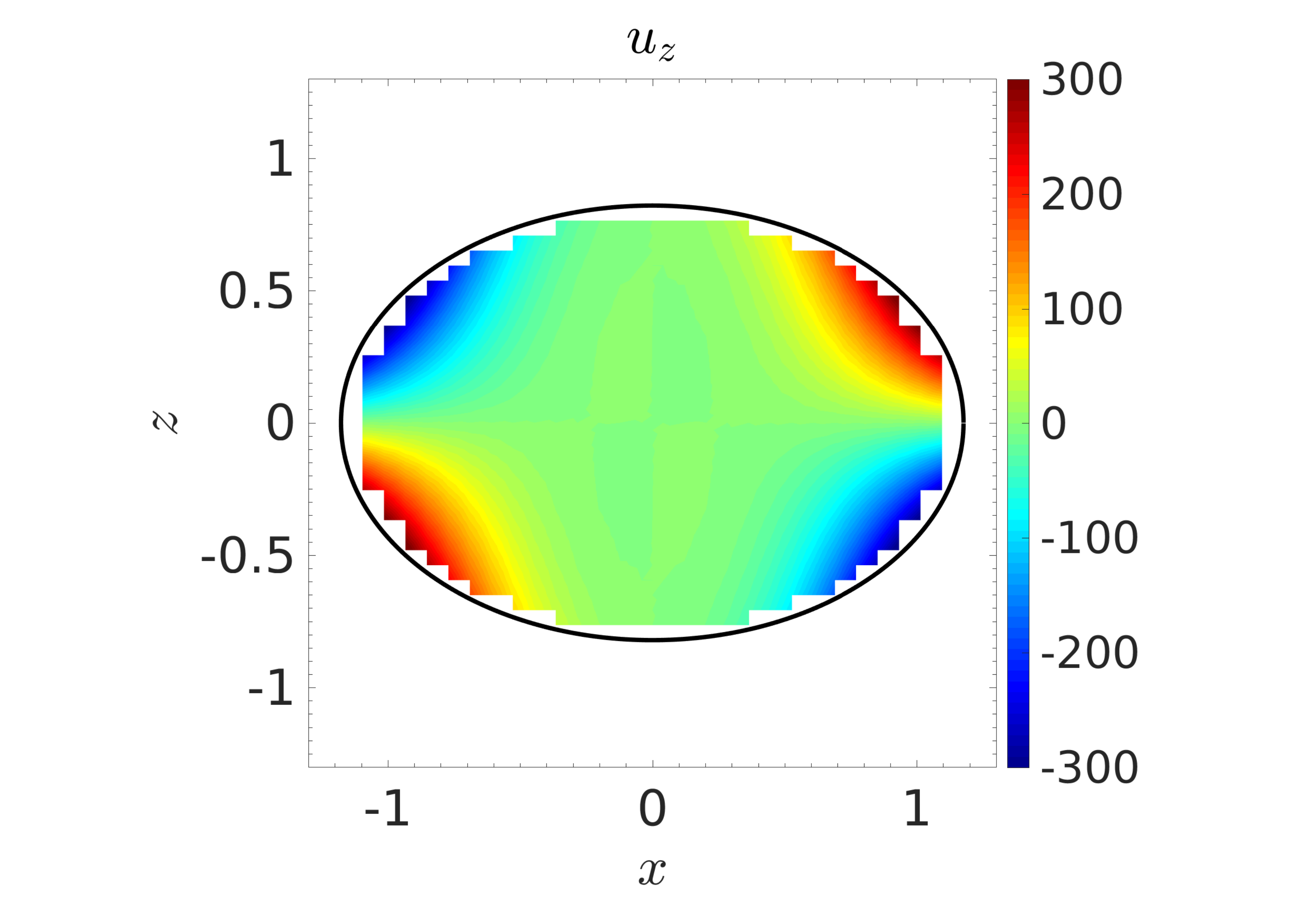} }
      \end{center}
       \caption{Eigenfunction of the most unstable mode with $\ell \leq 5$, $\Omega=0.2,n=-0.3$ and $A=0.15$. This mode has has $\omega=0$ (in the bulge frame) and $\sigma=0.1688$. Left: $|\boldsymbol{u}|$ on the $xy$-plane. Right: $u_z$ on the $xz$-plane. Only the relative mode amplitudes are meaningful because this is a linear calculation.}
  \label{4d}
\end{figure}

The third row of Fig.~\ref{7} shows that further instability bands appear when modes with $\ell\leq 5$ are considered. One such example is plotted in Fig.~\ref{4c} for an aligned spin with $\Omega=0.4, n=0.1$ and $A=0.1$. This instability is one with $\ell\leq 4$. Finally, we plot an example of the most unstable mode in a case with $n\lesssim -\Omega$ in Fig.~\ref{4d}, for a case with $\Omega=0.2, n=-0.3$ and $A=0.15$. This is more complicated than the purely horizontal epicyclic motion that is predicted by the local WKB analysis of Appendix \ref{WKB}, and has a nonzero vertical velocity field. Nevertheless, this mode is the global manifestation of the ``stack of pancakes"-type instability analysed in Appendix \ref{WKB}, which is centred on $n=-\Omega$ (we have confirmed that the most unstable mode for the corresponding case with $n=-0.2$ has the same form).

\subsection{Elliptical instability in a container with a rigid outer boundary}
\label{RB}

Our formalism allows the instability in a container that has a rigid boundary to be straightforwardly analysed. Such calculations are relevant to understand previous numerical simulations (which adopt a rigid boundary for computational convenience), in addition to laboratory experiments (where adopting a rigid boundary is inevitable). Our calculations in this section are accomplished by considering the polynomial basis to consist solely of the basis for the V$_\ell$-subspace. This eliminates SGMs but retains the IMs. We have verified that our code accurately reproduces the IMs with $\ell=2$ of a spheroidal \citep{Kerswell1994} and ellipsoidal \citep{Vant2014} container (this is done by setting $\gamma=0$, and varying $n$ with $A=10^{-3}$ to approximate $A=0$) -- furthermore, in \S~\ref{MS} we will compare the frequencies of IMs (and SGMs) with $\ell\leq 4$ with those computed from an independent analysis for a ``Maclaurin-like" spheroid, which, together, gives us confidence that we are correctly computing the IM frequencies. (Note that unlike the case with a free surface, the $\ell=2$ ``spin-over" mode of a spheroid is no longer a ``trivial" mode if the boundary is rigid.)

The elliptical instability in a rigid ellipsoidal container has previously been studied for modes with $\ell\leq3$ \citep{GP1992,Kerswell2002}. Here, we undertake calculations for modes with $\ell> 3$ for the first time. The fourth row of Fig.~\ref{7} shows the maximum growth rate of the resulting instability on the $(n,\Omega)$-plane, assuming that the shape of the ellipsoid is still that predicted by Eqs.~\ref{shape1} and \ref{shape2}. The strong correspondence between the third (free surface) and fourth (rigid boundary) rows, indicates that freedom of the outer boundary is unimportant for the elliptical instability. This makes sense, because this instability is one of IMs, and these only weakly perturb the surface. At large $\Omega$, IMs weakly move the surface if it is free (so cannot be exactly represented solely using the V$_\ell$-basis), which probably explains the minor differences between the third and fourth rows when $\Omega\gtrsim 0.5$.

\begin{figure}{H}
  \begin{center}
  \subfigure{\includegraphics[trim=4cm 0cm 8cm 0cm, clip=true,width=0.32\textwidth]{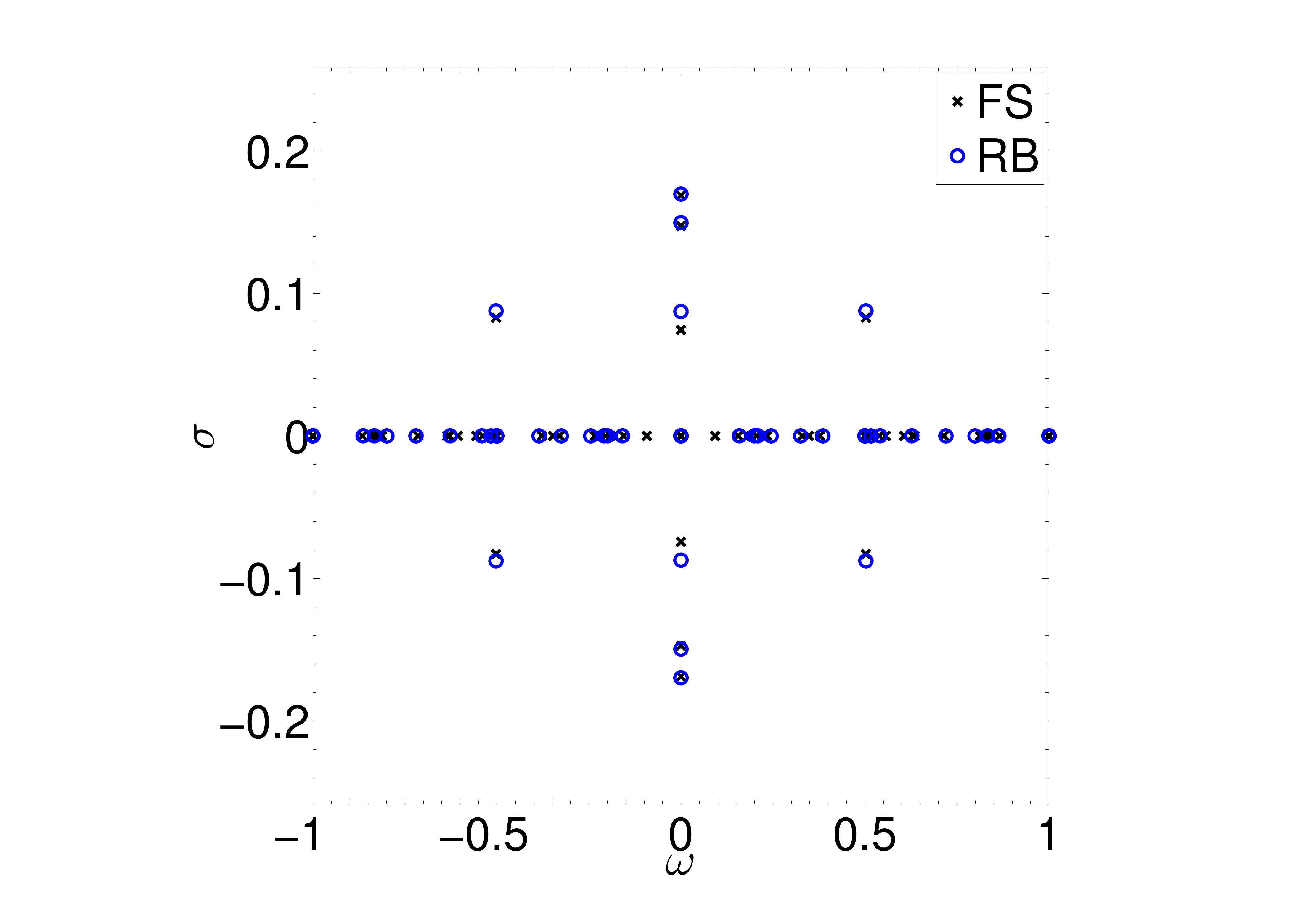} }
   \subfigure{\includegraphics[trim=4cm 0cm 8cm 0cm, clip=true,width=0.32\textwidth]{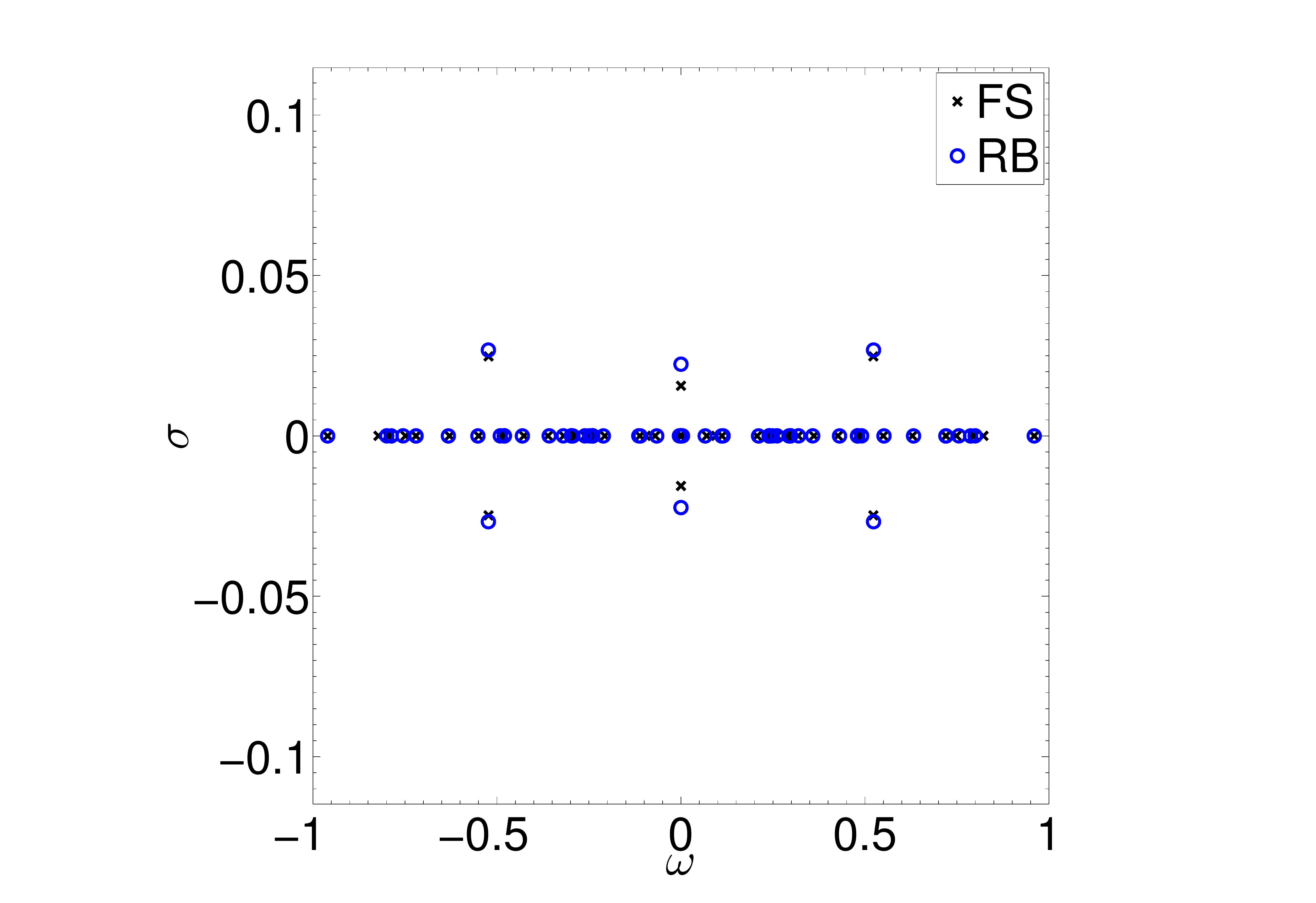} }
    \subfigure{\includegraphics[trim=4cm 0cm 8cm 0cm, clip=true,width=0.32\textwidth]{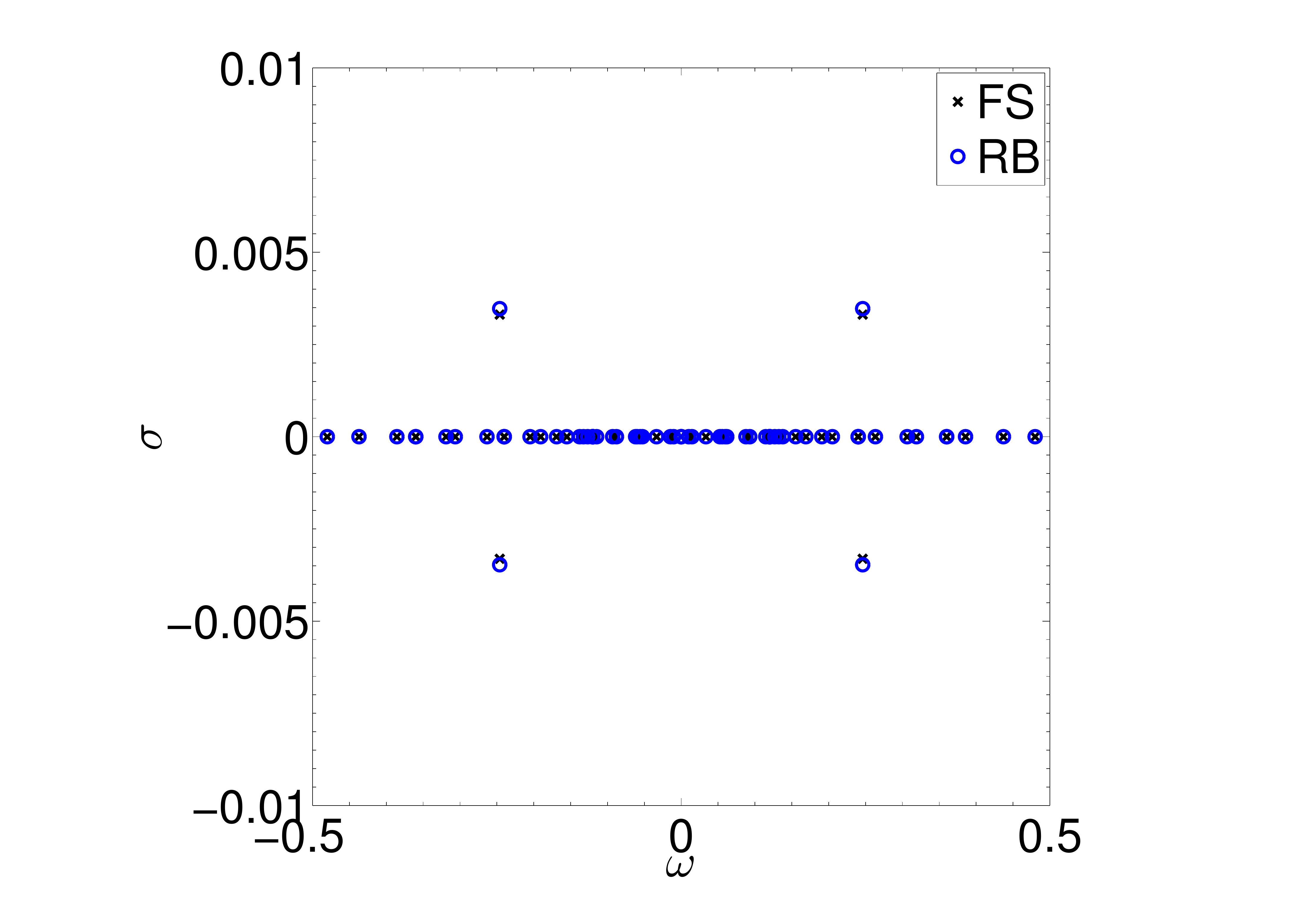} }
       \end{center}
  \caption{Comparison of the spectrum (frequency $\omega$ in the bulge-frame vs growth rate $\sigma$ for all modes) of an ellipsoid with a free surface (FS) and a rigid boundary (RB) for the same container shape for three example cases solving for modes with $\ell\leq5$. Top: $\Omega=0.2, n=-0.3,A=0.15$, where we obtain instability for $n\leq -\Omega$ (where the elliptical instability is not normally thought to operate). Middle: $\Omega=0.3, n=0.06, A=0.15$. Bottom: $\Omega=0.1,n=-0.02,A=0.05$. This demonstrates that the elliptical instability has similar properties in a container with a rigid boundary. (Note that some SGMs appear in this frequency range in the top two panels, corresponding with black crosses that do not match blue circles).}
  \label{8}
\end{figure}

In Fig.~\ref{8}, we plot the global spectrum on the $(\omega,\sigma)$-plane for three illustrative examples where results computed that assume a free surface and a rigid boundary are compared (with $\ell_\mathrm{max}=5$). The frequencies of the IMs (shown here in the inertial frame) are negligibly affected by constraining the boundary to be rigid. The instability growth rates are only weakly affected, typically being slightly stronger if the boundary is rigid. We have verified that the eigenfunctions in each case are also similar.

Our results suggest that the elliptical instability in a body with a free surface is well described by considering the instability in a container with a rigid boundary. This provides support for the use of numerical simulations (and laboratory experiments) that adopt a rigid boundary to study the instability. In \cite{Barker2015b}, we will show that the nonlinear evolution is also similar in both cases (except that the shape of the ellipsoid cannot self-consistently evolve unless it is modelled as a free surface).

In the next section we turn to study the effect of rotation and tidal deformation on the mode frequencies, before finishing with our conclusions.

\section{Effects of rotation and tidal deformation on mode frequencies}
\label{MS}

\begin{figure}
  \begin{center}
      \subfigure{\includegraphics[trim=7cm 0cm 8cm 2cm, clip=true,width=0.35\textwidth]{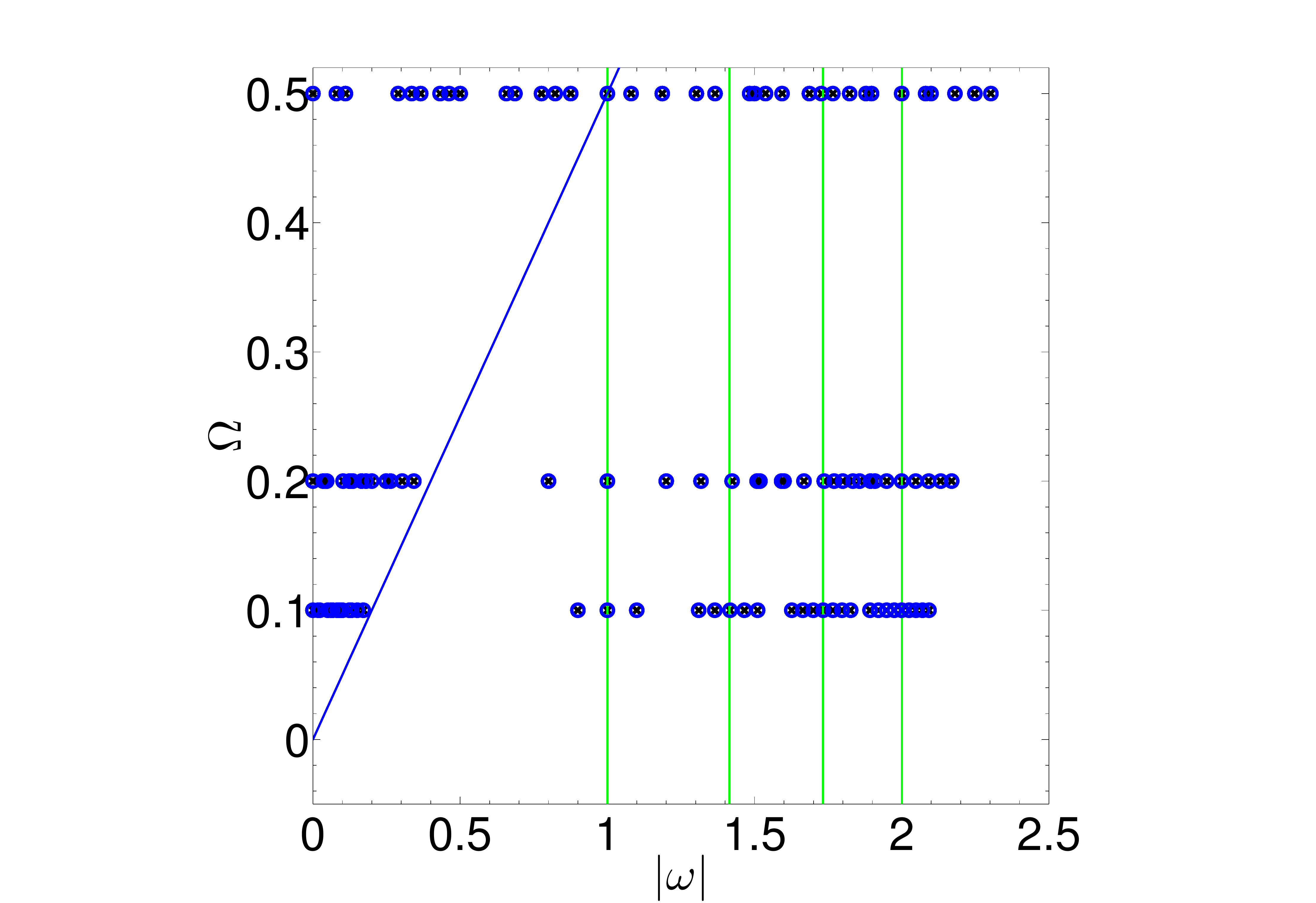} }
       \end{center}
       \caption{Code test: magnitude of the mode frequencies in the fluid frame (black crosses) compared with predictions from an independent analysis for a Maclaurin spheroid (blue circles) up to $\ell=4$ \citep{Braviner2015}. The agreement is excellent. For reference, we have plotted the SGM frequencies for a non-rotating sphere as green vertical lines, and $|\omega|=2\Omega$ as the blue line.} 
      \label{1}
\end{figure}

If we consider an isolated non-rotating planet (i.e.~$\Omega=n=A=0$), its shape would be spherical, and the only free oscillation modes would be SGMs. These have frequencies\footnote{Modes with $\ell=1$ are physically unrealistic modes in which the whole body oscillates in the fixed background potential, which would be trivial modes (with zero frequency) if we solved for the self-gravity of the body.} $\omega_{\ell}= \pm\sqrt{\ell}\omega_d$ for a given harmonic degree $\ell$ (e.g.~\citealt{Cowling1941}). The frequencies of modes with $\ell\gg 1$ ultimately match those of a self-gravitating fluid, but their frequencies are somewhat larger than those of a self-gravitating fluid for small $\ell$ -- however, we expect the qualitative behaviour of these modes to be similar even when $\ell$ is small.

\begin{figure}
  \begin{center}
      \subfigure{\includegraphics[trim=7cm 0cm 9cm 2cm, clip=true,width=0.35\textwidth]{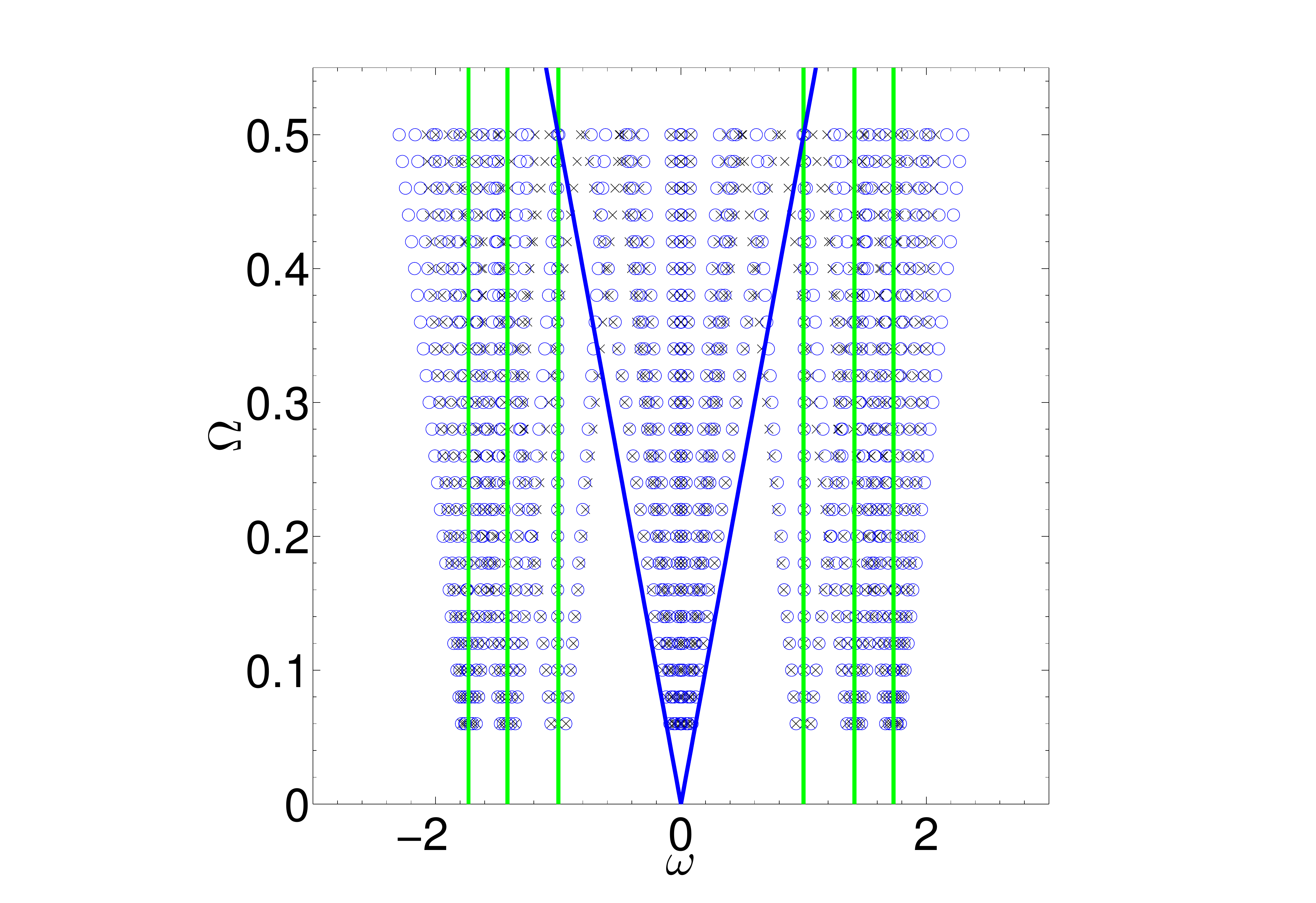} }
       \end{center}
  \caption{Magnitude of (real) frequencies in the fluid frame as a function of rotation ($\Omega$) for modes with $\ell\leq3$. The vertical green dashed lines are at $\omega_\ell=\pm \sqrt{\ell}\omega_d$ (the SGMs of a non-rotating sphere) and the blue lines demarcate the IM part of the spectrum $\omega=\pm2\Omega$. For the prograde sectoral SGMs, rotation reduces their frequencies, moving them closer to the IM spectrum. The black crosses indicate the full solutions and the blue circles indicate solutions that have been computed by minimising the centrifugal deformation by multiplying the centrifugal acceleration by $0.01$ for the basic state -- the agreement between these for most modes indicates that the Coriolis force is responsible for the frequency variation of lowest frequency SGMs.}
  \label{2}
\end{figure}

A planet that is rotating in the absence of a tidal deformation is an oblate spheroid which is dynamically stable (unless $\Omega \gtrsim \omega_d$). Its free modes of oscillation consist of the IMs and SGMs of a ``Maclaurin-like" spheroid. To verify that our code correctly calculates the modes with $\ell\leq 4$, we compare the magnitude of the computed mode frequencies in the fluid frame (black crosses) with predictions from an independent analysis for a ``Maclaurin-like" spheroid (\citealt{Braviner2015}, based on \citealt{Braviner2014}; shown as blue circles) for several $\Omega$ in Fig.~\ref{1}. To make such a comparison we work in the frame that rotates at the rate $\Omega$, choosing $\Omega=n$ and\footnote{To approximate $A=0$ we must choose a small nonzero value for $A$ because solid ellipsoidal harmonics are singular when $a=b$.} $A=10^{-3}$. (For this problem all particle relabelling modes have zero frequency.) Inspection of Fig.~\ref{1} shows that our code accurately reproduces all of the modes. For reference, the frequencies of non-rotating SGMs are plotted as the green vertical lines and the IM part of the spectrum ($|\omega|\leq 2\Omega$) is demarcated by the blue lines.

\cite{Braviner2014} demonstrated that rotation affects the frequencies of SGMs in a Maclaurin spheroid. This can already be seen in Fig.~\ref{1}, but we further explore this behaviour in Fig.~\ref{2}, where modes with $\ell\leq 3$ have been computed by choosing $\Omega=n$ and $A=0.01$ (to approximate $A=0$), plotted as black crosses. The IM part of the spectrum ($|\omega|\leq 2\Omega$) is demarcated by the blue lines, and the SGM frequencies of a non-rotating sphere are again plotted as green vertical lines. Fig.~\ref{2} illustrates that the frequency of the lowest frequency SGM for each $\ell$ decreases as $\Omega$ is increased, entering the IM part of the spectrum when $\Omega\gtrsim 0.35$ for $\ell=1$ and $\Omega\gtrsim 0.45$ for $\ell=2$. These modes are the prograde sectoral SGMs, which have azimuthal wavenumbers $m$ that satisfy $\ell=m$.

Fig.~\ref{2} shows that the Coriolis force is the primary cause for this frequency variation, and that the surface deformation plays a weaker (but non-negligible) role. This is shown by the reasonable agreement between the black crosses and the blue circles for these modes, where the latter represent solutions in which the centrifugal acceleration is artificially reduced by a multiplicative factor of $0.01$ in the construction of the basic state of the ellipsoid (the centrifugal acceleration does not directly affect the perturbations), so that the body is approximately spherical\footnote{Note that several of the modes illustrated by blue circles are unstable (``spin-down") modes 
because the ellipsoidal shape is not an equilibrium in this case.}. (We comment that a smaller centrifugal deformation may be in fact be more relevant for realistic bodies with strong central condensations.) The prograde sectoral modes are the lowest frequency SGMs for each $\ell$, and these are shifted to lower frequency as $\Omega$ is increased. For these modes with $\ell\leq2$, the centrifugal deformation is not primarily responsible for their frequency variation. However, modes with $\ell=3$ are much more strongly affected by the deformation. This can be understood if we realise that modes with larger $\ell$ are more strongly localised near to the surface (in a sphere the magnitudes of their radial displacements scale as $r^{\ell-1}$), so we expect them to be more strongly affected by the shape of the surface than lower $\ell$ modes. Fig.~\ref{2}  also shows that IMs are only weakly affected by the centrifugal deformation until $\Omega\gtrsim 0.4$.

\begin{figure}
  \begin{center}
      \subfigure{\includegraphics[trim=7cm 0cm 9cm 0cm, clip=true,width=0.3\textwidth]{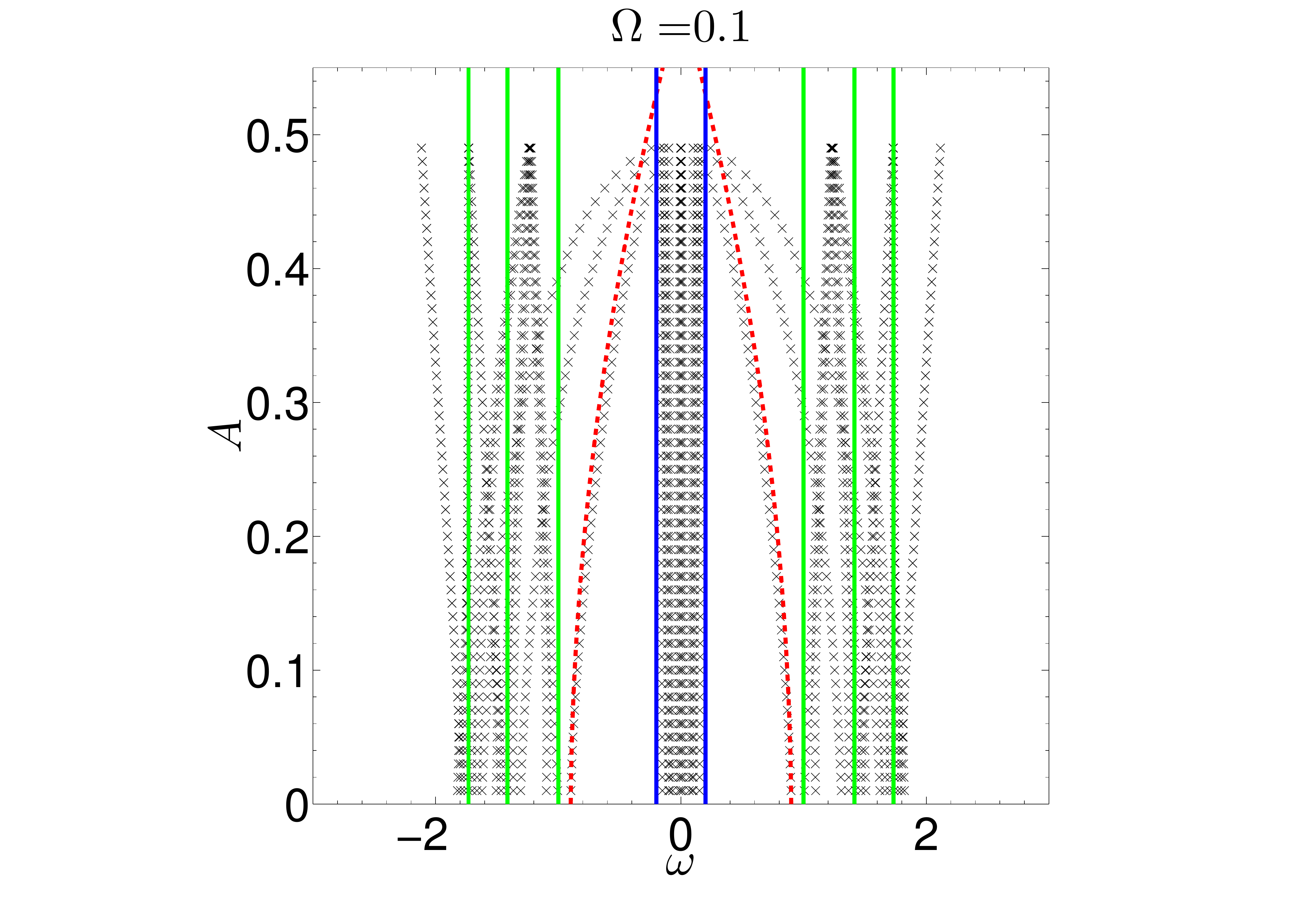} }
      \subfigure{\includegraphics[trim=7cm 0cm 9cm 0cm, clip=true,width=0.3\textwidth]{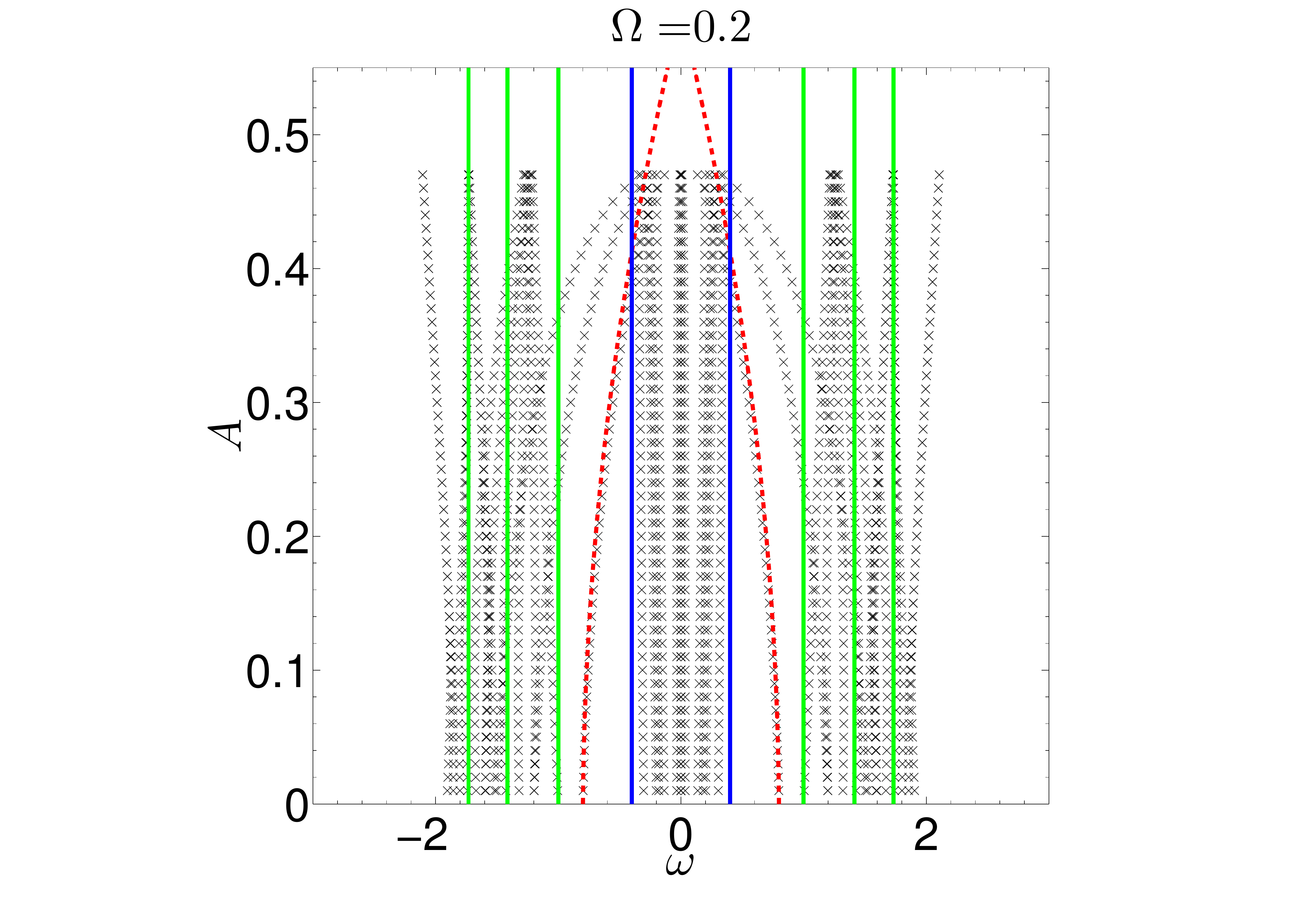} }
      \subfigure{\includegraphics[trim=7cm 0cm 9cm 0cm, clip=true,width=0.3\textwidth]{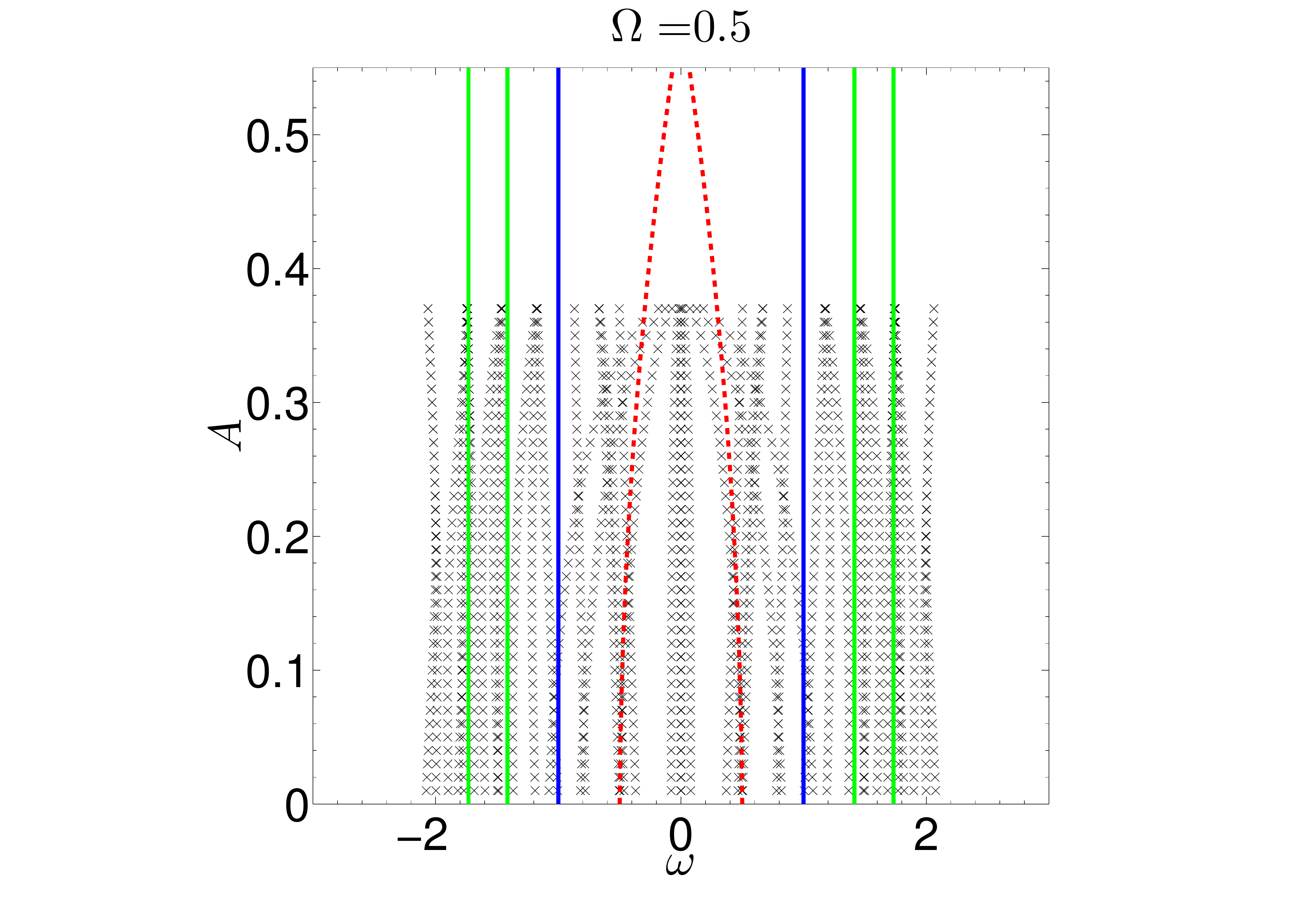} }
       \end{center}
  \caption{Magnitude of (real) frequencies as a function of tidal amplitude in the rotating frame with $\Omega=n$ for modes with $\ell_\mathrm{max}=3$. The vertical green lines show $\omega_\ell=\pm \sqrt{\ell}\omega_d$ (the SGMs of a non-rotating sphere) and the blue lines show $\omega=\pm 2\Omega$, which demarcates the IM part of the spectrum. The lowest frequency SGM (the prograde sectoral mode with $\ell=m$) decreases as the tidal amplitude is increased, until its frequency becomes comparable with the IMs. Note that $A>0.1$ or so are more extreme than observed planets. Note that when the prograde sectoral SGM frequencies pass through zero, an  ellipsoidal shape can no longer be found in equilibrium \citep{C1987}.}
  \label{3}
\end{figure}

Finally, we briefly study the effects of a tidal deformation on the mode frequencies. Fig.~\ref{3} shows the computed mode frequencies (up to $\ell=3$) in the rotating frame as a function of $A$ for $\Omega=n=0.1,0.2$ and $0.5$. The frequency of the lowest frequency SGM (the prograde sectoral mode with $\ell=m$) decreases with increasing tidal amplitude, eventually entering the IM part of the spectrum (demarcated by blue lines) for a critical $A$ that depends on $\Omega$. Based on expectations from a weakly nonlinear oscillator, the frequency shift for a mode not undergoing parametric resonance (such as the SGMs) should be $\delta \omega\propto A^2$ \citep{Landau1969}. To check this, we have plotted $\omega=\pm(1-\Omega+10 A^2)$ as the red dashed line in each panel, which shows that for $A\lesssim 0.3$, a quadratic dependence on $A$ (the factor of 10 is chosen to fit the data) indeed approximately describes the frequency shift of the $\ell=1$ prograde sectoral SGM (the adopted dependence on $\Omega$ results from the Coriolis force and can be obtained analytically for slow rotation in the case of a sphere). The frequencies of the global IMs are also weakly affected by $A$ (e.g.~\citealt{Vant2014}), but to a much lesser extent than SGMs (which are more sensitive to the shape of the ellipsoid), and this effect appears negligible for realistic values of $A\lesssim 0.1$. 

The combination of a larger tidal deformation and faster rotation work together to decrease the frequencies of the prograde sectoral SGMs and to shift them towards the IM part of the spectrum. When these frequencies go through zero this represents the classical dynamical instability of the Roche-Riemann ellipsoids \citep{C1987}, and is related to the Roche limit. Beyond this point, tidal and centrifugal forces are too strong to be balanced by the planet's gravity, so a stable equilibrium does not exist for such a configuration.

Consider the tidal circularisation of a planet on an eccentric orbit that is already spin-synchronised ($\Omega=n$). In this case, the dominant tidal frequency typically has a magnitude equal to its orbital frequency $\Omega\leq 2\Omega$. Similarly, for the tidal  synchronisation of a planet that is initially rotating faster than its orbit ($\Omega>n$), the magnitude of the dominant tidal frequency (usually 2$\gamma$) eventually decreases to become smaller than $2\Omega$. The shift in the frequencies of the prograde sectoral SGMs (particularly the one that corresponds with $\ell=m=2$) towards the IM part of the spectrum increases the prospect of their resonant (or non-resonant) excitation by realistic tidal forcing. This suggests that the amplitude of the $\ell=m=2$ ``equilibrium tide" deformation could be amplified by rotation and tidal gravity (over expectations from linear theory). We relegate further investigation of this possibility to future work. To give some numbers, for observed hot Jupiters, it is reasonable to consider $A\sim 0.05$ and $\Omega\sim 0.3$ (assuming spin synchronisation) to provide upper bounds, so we might expect SGM frequencies to be shifted by up to $30\%$ over their values in a non-rotating sphere. Note that while we expect self-gravity to quantitatively affect the mode frequencies, the qualitative behaviour presented here should carry over (this has already been demonstrated for the effect of rotation on SGMs; \citealt{Braviner2014}). 

\section{Conclusions}
 
We have revisited the global modes and instabilities of homogeneous rotating ellipsoidal fluid masses. These are the simplest models of gaseous planets or stars with finite tidal and rotational deformations, in which nonlinear tidal effects can be studied.
Our primary aim was to study global aspects of the elliptical instability, which is a fluid instability of elliptical streamlines \citep{Kerswell2002} that could be important for the circularisation, synchronisation and spin-orbit alignment for the shortest-period hot Jupiters. 

We have applied the elegant formalism of \cite{L1989a,L1989b} and \cite{LL1996} to analyse the global (linear) stability of tidal flows in (aligned or anti-aligned) non-synchronously rotating planets or stars with finite tidal and rotational deformations. We have also complemented our global stability analysis with a local WKB analysis (Appendix \ref{WKB}). This paper has primarily focused on the longest-wavelength global modes. This is because, if these are excited, they might be expected to dominate the instability-driven turbulence and its resulting tidal dissipation (as indicated by the previous local simulations of \citealt{BL2013,BL2014}).  In addition, the non-negligible tidal amplitude for some hot Jupiters increases the prospect of global modes being excited inside these planets (since these modes can be excited out of exact resonance). In a companion paper \citep{Barker2015b}, we use our results to help us understand global simulations of the elliptical instability.

Our main result is Fig.~\ref{7}, which maps the growth rate of the fastest growing global modes (with $\ell \leq 5$) as a function of the orbital and spin angular frequencies for realistic tidal amplitudes appropriate for observed hot Jupiters. In general, the elliptical instability has its largest growth rates for anti-aligned (retrograde) spins. This suggests that the resulting tidal dissipation could be more efficient for retrograde spins, so that tidal evolution may preferentially drive systems out of anti-aligned configurations. We have also identified a violent instability which occurs outside of the usual frequency range for the elliptical instability. This occurs for anti-aligned spins if the tidal amplitude is sufficiently large, but nevertheless comparable with the maximum values observed for hot Jupiters, allowing the elliptical instability to be excited for a wider range of orbital frequencies. We explain this result in Appendix \ref{WKB} as being due to the finite widths of the ``stack of pancakes"-type instability bands, which are centred on $n=-\Omega$. We simulate several anti-aligned cases in the companion paper, finding that the elliptical instability could play an important role in driving tidal synchronisation and spin-orbit alignment. 

On the other hand, we expect the elliptical instability to be much less effective in driving evolution of the stellar obliquity and spin (e.g.~\citealt{Albrecht2012}) since the tidal amplitude in the star is typically much smaller than in the planet. In addition, stellar spins are typically much slower than the orbits of observed hot Jupiters, so these systems primarily lie outside the frequency range in which the elliptical instability could be excited (at least at their current ages).
It has also been proposed that the excitation of the ``spin-over" mode (effectively a rigid tilting on the spin axis of the body) by the elliptical instability could produce the spin-orbit misalignments observed for some hot Jupiter host stars. However, this mode is not excited for prograde rotations (see the top panels of Fig.~\ref{7}), so it cannot \textit{produce} spin-obit misalignments in exoplanet host stars (cf.~\citealt{Cebron2013}). However, the excitation of this mode, and its dissipation through secondary instabilities, could play a role in driving tidal spin-orbit alignment in systems that are already misaligned (e.g.~\citealt{Lai2012}). 

We have also analysed the elliptical instability in a rigid ellipsoidal container for global modes with $\ell\leq5$, which extends previous work where this was accomplished for $\ell\leq3$ \citep{GP1992,Kerswell2002}. We find that the freedom of the boundary is unimportant for the instability, and our results are quantitatively very similar to the realistic case with a free surface. This makes sense, because the instability excites inertial modes, and these only weakly perturb the surface if we allow it to be free. This agreement is helpful, because numerical simulations (and laboratory experiments) which adopt a rigid boundary are much more straightforward and computationally less expensive (e.g.~\citealt{LeBars2010,Cebron2010,Cebron2013}). We will compare the nonlinear outcome of the instability in both cases in our companion paper \citep{Barker2015b}.

Finally, we have analysed the rotation and tidal amplitude variation of the frequencies of SGMs, extending \cite{Braviner2014} to study finite tidal deformations. We find that larger rotations and larger tidal deformations both act to decrease the frequencies of the prograde sectoral surface gravity modes (those with $\ell=m$) towards the inertial mode part of the spectrum (which has $|\omega|\leq 2|\Omega|$). This increases the prospect of their resonant or non-resonant excitation by realistic tidal forcing, since tidal frequencies are typically comparable in magnitude with $|\Omega|$. This behaviour could act to amplify the equilibrium tide response over expectations from linear theory.

To determine the astrophysical importance of the elliptical instability for tidal dissipation, we must perform global numerical simulations to explore its nonlinear outcome. The results of such calculations are presented in the companion paper \citep{Barker2015b}.

\section*{Acknowledgements}
AJB is supported by the Leverhulme Trust and Isaac Newton Trust through the award of an Early Career Fellowship. The early stages of this research were supported by STFC through grants ST/J001570/1 and ST/L000636/1. HB was supported by a studentship funded by STFC and Trinity College, Cambridge. We would like to thank the referee, Jeremy Goodman, for several suggestions which have allowed us to improve the paper.

\appendix

\section{Effects of self-gravity}
\label{SG}

The effects of self-gravity on the shape of an ellipsoid for a given $A$, $\Omega$ and $n$ can be understood most simply for a Maclaurin spheroid (with $a=b$). The magnitude of the rotation rate of a Maclaurin spheroid (with self-gravity) in equilibrium can be determined from its eccentricity 
\begin{eqnarray}
e=\sqrt{1-\frac{c^2}{a^2}},
\end{eqnarray}
by \citep{C1987}
\begin{eqnarray}
\frac{\Omega}{\omega_d} &=&\sqrt{\frac{3}{2e^2} \left((3-2e^2)\sqrt{1-e^2}\frac{\sin^{-1} e}{e}-3(1-e^2)\right)}, \\
&\approx& 0.633 e,
\end{eqnarray} 
when $e\ll 1$. On the other hand, our case with a fixed potential has 
\begin{eqnarray}
\frac{\Omega}{\omega_d} = e.
\end{eqnarray}
Hence, if self-gravity is included, the body need not rotate as rapidly to produce a given eccentricity than if we adopt a fixed potential. This demonstrates that the inclusion of self-gravity will quantitatively modify the shape of the ellipsoid for a given $A$, $\Omega$ and $n$. However, the properties of the elliptical instability are unlikely to be significantly modified, since IMs are only weakly affected by self-gravity. To justify this, we use Tables 2.1 and 2.2 in \cite{Braviner2015} to compute the maximum percentage difference in the IM frequencies (normalised by the rotation rate) calculated with and without self-gravity. For a rapidly rotating spheroid with an eccentricity of $e=0.5$, we found that neglecting self-gravity shifts the $\ell=3$ IM frequencies by no more than $3.5\%$. For the $\ell=4$ modes the greatest shift is $2.9\%$ (in each case removing self-gravity decreases the mode frequency). This effect is expected to increase with the eccentricity of the centrifugal bulge, and we did indeed see the shifts to be less at lower values of $e$. Hence, IM frequencies are only very weakly affected by self-gravity. In addition, the amplitude of the surface perturbations caused by IMs (which are nonzero when $e\ne 0$) will be weakly modified by neglecting self-gravity \citep{Bryan1889,Braviner2014}, but this is primarily relevant for the direct tidal excitation of these waves \citep{BravinerOgilvie2015}, which we do not set out to consider (we restrict the tidal potential to $\ell=2$, which does not directly excite IMs). Instead, we wish to focus on the elliptical instability, where this modification can safely be neglected. Finally, given that a realistic gaseous planet is not homogeneous and is in fact centrally condensed, we might expect self-gravity to be much weaker in reality than would be suggested by its effects in a Maclaurin spheroid. This is one of the reasons that we have neglected self-gravity in this work, together with the fact that our model is much simpler to analyse.

\section{Tidal flows in a homogeneous ellipsoid}
\label{ST}
\cite{ST1992} derived a set of ODEs that describe the evolution of the shape and simplest internal flows of an incompressible homogeneous fluid body subjected to a tidal potential. We outline their formalism in this section for the case of an aligned (or anti-aligned) spin and orbit. The free surface of the ellipsoid is defined by the quadratic form $\boldsymbol{x}^{T}\mathrm{S} \boldsymbol{x}$=1 (cf.~Eq.~\ref{bdry}), where $\mathrm{S}$ is a symmetric matrix that describes the shape of the body. The internal velocity field is assumed to be of the form
\begin{eqnarray}
\label{STflow}
\boldsymbol{U}=\mathrm{A_F}\boldsymbol{x},
\end{eqnarray}
where
\begin{eqnarray}
\mathrm{A_F}=\gamma\left(\begin{array}{c c c}
 0 & -\frac{a}{b} & 0 \\
 \frac{b}{a} & 0 & 0 \\
 0 & 0 & 0
 \end{array}\right),
\end{eqnarray}
with $\mathrm{Tr(A_F})=0$. The pressure vanishes on the surface and has the form 
\begin{eqnarray}
p=p_{0}(t)(1-\boldsymbol{x}^{T}\mathrm{S} \boldsymbol{x}).
\end{eqnarray}
The combined gravitational and centrifugal potential is $\boldsymbol{x}^{T}\left(\mathrm{B_{B}+B_{T}+B_{C}}\right)\boldsymbol{x}$,
where the fixed background gravitational potential is represented by the matrix 
\begin{eqnarray}
\mathrm{B_{B}}=\frac{\omega_{d}^{2}}{2}\mathrm{I},
\end{eqnarray}
the centrifugal potential ($-\frac{1}{2}|\boldsymbol{n}\times\boldsymbol{x}|^{2}$) is represented by 
\begin{eqnarray}
\mathrm{B_{C}}=-\frac{n^{2}}{2}\left(\begin{array}{c c c}
1 & 0 & 0\\
 0 & 1 & 0 \\
 0 & 0 & 0
 \end{array}\right),
\end{eqnarray}
and the (quadrupolar) tidal potential is represented by
\begin{eqnarray}
\mathrm{B_{T}}=-\frac{A}{2}\left(\begin{array}{c c c}
 2 & 0 & 0 \\
 0 & -1 & 0 \\
 0 & 0 & -1
 \end{array}\right).
\end{eqnarray}
The Coriolis acceleration is taken into account by defining
\begin{eqnarray}
\mathrm{N}=n\left(\begin{array}{c c c}
  0 & -1 & 0 \\
 1 & 0 & 0 \\
 0 & 0 & 0
 \end{array}\right)
\end{eqnarray}
Starting from Eqs.~\ref{Eq1}--\ref{Eq3}, we can show that the shape and internal fluid motions evolve according to
\begin{eqnarray}
\frac{\mathrm{d}\, \mathrm{A_F}}{\mathrm{d}\, t}&=&-\mathrm{A_F^2}+2 p_{0}\mathrm{S} - 2\left(\mathrm{B_B}+\mathrm{B_T}+\mathrm{B_C}+\mathrm{N}\mathrm{A_F}\right), \\
\frac{\mathrm{d} \, \mathrm{S}}{\mathrm{d}\, t}&=&-\mathrm{A_F^T}\mathrm{S}-\mathrm{S}\mathrm{A_F},
\end{eqnarray}
where 
\begin{eqnarray}
p_{0}=\frac{\mathrm{Tr\left(A^2_F+2 (B_{B} +B_{C}+NA_F)\right)}}{2\mathrm{Tr(S)}},
\end{eqnarray}
since $\mathrm{Tr(B_{T})}=0$. These equations can be solved to find the steady shape and internal flow of the ellipsoid if the LHS is set to zero. Alternatively, the same result can be obtained via the second-order virial equations \citep{C1987}. In addition, for a given initial $\mathrm{A_F}$ and $\mathrm{S}$, we can determine how these should evolve, which allows us to test our global numerical simulations \citep{Barker2015b}.

\section{Construction of the basis vectors}
\label{basisfunctions}

In this section we briefly describe how to construct a basis for the space of allowable perturbations of our ellipsoidal planet that are solenoidal
vector polynomials up to degree $\ell_\mathrm{max}$. To do this, we consider two disjoint subspaces whose direct sum represents the whole space: U$_\ell$ and V$_\ell$.
U$_\ell$ is the subspace of irrotational motions that represent boundary perturbations up to degree $\ell$, and V$_\ell$ is the subspace of purely vortical perturbations up to degree $\ell$ that do not move the boundary.  We have written a Matlab script that computes both sets of basis functions for any degree\footnote{Though in practice we have found it difficult to \textit{accurately} construct a basis for U$_\ell$ if $\ell>5$.} $\ell$.

\subsection{Construction of a basis for U$_\ell$: irrotational motions that perturb the boundary}

We construct a basis for U$_\ell$ using
\begin{eqnarray}
\boldsymbol{\xi}=\nabla \phi,
\end{eqnarray}
where $\phi$ is a Cartesian polynomial of degree $\ell$ that satisfies $\nabla^2\phi=0$. The functions $\phi$ are proportional to the Cartesian representations of the solid ellipsoidal harmonics, which form a linearly independent set for degree $\ell$ \citep{L1953,Dassios2012}. A set of $N_U=\ell(\ell+2)$ of these functions is required to capture any perturbations of the boundary. SGMs up to degree $\ell$ are well represented by considering only this basis (at least for small values of $\Omega$ and $A$). An example set for $\ell=1$ are the vectors
\begin{eqnarray}
(1,0,0)^T, \;\;\; (0,1,0)^T, \;\;\; (0,0,1)^T,
\end{eqnarray}
and for $\ell=2$ are the three vectors
\begin{eqnarray}
(y,x,0)^T, \;\;\; (0,z,y)^T, \;\;\; (z,0,x)^T,
\end{eqnarray}
together with the two
\begin{eqnarray}
 \left(\frac{x}{a^2+\theta_i},\frac{y}{b^2+\theta_i},\frac{z}{c^2+\theta_i}\right)^T,
\end{eqnarray}
for $i=1$ and $2$, where $\theta_i$ are the roots of the quadratic
\begin{eqnarray}
\frac{1}{a^2+\theta}+\frac{1}{b^2+\theta}+\frac{1}{c^2+\theta}=0.
\end{eqnarray}
A similar construction can be carried out for $\ell>2$, for which we refer the reader to \cite{LL1996} for further details.

\subsection{Construction of a basis for $V_\ell$: vortical motions that do not perturb the boundary}

We construct a basis for V$_\ell$ by constructing the vectors (which are tangential to the boundary)
\begin{eqnarray}
\boldsymbol{\xi}_j=\begin{cases}
                                \nabla \left(p_j \left(1-\frac{x^2}{a^2}-\frac{y^2}{b^2}-\frac{z^2}{c^2}\right)\right)\times \boldsymbol{e}_x, \;\;\; j\in[1,N_2],   \\
                                \nabla \left(p_j \left(1-\frac{x^2}{a^2}-\frac{y^2}{b^2}-\frac{z^2}{c^2}\right)\right)\times \boldsymbol{e}_y, \;\;\; j\in[1,N_2],   \\
                                \nabla \left(p_j \left(1-\frac{x^2}{a^2}-\frac{y^2}{b^2}-\frac{z^2}{c^2}\right)\right)\times \boldsymbol{e}_z, \;\;\; j\in[1,N_1], 
            \end{cases}
\end{eqnarray}
where $p_j$ is a polynomial of degree $\ell-2$ or less, $N_1=\frac{1}{2}\ell(\ell-1)$ and $N_{2}=\frac{1}{6}\ell(\ell-1)(\ell+1)$. The total number of these vectors is $N_V=\frac{1}{6}\ell(\ell-1)(2\ell+5)$.
Specifically, we choose 
\begin{eqnarray}
\left\{p_j\right\}=\left\{\underbrace{1,x,y,\dots,x^{\ell-2},y^{\ell-2}}_{j=[1,N_1]},\underbrace{z,\dots,z^{\ell-2}}_{j=[N_1+1,N_{2}]}  \right\}.
\end{eqnarray}
The set of vectors constructed in this way is linearly independent \citep{L1989a}, and represents vortical perturbations that are tangential to the boundary of the unperturbed ellipsoid. These can be computed analytically for any degree $\ell$, and IMs in the limit $\Omega^2\ll \omega_d^2$ are well represented by considering only this set of basis vectors.

The set for $\ell=1$ is empty, whereas for $\ell=2$ we have the three vectors
\begin{eqnarray}
\left(0,\frac{z}{c^2},-\frac{y}{b^2}\right)^T, \;\;\; \left(-\frac{z}{c^2},0,\frac{x}{a^2}\right)^T, \;\;\; \left(\frac{y}{b^2},-\frac{x}{a^2},0\right)^T.
\end{eqnarray}
A similar construction can be carried out for $\ell>2$, and the set of vectors constructed is equivalent to the basis used by \cite{GP1992} for studying the elliptical instability (of IMs) in a rigid container. An equivalent set has also been adopted by \cite{WuRoberts2011} (where the basis vectors up to $\ell=5$ are written down explicitly) for studying instabilities of precessional motions in the Earth's core, and by \cite{Vant2014} for studying the free IMs of a triaxial ellipsoid in a rigid container (which was subsequently applied to study latitudinal libration-driven elliptical instability by \citealt{Vant2015}).

\section{Local WKB analysis of the elliptical instability}
\label{WKB}
In this section, we perform a complementary local WKB analysis of the elliptical instability to further aid our understanding, following \cite{Craik1989} and \cite{LL1996a}. This analysis aims to determine the stability of the base flow (Eq.~\ref{basicflow} or Eq.~\ref{STflow}) in the absence of boundaries, and its results should correspond with those of the global stability analysis in the limit $\ell\rightarrow \infty$. We study linear perturbations (with velocities $\boldsymbol{u}^{\prime}$) that satisfy Eq.~\ref{Eq1} with $\boldsymbol{u}=\boldsymbol{U}_{0}+\boldsymbol{u}^{\prime}$, and seek solutions that are plane-waves with time-dependent wavevectors, such that
\begin{eqnarray}
\boldsymbol{u}^{\prime}(\boldsymbol{x},t)=\mathrm{Re}\left[\hat{\boldsymbol{u}}(t)\mathrm{e}^{\mathrm{i}\boldsymbol{k}(t)\cdot \boldsymbol{x}}\right],
\end{eqnarray}
and similarly for $\Pi^{\prime}$ (a time-dependent wavevector is required to eliminate the term $\boldsymbol{U}_{0}\cdot \nabla \boldsymbol{u}^{\prime}$, which is linear in the spatial coordinates, thereby preventing analysis using time-independent plane-waves). The Fourier amplitudes satisfy Eq.~\ref{Eq1} if
\begin{eqnarray}
\label{WKBeq1}
&& \partial_{t}\hat{\boldsymbol{u}} + 2\boldsymbol{n}\times\hat{\boldsymbol{u}} + \mathrm{A_F}\hat{\boldsymbol{u}}+ \mathrm{i}\boldsymbol{k} \hat{\Pi}=0, \\
\label{WKBeq2}
&& \boldsymbol{k}\cdot \hat{\boldsymbol{u}}=0, \\
\label{WKBeq3}
&& \partial_{t} \boldsymbol{k} + \mathrm{A^T_F}\boldsymbol{k}=0,
\end{eqnarray}
where a superscript T denotes a transpose operation.
The latter implies $k_x=k_{x,0}\cos \gamma t-\frac{bk_{y,0}}{a}\sin \gamma t$, $k_y=k_{y,0}\cos \gamma t+\frac{ak_{x,0}}{b}\sin \gamma t$, and $k_z=k_{z,0}$, where $\boldsymbol{k}(0)=\left(k_{x,0},k_{y,0},k_{z,0}\right)$. 

We can eliminate $\hat{\Pi}$ in Eq.~\ref{WKBeq1} through the use of Eq.~\ref{WKBeq2}. This leaves a system of three first-order linear differential equations with coefficients that are periodic functions of time. Our aim is to determine the maximum growth rate of the instability as a function of $\Omega$, $n$ and $A$, analysing this system numerically using a Floquet method. For each choice of $\Omega$, $n$ and $A$ (allowing us to compute $a$ and $b$ as in \S~\ref{model}), we compute the maximum growth rate as the fastest growing solution from a range of initial wavevectors $\boldsymbol{k}(0)=(\sin\theta,0,\cos\theta)$ with unit degree spacing in $\theta\in[0^{\circ},90^{\circ}]$. For each calculation, we construct the monodromy matrix of linearly independent solutions by integrating the ODEs over one period ($2\pi/\gamma$) for initial conditions such that all velocity components except one are set to zero. The eigenvalues of the monodromy matrix allow us to obtain the complex growth rates of the instability.

\begin{figure*}
  \begin{center}
      \subfigure[$\ell\rightarrow \infty$]{\includegraphics[trim=3cm 0cm 6cm 0cm, clip=true,width=0.24\textwidth]{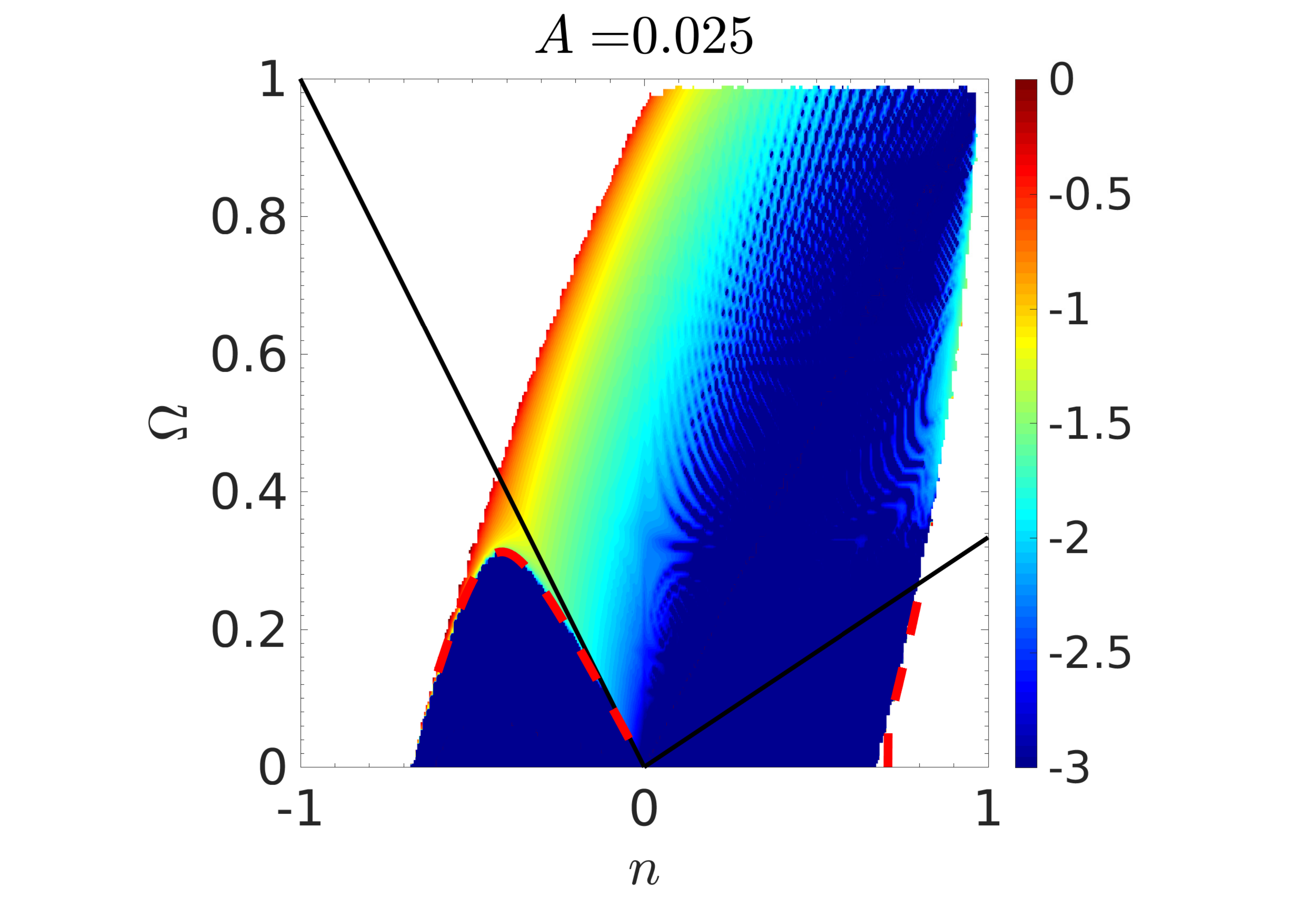} } 
      \subfigure[$\ell\rightarrow \infty$]{\includegraphics[trim=3cm 0cm 6cm 0cm, clip=true,width=0.24\textwidth]{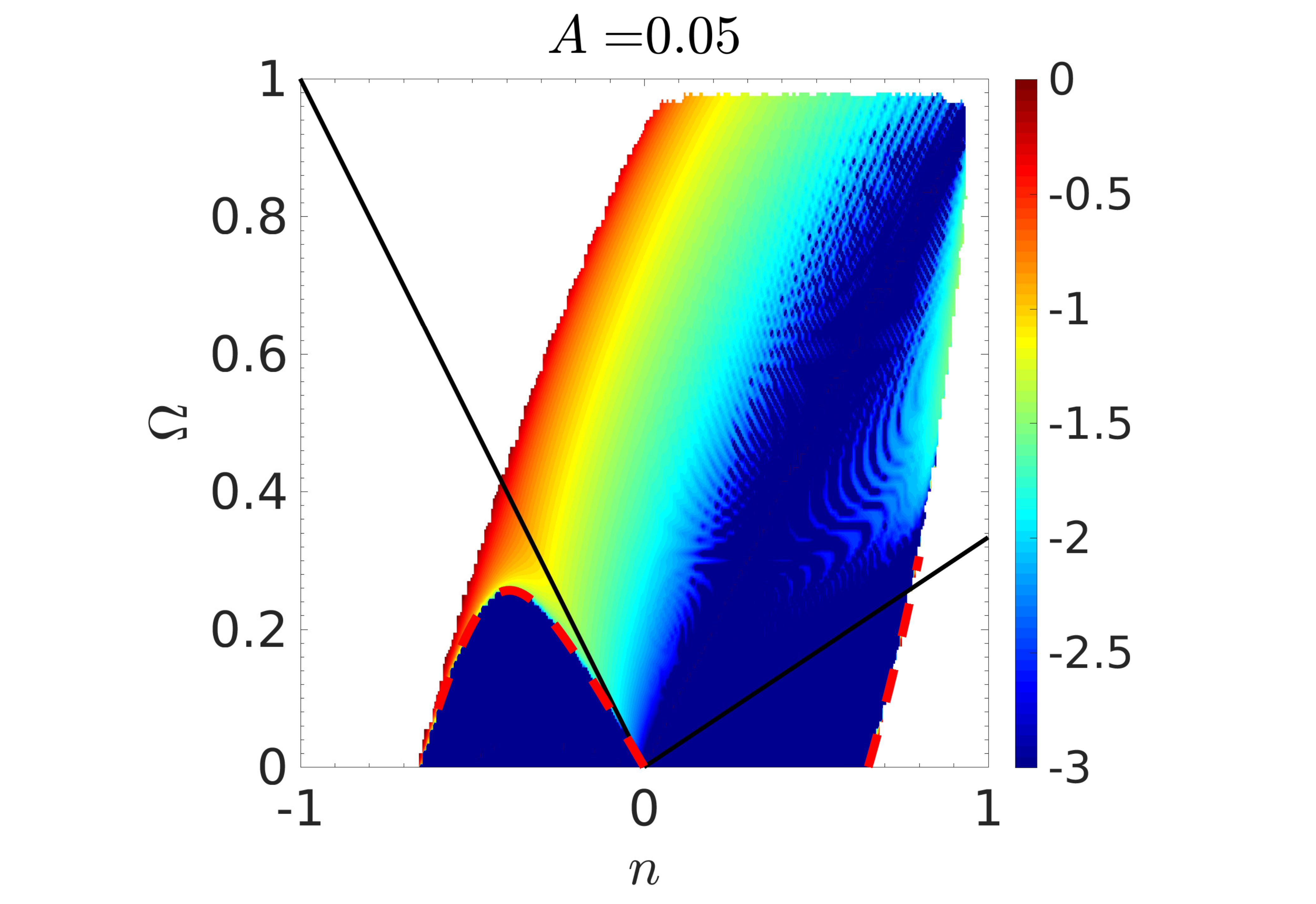} } 
      \subfigure[$\ell\rightarrow \infty$]{\includegraphics[trim=3cm 0cm 6cm 0cm, clip=true,width=0.24\textwidth]{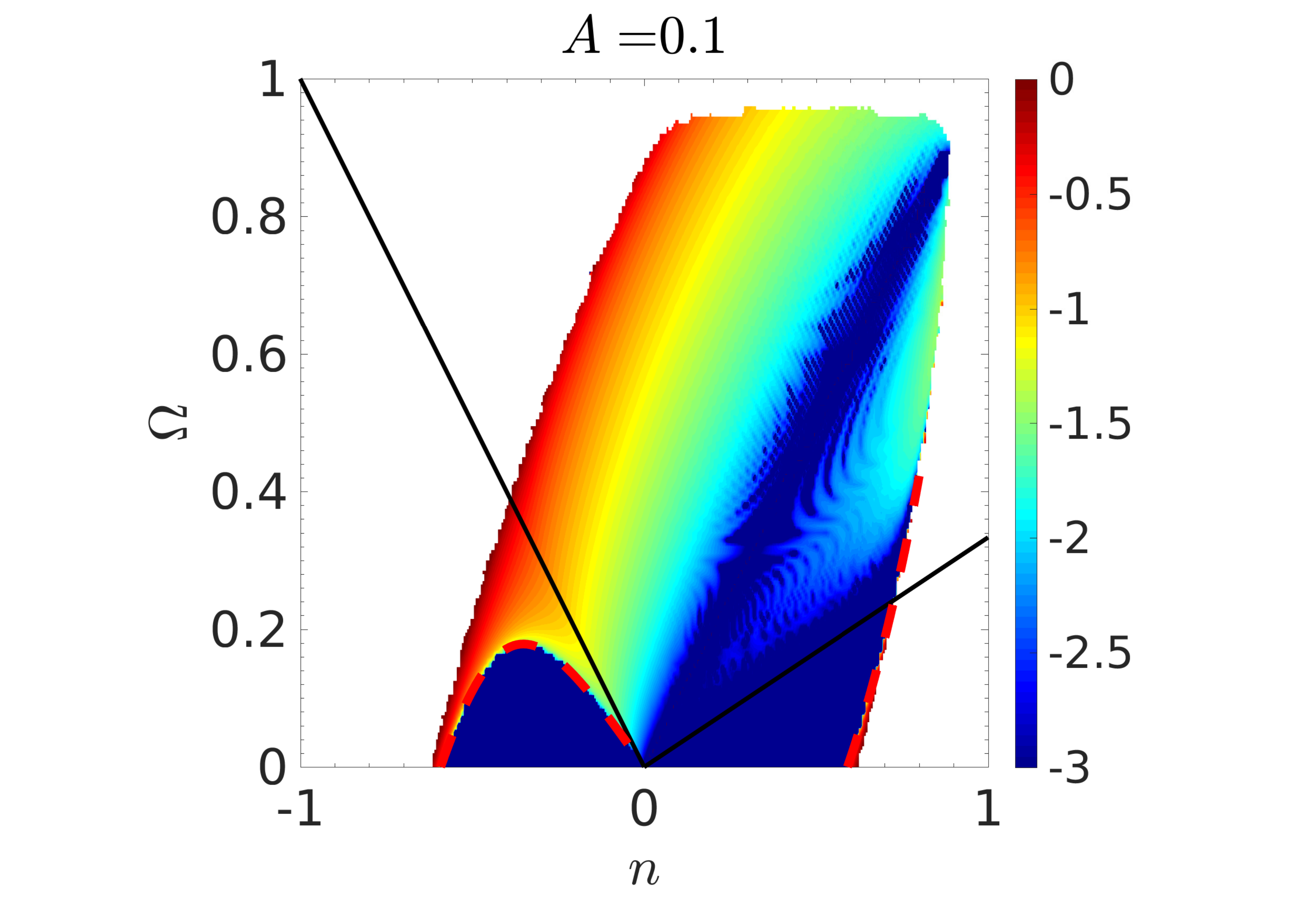} } 
      \subfigure[$\ell\rightarrow \infty$]{\includegraphics[trim=3cm 0cm 6cm 0cm, clip=true,width=0.24\textwidth]{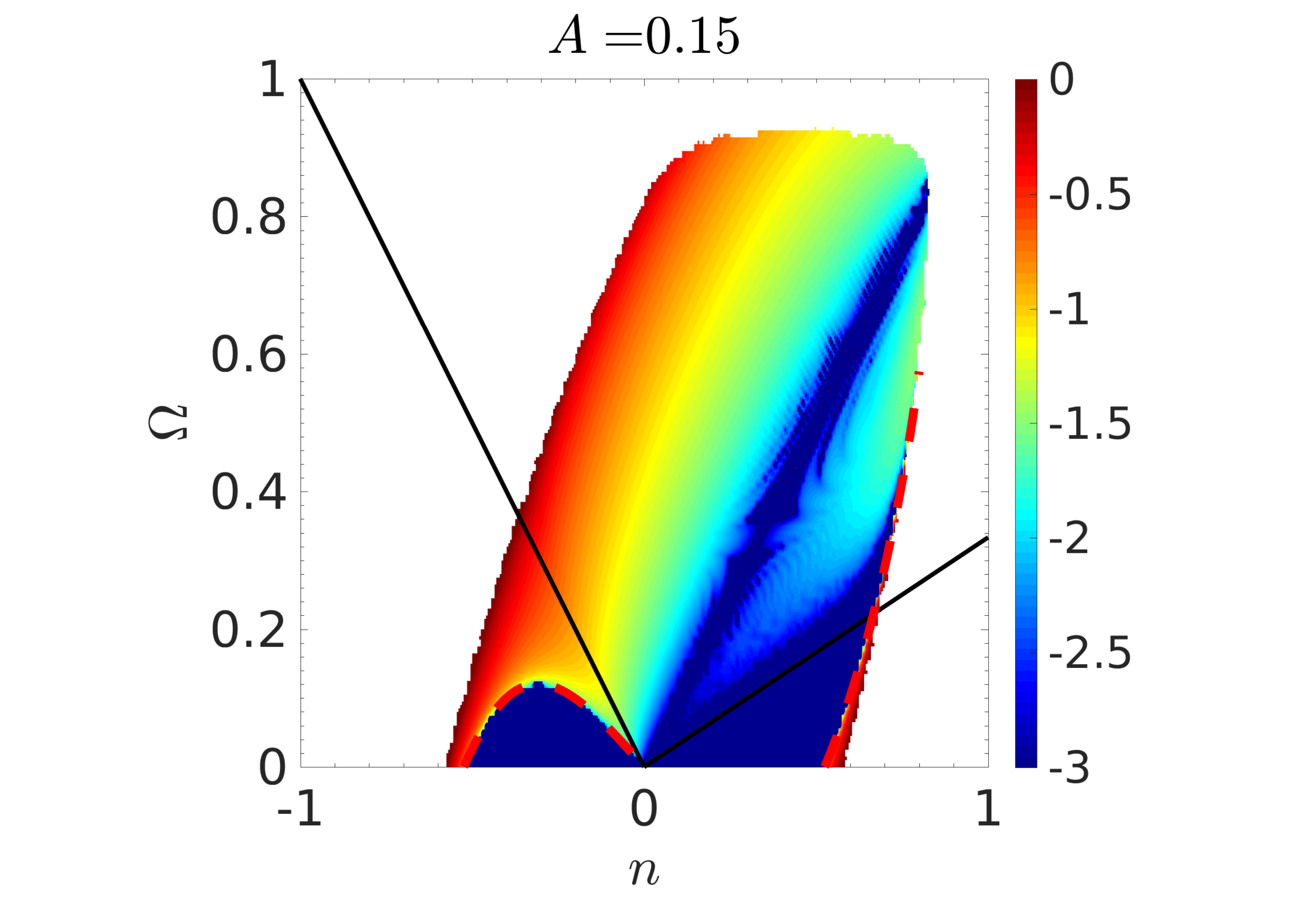} } 
       \end{center}
  \caption{To be compared with Fig.~\ref{7}, this shows results from the local WKB analysis, plotting $\log\sigma$ for the maximum growth rate on the $(n,\Omega)$-plane for several $A$ from the solution of Eqs.~\ref{WKBeq1}--\ref{WKBeq3} using a Floquet method. The usual elliptical instability of IMs is obtained above the solid black lines for $n\in[-\Omega,3\Omega]$. Instability is also observed for $n\lesssim -\Omega$ for sufficiently large $A$, and the red dashed lines represent solutions to $\Omega=-n\left(\frac{2b}{a}-1\right)$. This represents the range of parameters in which the  ``stack of pancakes"-like instability occurs, which explains the instability that is observed for $n\lesssim -\Omega$. White regions represent where the basic ellipsoid becomes undefined. (Note that the small-scale patterns result from the finite number of points for which the instability was computed.)}
  \label{D1}
\end{figure*}

In Fig.~\ref{D1}, we plot the resulting maximum growth rate for 100 points for $\Omega\in[0,1]$ and 200 points for $n\in[-1,1]$ for $A=0.025, 0.05, 0.1$ and $0.15$, respectively, for which the basic shape is ellipsoidal (these results have been interpolated and smoothed on a uniform grid with equal spacing 0.005 for the purposes of this figure). This can be compared with Fig.~\ref{7}, and shows what we would expect our global analysis to show in the limit $\ell\rightarrow \infty$.

\subsection{Explanation of the instability when $n\lesssim -\Omega$: ``stack of pancakes"-like perturbations}
\label{pancakes}
We can solve Eqs.~\ref{WKBeq1}--\ref{WKBeq3} analytically for the set of perturbations which have $k_x=k_y=0$, so that all spatial variation is along $z$ \citep{Craik1989,LL1996a}. This case corresponds with ``stack of pancakes"-like perturbations, where the solutions at each height undergo purely horizontal epicyclic oscillations independently of all other heights. Eqs.~\ref{WKBeq1}--\ref{WKBeq3} can be straightforwardly solved in this case for solutions with time-dependence $\propto \mathrm{e}^{-\mathrm{i}\hat{\omega} t}$, to obtain the complex frequency
\begin{eqnarray}
\hat{\omega} = \left(\frac{\gamma a}{b}+2n\right)^{\frac{1}{2}}\left(\frac{\gamma b}{a}+2n\right)^{\frac{1}{2}}.
\end{eqnarray}
Instability occurs when $\sigma=\mathrm{Re}\left[-\mathrm{i}\hat{\omega}\right]>0$, which occurs in the interval (cf. Eq.~5.12 in \citealt{LL1996a})
\begin{eqnarray}
\label{nOmbdry}
\frac{-\Omega}{\frac{2b}{a}-1} \leq n \leq \frac{-\Omega}{\frac{2a}{b}-1},
\end{eqnarray}
which is centred on $n=-\Omega$, with resonant width $O(\epsilon\gamma)$ in the limit $\epsilon\gamma \ll 1$. Note that the real frequency of the mode is zero in the bulge frame (which is consistent with what we have observed in our global calculations in the top panel of Fig.~\ref{8}). The lower limit to this instability band is plotted in Fig.~\ref{D1} as the red dashed lines (this is not a straight line because $a$ and $b$ are functions of $\Omega$ and $n$). This clearly demonstrates that the instability that we have observed when $\frac{n}{\Omega}\lesssim -1$ is due to the finite width of the instability centred on $n=-\Omega$ (and is not due to higher-order wave couplings for large $A$). This (and comparison of Figs.~\ref{7} and \ref{D1}) also illustrates that the region of instability for $n\lesssim -\Omega$ does not continue to grow indefinitely as $\ell$ is increased, and is in fact bounded by Eq.~\ref{nOmbdry}.

\bibliography{tid}
\bibliographystyle{mn2e}
\label{lastpage}
\end{document}